\documentclass{aa}  

\usepackage{graphicx}
\usepackage{natbib}
\bibpunct{(}{)}{,}{a}{}{;}
\usepackage{amssymb}
\usepackage{amsmath}
\usepackage{txfonts}
\begin{document}
   \title{The nature of the Class I population in Ophiuchus as revealed through gas and dust mapping}


   \author{T.A. van Kempen
          \inst{1,2}
	  \and
	  E.F. van Dishoeck\inst{1,3}
	  \and
          D.M. Salter\inst{1}
          \and
          M.R. Hogerheijde\inst{1}
          \and
          J.K. J{\o}rgensen\inst{4}
          \and
	  A.C.A. Boogert\inst{5}         
                   }
   \offprints{T. A. van Kempen\\ 
              \email{tvankempen@cfa.harvard.edu} }

   \institute{$^1$ Leiden University, Leiden Observatory, P.O. Box 9513,
              2300 RA Leiden, The Netherlands\\	
			$^2$ Center for Astrophysics, 60 Garden Street, Cambridge, MA 02138, USA      \\
              $^3$ Max-Planck Institut f\"ur Extraterrestrische 
             Physik (MPE), Giessenbachstr.\ 1, 85748 Garching, Germany \\
              $^4$ Argelander-Institut f{\"u}r Astronomie, University of Bonn, Auf dem H{\"u}gel 71, 53121, Bonn, Germany \\
              $^5$ Infrared Processing and Analysis Center (IPAC), NASA Herschel Science Center, Mail Code 100-22, California Institute of Technology, Pasadena, CA 91125, USA\\
             }

   \date{Received June 21st 2008}
 \titlerunning{The Class I population in the Ophiuchus clouds}
 
  \abstract 
{}
 {Our aim is to characterize the structure of protostellar envelopes
 on an individual basis and to correctly identify the embedded YSO
 population of L~1688.}
{ Spectral maps of the HCO$^+$ $J=$4--3 and C$^{18}$O $J=$3--2 lines
using the HARP-B array on the James Clerk Maxwell Telescope and SCUBA
850 $\mu$m dust maps are obtained of all sources in the L~1688 region
with infrared spectral slopes consistent with, or close to, that of
embedded YSOs. Selected 350 $\mu$m maps obtained with the Caltech
Submillimeter Observatory are presented as well. The properties,
extent and variation of dense gas, column density and dust 
up to 1$'$ ($\sim$ 7,500 AU) are probed at 15$''$ resolution. Using
the spatial variation of the gas and dust, together with the intensity
of the HCO$^+$ $J=$4--3 line,
we are  able to
accurately identify the truly embedded YSOs and determine their
properties. }
{The protostellar envelopes range from 0.05 to 0.5 M$_{\rm{\odot}}$ in
 mass. The concentration of HCO$^+$ emission ($\sim$ 0.5 to 0.9) is
 generally higher than that of the dust concentration. Combined with
 absolute intensities, HCO$^+$ proves to be a better tracer of
 protostellar envelopes than dust, which can contain disk and cloud
 contributions. Our total sample of 45 sources, including all
 previously classified Class I sources, several flat-spectrum sources
 and some known disks, was re-classified using the molecular
 emission. Of these, only 17 sources are definitely embedded YSOs.
 Four of these embedded YSOs have little (0.1--0.2 M$_{\rm{\odot}}$)
 envelope material remaining and are likely at the interesting
 transitional stage from embedded YSO to T Tauri star. About half of the flat-spectrum sources are found to be  embedded YSOs and about half are disks.}
   {The presented classification method is successful in separating
embedded YSOs from edge-on disks and confused sources. The total
embedded population of the Ophiuchus L~1688 cloud is found almost
exclusively in Oph-A, Oph-B2 and the Ophiuchus ridge with only three
embedded YSOs not related to these regions. 
The detailed characterization presented will be necessary to interpret
deep interferometric ALMA and future Herschel observations. }


   \maketitle
%
\section{Introduction}
Low-mass young stellar objects (YSOs) have traditionally been
classified using their observed infrared (IR) spectral slope,
$\alpha_{\rm{IR}}$, from 2 to $\sim$20 $\mu$m \citep{Lada84,Adams87} or
their bolometric temperature, $T_{\rm{bol}}$ \citep{Myers93}. 
Together with the subsequent
discovery of the Class 0 stage \citep{Andre93} this
led \citet{Greene94} 
to identify 5 classes of YSOs:
\begin{itemize}
\item Class 0 , no $\alpha_{2-14}$, high $L_{\rm{sub-mm}}$/$L_{\rm{bol}}$
\item Class I , $\alpha_{2-14}$ $>$ 0.3,  $T_{\rm{bol}} <$ 650 K 
\item Flat Spectrum, -0.3 $<  \alpha_{2-14}$ $<$ 0.3, $T_{\rm{bol}}$ $\sim$ 400-800 K
\item Class II , -2 $< \alpha_{2-14}$ $<$ -0.3, 650$<$ $T_{\rm{bol}} <$ 2800 K 
\item Class III , $\alpha_{2-14} <$ -2, $T_{\rm{bol}} >$ 2800 K 
\end{itemize}
Each class is thought to be represent a different category, and
probably evolutionary stage, of YSOs. Class 0 sources are the
earliest, deeply embedded YSO stage; Class I sources are thought to be
more evolved embedded YSOs, Class II the T Tauri stars with gas-rich
circumstellar disks and Class III the pre-main sequence stars
surrounded by tenuous or debris disks.  Deep mid-IR photometry
introduced the 'flat-spectrum' (FS) sources \citep[e.g.][]{Greene94},
with IR spectral slope close to 0 and which may represent a stage
intermediate between Class I and II.
An accurate classification and physical characterization of YSOs is
important for constraining the time scales of each of the phases and
for determining the processes through which an object transitions from
one phase to the next. In this study, we focus on the embedded YSO
population and their transition to the T Tauri phase. \\

The $\rho$ Ophiuchus molecular clouds, part of the Gould Belt, are
some of the nearest star-forming regions and contain many Class I and
II sources.  Consisting of two main clouds, L~1688 and L~1689, the star
formation history and protostellar population of these regions have
been studied extensively. Although the distance to Ophiuchus has long
been debated \citep{Knude98}, recent work constrains it to 120$\pm$4 pc
for L~1688 \citep{Loinard08}. 
The large-scale structure of the Ophiuchus clouds at millimeter
wavelengths was first mapped by \citet{Loren89} using the $^{13}$CO
molecular line emission with 2.4$'$ resolution. It was found that much
of the cloud is in filamentary structures, but also that a diffuse
foreground layer is present in Ophiuchus, resulting in a higher
average extinction toward YSOs than in other clouds such as Taurus
\citep{Dickman90}.  
Subsequent continuum observations at millimeter (mm) wavelengths
mapped most of the large-scale structure present  in
detail, distinguishing the Oph-A through Oph-F regions within L~1688
\citep{Mezger92,Motte98,Johnstone00,Stanke06,Young06}.  \\

\begin{figure*}[!htp]
\begin{center}
\includegraphics[angle=90,width=470pt]{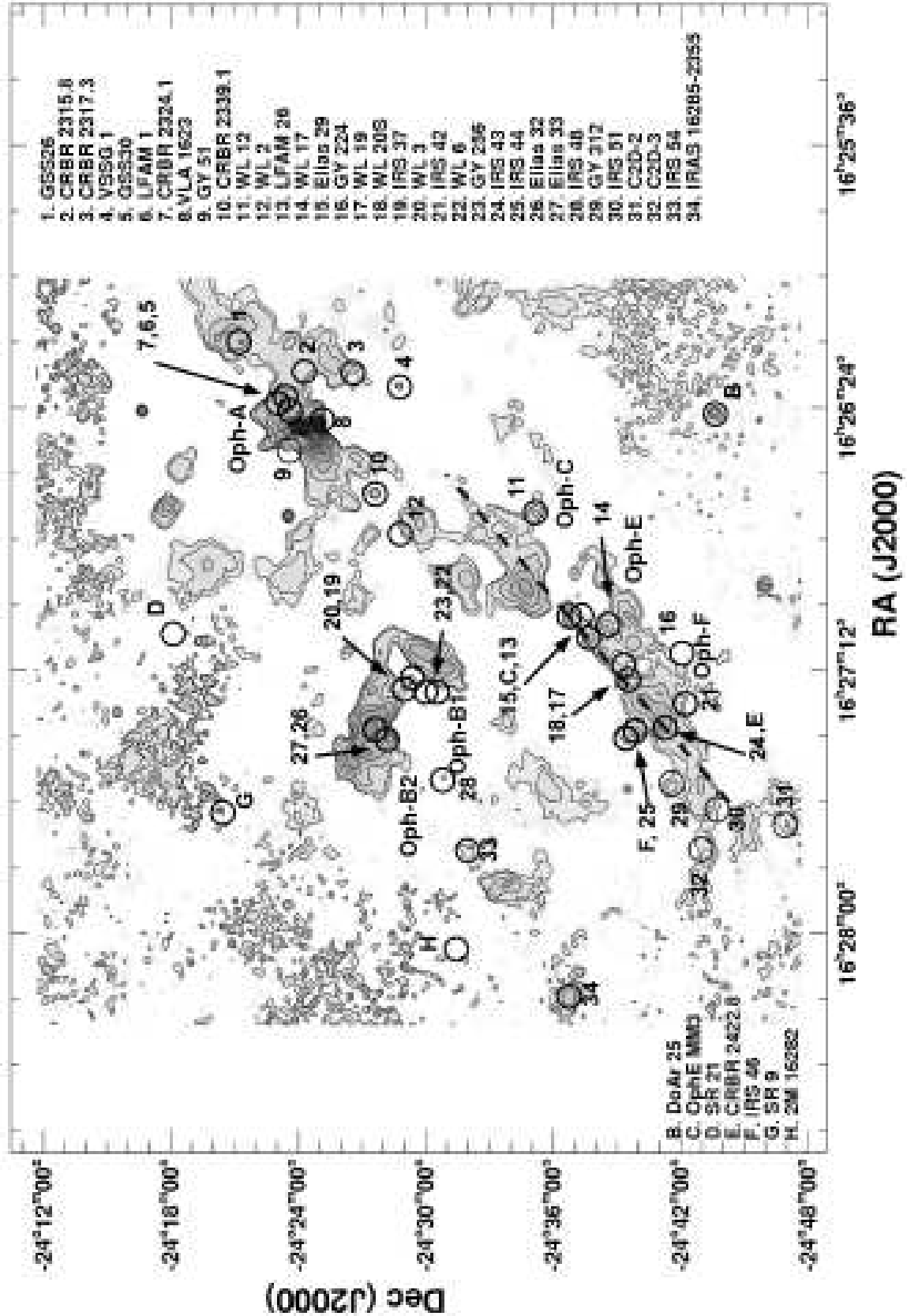}
\end{center}
\label{fig:complete}
\caption{The L~1688 core in Ophiuchus. In grey-scale and contours the
850 $\mu$m SCUBA map as published by \citet{Johnstone00} and
\citet{diFrancesco08} is shown. The locations of all the sources as
observed in this study are shown, except for C2D-162527.6, Haro 1-4,
C2D-162748.2 and IRS 63. C2D-16274.1 is marked as C2D-2 and
C2D-162748.2 as C2D-3. The dashed line indicates the Oph ridge. }
\end{figure*}

\begin{table*}[htp]
\caption{Sample of embedded sources in L~1688, with additional known disks included.}
\small
\begin{center}
\begin{tabular}{l l l l c r}
\hline \hline 
Source  	& Other names	&\multicolumn{2}{c}{Coordinates (J2000)}  & Ref.$^a$ & $\alpha_{\rm{2-24 \mu m}}$$^b$\\
	        &  &	RA	&	Dec  \\\hline 
C2D-162527.6& SSTc2d J162527.6-243648  &16:25:27.6	&	-24:36:48.4& 2&0.36\\
GSS 26	        &              &	16:26:10.4	&	-24:20:58& 1&-0.46\\
CRBR 2315.8-1700&       	 &	16:26:17.2	&	-24:23:45.1& 2&0.69 \\
CRBR 2317.3-1925.3&	SKS 1-10 &	16:26:18.8	&	-24:26:13& 1 &-0.56\\
VSSG 1	        & Elias 20	 &      16:26:18.9	&	-24:28:22& 1 &-0.73\\
GSS 30	& Elias 21/GY 6/GSS 30-IRS1   & 16:26:21.4	&	-24:23:04.1& 2 &1.46\\
LFAM 1	        & GSS 30-IRS3	 &      16:26:21.7      &       -24 22 51.4& 2 &0.73\\
CRBR 2324.1-1619&              &	16:26:25.5	&	-24:23:01.6& 2&0.87\\
VLA 1623        & VLA 1623.4-2418&      16:26:26.4	&	-24:24:30.3& 2& no\\
GY 51            & VSSG 27	&       16:26:30.5	&	-24:22:59  & 1&0.05\\
CRBR 2339.1-2032 &          GY 91&	16:26:40.5	&       -24:27:14.3& 2&0.45\\
WL 12	        &  GY 111	 &      16:26:44.0	&	-24:34:48& 1 &2.49\\
WL 2            & GY 128         &      16:26:48.6	&	-24:28:39 &  1&0.02\\
LFAM 26&         CRBR 2403.7/GY 197    &16:27:05.3      &       -24:36:29.8& 2&1.27\\
WL 17	        &GY 205	         &      16:27:07.0	&       -24:38:16.0& 1&0.61\\
Elias 29        & WL15/GY 214	 &      16:27:09.6	&       -24:37:21.0& 1&1.69 \\
GY 224	        &	         &      16:27:11.4	&	-24:40:46  & 1 &-0.05\\
WL 19            &GY 227           & 	16:27:11.9	&	-24:38:31.0& 1 &-0.43\\
WL 20S	        &   GY 240	 &      16:27:15.9	&	-24:38:46  & 1 &2.75\\
IRS 37	        & GY 244	 &      16:27:17.6	&	-24:28:58  & 1 &0.25\\
WL 3            & GY 249            &	16:27:19.3	&	-24:28:45  & 1 &-0.03\\
IRS 42	        & GY 252	 &      16:27:21.6	&	-24:41:42  & 1 &-0.03\\
WL 6	        & GY 254	 &      16:27:21.8	&	-24:29:55  & 1 &0.72\\
GY 256	        &       &	        16:27:22.0	&	-24:29:39.9& 2 &-0.05\\
IRS 43	        & GY 265	 &      16:27:27.1	&	-24:40:51  & 1 &1.17\\
IRS 44	        & GY 269	 &      16:27:28.3	&	-24:39:33.0& 1&2.29\\
Elias 32        & IRS 45/GY 273/VSSG 18&16:27:28.6	&       -24:27:19.8& 2 &-0.03\\
Elias 33        & IRS 47/GY 279/VSSG 17&16:27:30.1	&       -24:27:43  & 1 &-0.12\\
IRS 48	        & GY 304	&       16:27:37.2	&	-24:30:34  & 1 &0.88\\
GY 312          &       	&	16:27:38.9	&	-24:40:20.5& 2&0.64\\
IRS 51	        &  GY3 15	 &      16:27:40.0	&	-24:43:13  & 1 &-0.15\\
C2D-162741.6  & SSTc2d J162741.6-244645&16:27:41.6	&	-24:46:44.6& 2&0.32\\
C2D-162748.2 &SSTc2d J162748.2-244225 &	16:27:48.2	&	-24:42:35.6& 2&1.55\\
IRS 54	        & GY 378	 &      16:27:51.7	&	-24:31:46.0& 1& 0.03\\
\multicolumn{2}{l}{IRAS 16285-2355}&    16:28:21.6     &       -24:36:23.7& 2& 1.23\\
C2D-162857.9 &  SSTc2d J162857.9-244055&16:28:57.9	&       -24:40:54.9& 2& 0.67\\
IRS 63	        &	         &      16:31:35.7	&	-24:01:29.5& 2 &0.14\\   \hline 
\multicolumn{4}{c}{\textbf{ Known Disks}}  \\ \hline
Haro 1-4        &	DoAr 16	&       16:25:10.5	&	-23:19:14.5& 2 &-0.89\\
DoAR 25	        & GY 17	        &       16:26:24.0	&       -24:43:09.0& 1 &-1.12\\
OphE MM3        &		 &      16:27:05.9	&	-24:37:08.2& 2 &-0.33\\
SR 21           &	Elias 30&	16:27:10.2	&	-24:19:16.0  & 1 &-0.79\\
CRBR 2422.8-3423.8&	CRBR 85  &	16:27:24.8	&	-24:41:03.0  & 1 &1.01\\
IRS 46	        &  GY 274         &	16:27:29.7	&	-24:39:16.0  & 1 &0.18\\
SR 9            &IRS 52/Elias 34&	16:27:40.5	&	-24:22:07.0  & 1 &-1.07\\
2MASS 16282     &		 &      16:28:13.7	&	-24:31:39.0  & 2 &-1.55\\\hline
\end{tabular} \\
\label{label:source}  
\end{center}
$^a$  References for positions : 1:\citet{Bontemps01}, 2: c2d delivery document\\
$^b$ From \citet{Evans08}. See also $\S$ 5.5
 
\end{table*}

The YSO population was first identified by
\citet{Elias78,Wilking83,Wilking89,Comeron93} and \citet{Greene94}
using IR observations.  With the arrival of (sub-)mm telescopes,
VLA~1623 in L~1688 was identified as the first deeply embedded YSO
\citep{Wootten89,Loren90,Andre93}. In more recent years, a large
population of Class I and Class II sources has  been found based
on their IR spectral slopes, using space-based observatories such as
the {\it Infrared Space Observatory} (ISO)
\citep[e.g.][]{Liseau99,Bontemps01} and ground-based IR
\citep[e.g.][]{Barsony97,Barsony05}. Most embedded sources in
L~1688 are clustered around the filaments of the Oph-A, Oph-B2, Oph-E
and Oph-F regions, while the Oph-C region only shows a single embedded
source and Oph-D does not have any embedded YSO \citep{Motte98}. In
Oph-E and Oph-F most sources are lined up along a relatively small
filament of material: for the purpose of this paper, we will adopt the
name `Ophiuchus ridge' for this region. \\

With the launch of the {\it Spitzer Space Telescope}, the Ophiuchus cloud
was included in the guaranteed time (GTO) and the `cores to disks'
(c2d) Legacy program \citep{Evans03}.  \citet{Padgett08} report on the
results at 24, 70 and 160 $\mu$m using the MIPS instrument, revealing
the emission from the large-scale structure at mid and far-IR
wavelengths. \citet{Jorgensen08} compared the results from the c2d
program with the COMPLETE 850 $\mu$m SCUBA sub-millimeter dust mapping
from \citet{Johnstone00} and \citet{Ridge06} to determine the association of YSOs
with dense cores. 

 The stellar ages of the Class II and III sources in Ophiuchus were
 found to be 0.1--1 Myr based on stellar spectroscopy compared with
 evolutionary tracks, indicating a relatively young age for the total
 cloud \citep{Greene95,Luhman99}. The Star-Formation Efficiency (SFE)
 was recently calculated with Spitzer and SCUBA photometry 
 to be of the order of 13\% within the cores and 4\% in the cloud
as found by \citet{Evans08} and \citet{Jorgensen08}, lower than
previous determinations
\citep{Wilking83}. 
\\

The relative timescales of the different phases are determined by the number of objects in each class of YSOs \citep[e.g.][]{Evans08}. Recent high resolution ground-based (near)-IR imaging show that some
of the Class I sources in Ophiuchus are physically different from an
embedded YSO,  confusing these timescales determinations.  For example, the Class I source CRBR 2422.8-3423 was
found to be an extincted edge-on disk from near-IR imaging
\citep{Brandner00,Pontoppidan05}. The source OphE MM3, classified as a
starless core by \citet{Motte98}, was also shown to be a edge-on disk
in the same study \citep{Brandner00}. The source IRS 46 has no
associated protostellar envelope and was re-classified based on {\it
Spitzer} and sub-mm data as a disk \citep{Lahuis06} . Much of the reddening seen
in the IR originates from the nearby envelope associated with IRS 44.
\\

Foreground material can also heavily influence the identification and
subsequent analysis of embedded sources \citep[e.g.,][]{Luhman99}. An
excellent example is provided by the Class I source Elias 29 in L~1688,
which has two foreground layers in addition to the ridge of
material in which the YSO is embedded \citep{Boogert02}. Only
a combination of molecular line emission at sub-millimeter and IR 
spectroscopy
could constrain the
protostellar envelope as well as the immediate environment around it
\citep{Boogert00, Boogert02}.  
Indeed, IR spectroscopy can be used as a complementary diagnostic
and spectra of many of the YSOs in Ophiuchus have been taken,
using ISO, {\it Spitzer} or groud-based telescopes
\citep[e.g.][]{Alexander03,Pontoppidan03,Boogert08}. 
Ice absorption features such as the
3 $\mu$m H$_2$O and 15.2 $\mu$m CO$_2$ bands are usually associated with
embedded sources whereas silicate emission at 10 and 20 $\mu$m is characteristic of 
Class II sources, but foreground absorption and edge-on disks
can confuse this classification \citep{Boogert02,Pontoppidan05}.
\begin{figure}[!htp]
\begin{center}
\includegraphics[width=250pt]{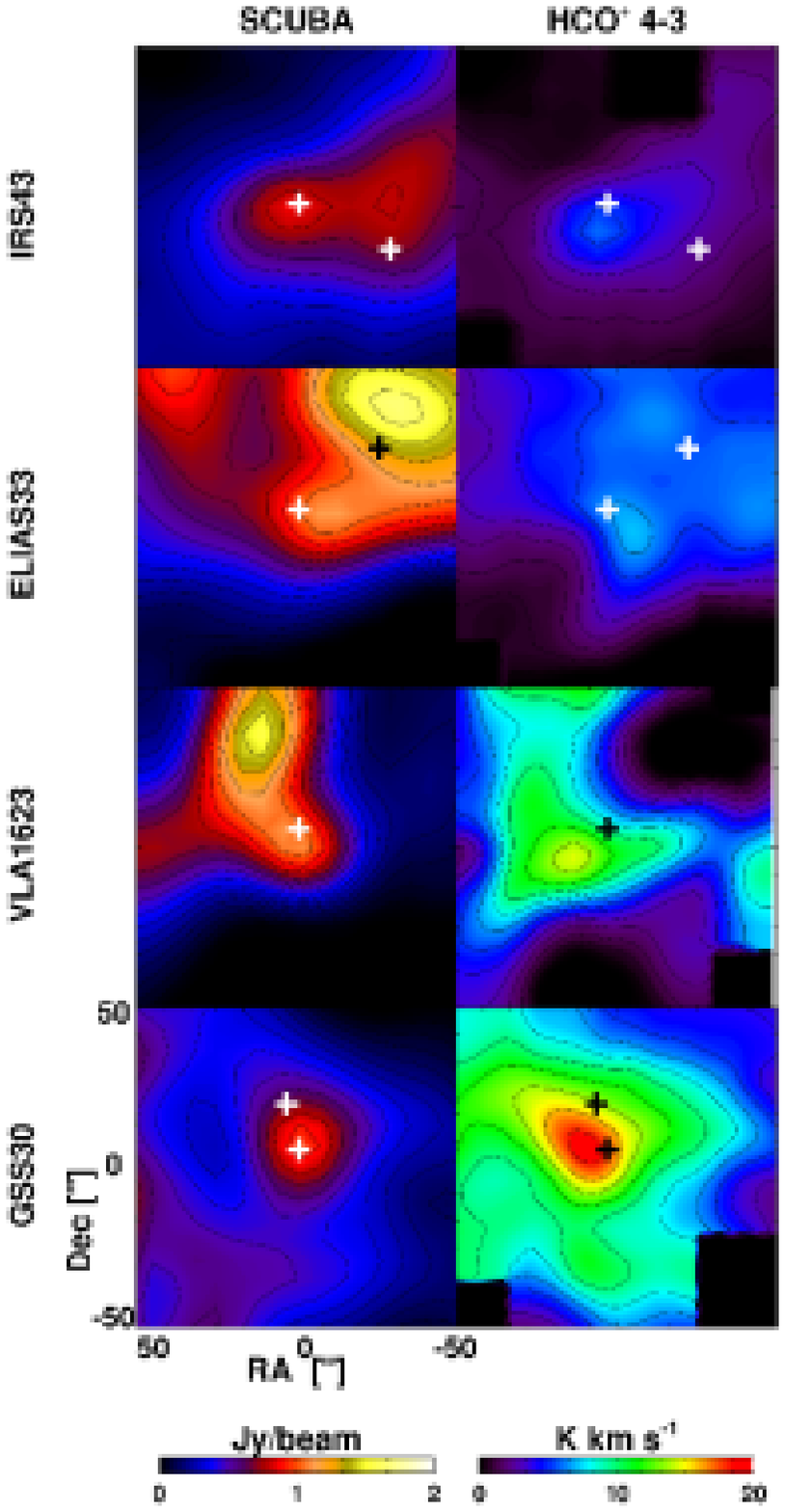}
\end{center}
\caption{SCUBA 850 $\mu$m continuum ({\it left}) and HCO$^+$ 4--3
({\it right}) integrated intensity maps plotted from bottom to top for GSS 30 IRS1, VLA 1623, Elias 33 and IRS 43. The IR
source positions are marked in the SCUBA maps with white crosses. Note
that the absolute scale of these maps is a factor 4 higher than those
in Fig. 3 and 4. The 850 $\mu$m continuum flux for VLA 1623 has been
scaled down by a factor 4. In the map of GSS 30, LFAM 1 is located
north of the GSS 30 IRS1 position. Elias 32 is located at the north-west
side of the Elias 33 map. CRBR 2422.8-3423 is located to the
south-west of IRS 43. All four main targets show strong HCO$^+$ and SCUBA
peaking within 5$''$ of  the IR position at (0,0), characteristic of truly embedded
sources.}
\label{Fig:harp1}
\end{figure}

In recent years, several detailed modelling efforts have been carried
out to study the relations between the observed spectral energy
distribution (SED) and the physical structure of embedded YSOs
\citep[e.g.][]{Jorgensen02,Whitney03a,Schoeier04,Young04,Robitaille06,Robitaille07,Crapsi08}. 
\citet{Whitney03a} show that it is possible for embedded YSOs with a
face-on projection to be classified as Class II. Due to their
orientation, these sources are viewed straight down the outflow cone,
directly onto the central star and disk system. \citet{Crapsi08} show
that a significant fraction of the Class I sources may be edge-on flaring disks,
that have already lost their protostellar envelope. The spectral slope
is much steeper than expected due to their structure.

A prime characteristic and component of embedded YSOs is the presence
of dense centrally condensed envelopes.  While dust maps at
sub-millimeter wavelengths have become very popular to trace the early
stages of star formation \citep[e.g.][]{Motte98,Shirley00,
Johnstone00,Stanke06}, the continuum emission at these wavelengths is
dominated by the cold outer envelope and cloud material, with disks
starting to contribute as the envelope disperses
\citep[e.g.][]{Hogerheijde97,Looney00,Young03}. Single-dish dust continuum
data by themselves are not able to distinguish between dense cores and
envelope or foreground material, nor quantify any disk
contributions. However, the dense gas ($\sim$10$^6$ cm$^{-3}$) located
in the inner regions of protostellar envelopes is uniquely probed by
molecular lines with high critical densities at sub-millimeter
wavelengths. Observations of deeply embedded Class 0 YSOs have indeed
revealed strong sub-millimeter lines of various molecules
\citep[e.g.][]{Blake94, Blake95,Schoeier02,
Jorgensen04,Maret04,Maret05}, but only a few studies have been carried
out on more evolved Class I embedded YSOs
\citep[e.g.][]{Hogerheijde97}.
 
A good high density tracer is the HCO$^+$ molecule, for which the
$J=$4--3 line both has a high critical density of $>$10$^6$
cm$^{-3}$ and is accessible from the ground.  
Dense gas is also
found in the circumstellar disk on scales of a few tens to hundreds
AU, but such regions are generally diluted by an order of magnitude in
single-dish observations. 
In contrast, molecular lines with much lower critical densities, such
as the low excitation C$^{18}$O transitions, contain much higher
contributions from low density material. This makes these lines
well-suited as column density tracers for large-scale cloud material
and the cold outer regions of the protostellar envelope
\citep[e.g.][]{Jorgensen02}. Both the HCO$^+$ and C$^{18}$O data have
velocity resolutions of 0.1 km s$^{-1}$ or better, thus allowing
foreground clouds to be identified. \\

Most sub-millimeter line data so far have been single pixel spectra
toward the YSO with at best a few positions around specific YSOs. The
recently commissioned HARP-B instrument is a 16-pixel receiver, operating in the 320
to 370 GHz atmospheric window   
allowing rapid mapping of small (2$'$) regions \citep{Smith00}.  HARP-B is mounted on the James Clerk
Maxwell Telescope (JCMT)\footnote{The James Clerk Maxwell
  Telescope is operated by The Joint Astronomy Centre on behalf of the
  Science and Technology Facilities Council of the United Kingdom, the
  Netherlands Organisation for Scientific Research, and the National
  Research Council of Canada.}.\\

We
present here HARP-B maps of all Class I sources in the L~1688 region in
C$^{18}$O 3--2 and HCO$^{+}$ 4--3. The combination of these two
molecular lines allows us to differentiate between protostellar envelopes,
dense cores and foreground cloud material, as well as edge-on
disks. The goal of this paper is to characterize the envelopes of the
embedded source population of L~1688, as well as present a new method
for identifying truly embedded sources and separate them from
(obscured) edge-on disks using the dense gas present in embedded YSOs.
In $\S$ 2, a sample of Class I sources in the L~1688 core is
selected. In $\S$ 3, we discuss the details of the heterodyne
observations carried out at the JCMT and the Atacama Pathfinder
EXperiment (APEX)\footnote{This publication is based on data acquired with the Atacama Pathfinder Experiment (APEX). APEX is a collaboration between the Max-Planck-Institut fur Radioastronomie, the European Southern Observatory, and the Onsala Space Observatory.}, as well as the supplementary observations obtained
in a continuum i.e. wideband mode.  $\S$ 4 presents the maps and spectra and in
$\S$ 5 we analyze the properties of the gas and dust of the sources in
the sample. The environment around the YSOs, column density, envelope
gas and concentration of the HCO$^+$ are discussed.
In $\S$ 6, we present a new method for identifying embedded YSOs from
(edge-on) disks and apply this method to the sample.  This
classification is then compared to traditional methods as well as
other recently proposed methods. The main conclusions of the paper are
given in $\S$ 7.

\begin{figure*}[!htp]
\begin{center}
\includegraphics[width=330pt]{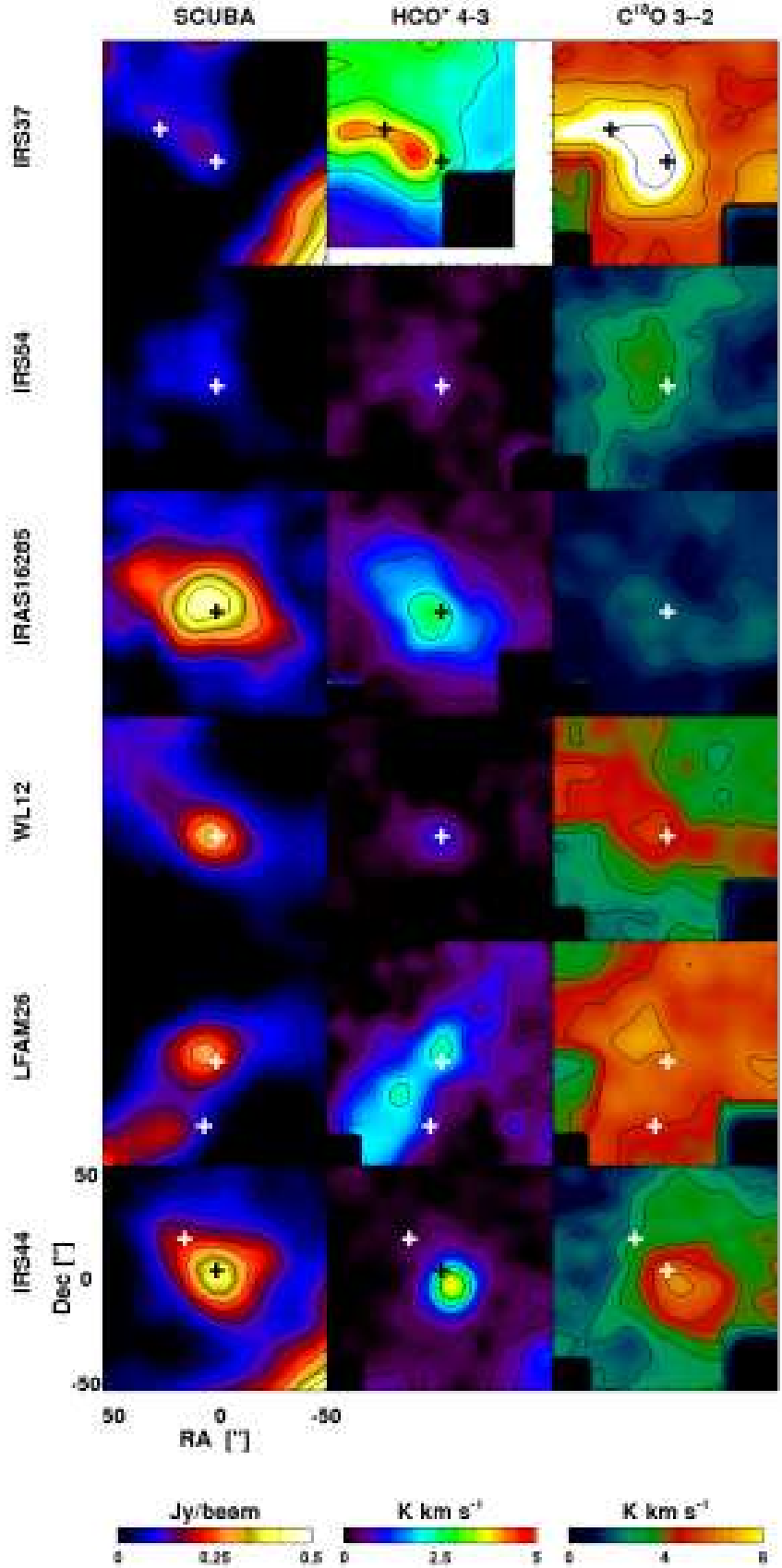}
\end{center}
\caption{SCUBA 850 $\mu$m continuum ({\it left}), HCO$^+$ 4--3 ({\it
middle}) and C$^{18}$O 3--2 ({\it right}) integrated intensity maps
plotted for (from {\it bottom} to {\it top}) IRS 44, LFAM 26, WL 12,
IRAS 16285, IRS 54 and IRS 37. The locations of the IR sources are
indicated in the SCUBA maps with white crosses. In the IRS 44 map, IRS
44 is indicated with a black cross, and IRS 46 with a white one. In
the LFAM 26 field, OphE MM3 is located south of LFAM 26. In the IRS 37
field, WL 3 is located west of IRS 37. The sources at (0,0) in this
figure show weaker HCO$^+$ images than in Fig. \ref{Fig:harp1}, but
are still peaking on the IR positions in both SCUBA and HCO$^+$, and
are thus embedded YSOs according to our classification.}
\label{Fig:harp2}
\end{figure*}

\begin{figure*}[!htp]
\begin{center}
\includegraphics[width=330pt]{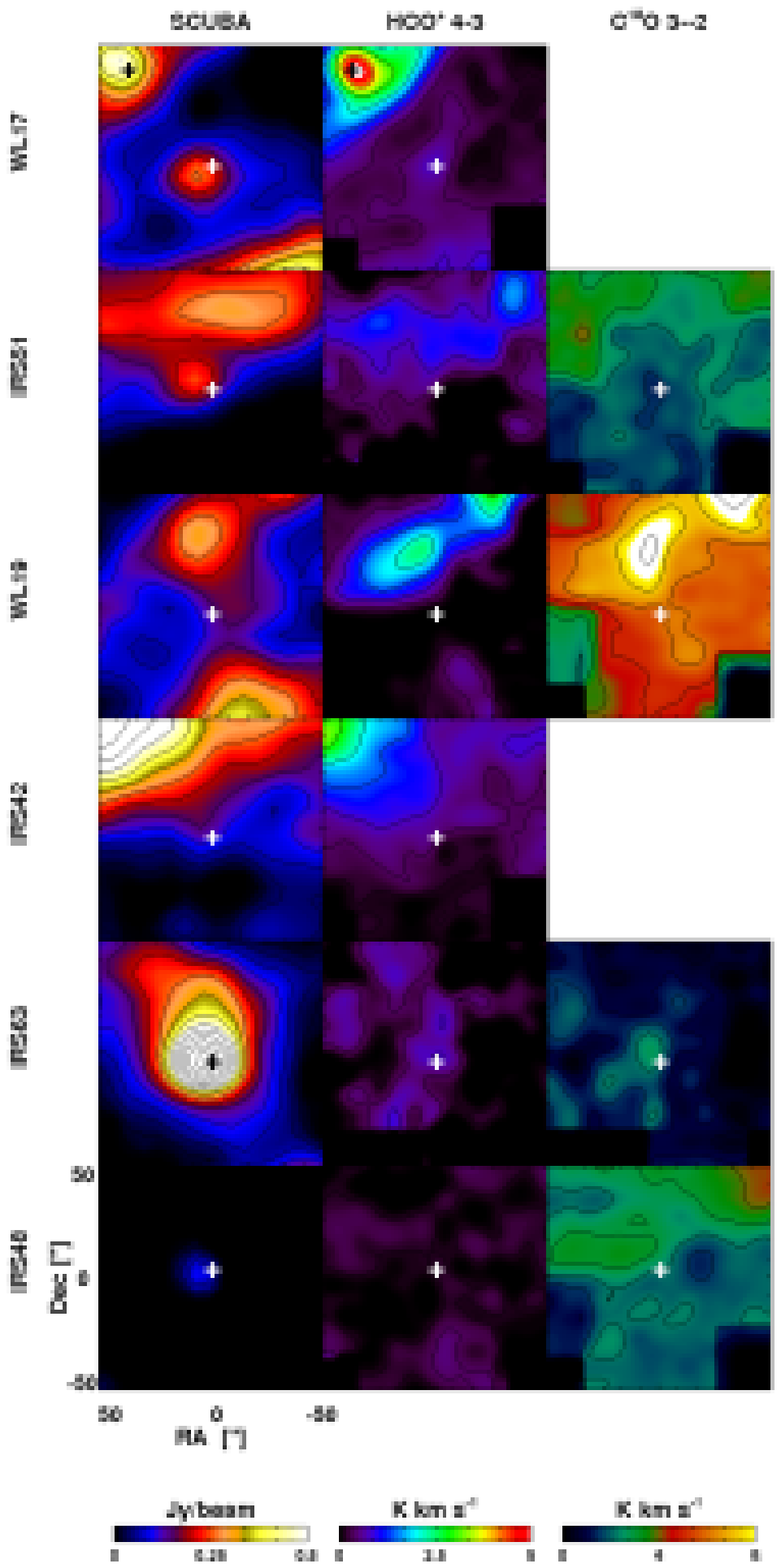}
\end{center}
\caption{SCUBA 850 $\mu$m continuum ({\it left}), HCO$^+$ 4--3 ({\it
middle}) and C$^{18}$O 3--2 ({\it right}) plotted for (from {\it
bottom} to {\it top}) IRS 48, IRS 63, IRS 42, WL 19, IRS 51 and
WL17. The locations of the IR sources are indicated in the SCUBA maps
with white crosses. The embedded source Elias 29 is shown with a black
cross in the field of WL 17. The peak 850 $\mu$m flux density for IRS
63 is 1.1 Jy beam$^{-1}$; it was not scaled to accentuate the
extended dust emission. The sources at (0,0) in this image are mostly
unresolved in SCUBA and show little or no HCO$^+$ emission related to
the source. They are either transitional or confused sources or
(edge-on) disks in our new classidfication, with the exception of IRS 63, which is classified as embedded (transitional).}
\label{Fig:harp3}
\end{figure*}

\section{Sample Selection}

Of the known (embedded) YSO population within L~1688, 45 objects were
selected for our sample using several criteria. First, we require all
sources be located within the 850 $\mu$m dust continuum map made by the
COMPLETE project of L~1688 using SCUBA on the JCMT
\citep{Johnstone00}. 

Second, sources must be included in the area covered by the
c2d program using IRAC and MIPS on {\it Spitzer}
\citep{Evans03,Padgett08}.  All sources
classified as Class I in either \citet{Andre94}, \citet{Barsony97},
\citet{Bontemps01} or the c2d delivery document
\citep{Evans07}\footnote{
http://ssc.spitzer.caltech.edu/} are included with luminosities $>$ 0.04 L$_\odot$.  Although these 41 objects have been
classified as Class I in one or more of these papers, only 6 of these 
have been classified consistently as Class I in all studies. Most
other sources are classified as either Class II or Flat spectrum
sources at least once.
All such sources are included in the analysis of this paper, but
conversely, our sample does not include all Flat spectrum or Class II
sources listed in the c2d survey. Known edge-on disks such as 2MASS
16282, IRS 46, OphE MM3 and CRBR 2422.2-3423, are among these 41
sources and are included within the sample to illustrate the results
of our method for such sources. The sample should not contain any
reddened background main-sequence stars which are readily identified
in the c2d analysis. However, other background sources, such
as AGB stars or background infrared galaxies, may be present.

\begin{figure}[!htp]
\begin{center}
\includegraphics[width=200pt]{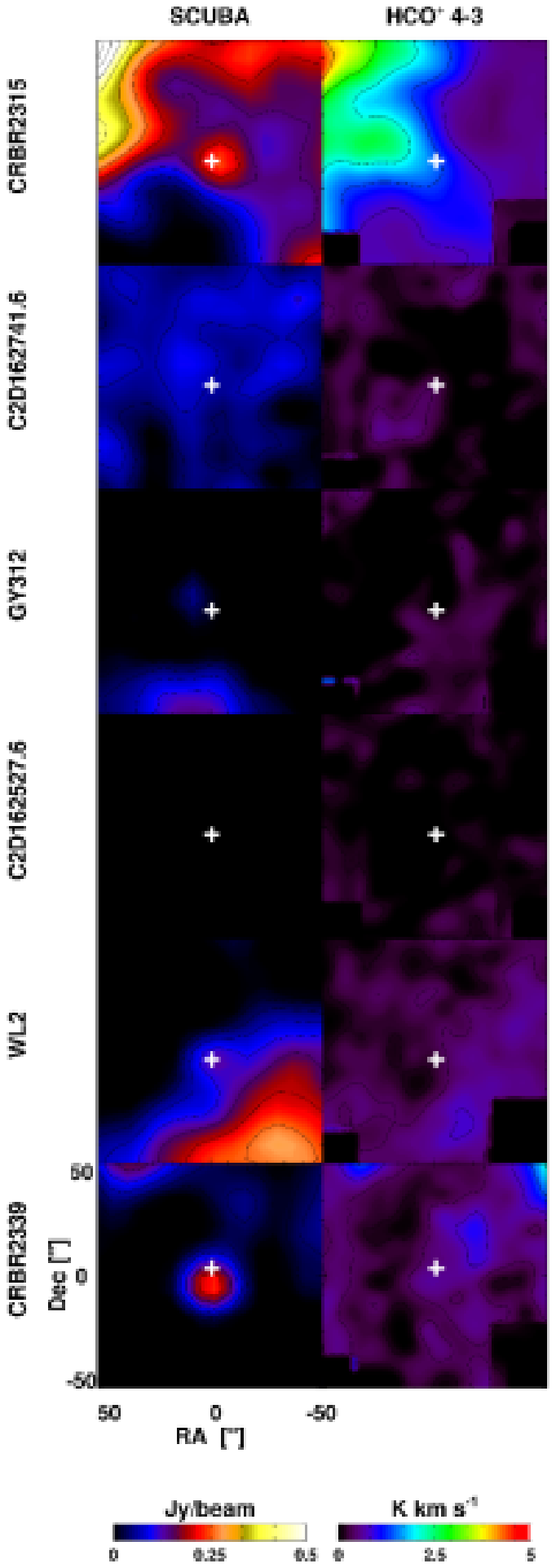}
\end{center}
\caption{SCUBA 850 $\mu$m continuum ({\it left}) and HCO$^+$ 4--3
({\it right}) plotted for (from {\it bottom} to {\it top}) CRBR
2339.1, WL 2, C2D 162527.6, GY 312, C2D 162741.6 and CRBR 2315.8. The
locations of the sources are indicated in the SCUBA maps with white
crosses. These sources have either unresolved or
little or no emission in the SCUBA 850
$\mu$m and HCO$^+$ 4--3 maps at (0,0), making them disk sources in our
classification. }
\label{Fig:harp4}
\end{figure}

Four sources with $\alpha_{\rm{2-24 \mu m}}$$>$ 0.0 were found only in
the recent {\it Spitzer} observations. These are SSTc2d
J162527.6-243648, SSTc2d J162741.6-244645, SSTc2d J162748.2-244225 and
SSTc2d J162857.9-24405. The names C2D-162527.6, C2D-162741.6 ,
C2D-162748.2 and C2D-162857.9 are adopted.  VLA 1623-2418 is included
as an embedded Class 0, but it is generally absent from the above
studies. In addition, IRS 63 was included, although it is not part of
the L~1688 core, due to its interesting characteristics as a Class I source as observed with the SubMillimeter Array (SMA). These recent interferometric results suggest that this
source is an embedded YSO with little envelope material left, and, as
such, presents an interesting test case for the proposed method of
identifying truly embedded sources \citep{Lommen08}. Four disk sources
within or near to the L~1688 region, Haro 1-4, DoAr 25, SR 9 and SR 21,
were included to serve as a sample of known disk sources.

The final source sample, which covers all potential Class I sources,
can be found in Table \ref{label:source}.  The positions as found in
\citet{Bontemps01} were used where available. If the source was not
included in \citet{Bontemps01}, or if confusion exists due to nearby
IR sources, the position as found in the c2d delivery document \citep{Evans07} was
adopted. For sources in common, most  positions agree within 3$''$.    
Fig. \ref{fig:complete} shows
the distribution of the sample as plotted on the SCUBA 850 $\mu$m map
of L~1688 \citep{Johnstone00,Ridge06,diFrancesco08}.

\begin{figure}[!thp]
\begin{center}
\includegraphics[width=250pt]{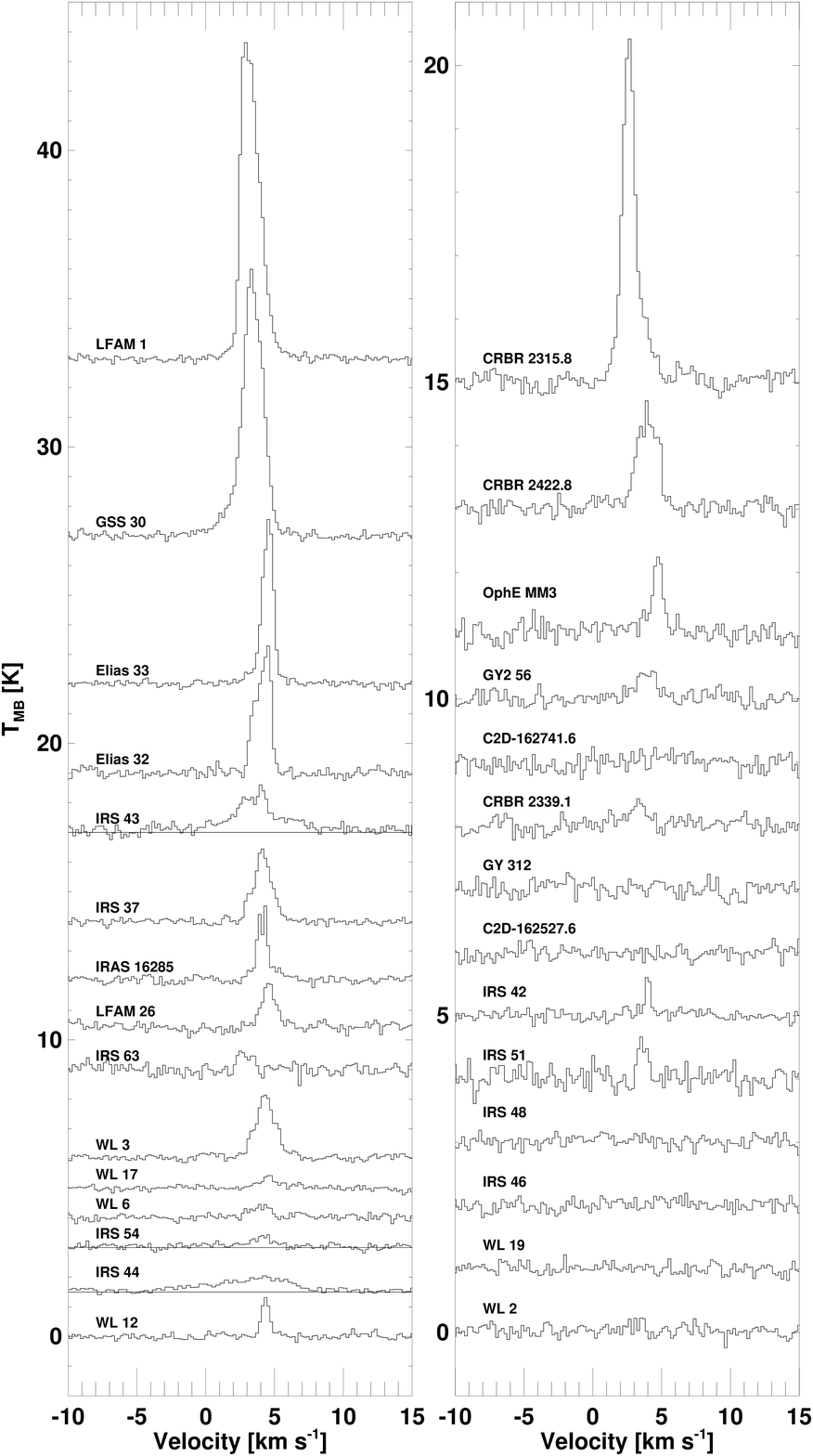}
\end{center}
\caption{HCO$^+$ 4--3 spectra extracted from the HARP-B maps in a 15$''$ around the IR positions. Sources classified as embedded in $\S$ 6.2 are shown in the left column. In the right column, the sources classified as confused or disks are plotted. }
\label{Fig:spec1}
\end{figure}

The table includes the spectral index $\alpha_{\rm{2-24 \mu m}}$ given
in \citet{Evans08}, calculated from
the 2MASS, IRAC and MIPS (24 $\mu$m only) fluxes.  No $\alpha_{\rm{IR}}$ could be determined from {\it
Spitzer} data for VLA 1623 (too faint)

\section{Observations}

\subsection{Gas line maps}
The majority of the sample was observed in HCO$^+$ 4--3 (356.7341 GHz)
 and C$^{18}$O 3--2 (329.3305 GHz) using the recently commissioned
 16-pixel heterodyne array receiver HARP-B on the James Clerk Maxwell
 Telescope (JCMT).  The high spectral resolution mode of 0.05 km s$^{-1}$
 available with the ACSIS back-end was used to 
 disentangle foreground material as well as any contributions from
 outflowing material.  Spectra were subsequently binned to 0.15 km
 s$^{-1}$.

HARP-B observations of 30 sources in 21 fields were carried out during
July and August 2007 in C$^{18}$O 3--2 and HCO$^+$ 4--3 under weather
conditions with an atmospheric optical depth, $\tau_{225\rm{GHz}}$
ranging from 0.035 to 0.08 (precipitable water vapor of 0.7 to 1.6
mm). The fields were observed down to a rms noise of 0.1 K in a 0.5 km
s$^{-1}$ bin. 
The HARP-B pixels have typical single side-band system
temperatures of 300--350 K. The 16 receivers are arranged in a
4$\times$4 pattern, separated by 30$''$. This gives a total foot print
of 2$'$ with a spatial resolution of 15$''$, the beam of the JCMT
at 345 GHz.  The 2$'\times2'$ fields were mapped using the specifically
designed jiggle mode HARP4\footnote{See JCMT website
http://www.jcmt.jach.hawaii.edu/}. 
A position switch of typically 30$'$ in azimuth was used, with larger throws if
needed.  The Class 0 source IRAS 16293-2422 was used as a line
calibrator and pointing source. Calibration errors are expected to
dominate the flux uncertainties, and are estimated at 20$\%$.
Pointing was checked every two hours and was generally found to be
within 2--3$''$. The map was re-sampled with a pixel size of 5$''$,
which is significantly larger than the pointing error.  The main-beam
efficiency was taken to be 0.67.  Data were reduced using the STARLINK
package GAIA and the CLASS reduction package.\\

\subsection{Gas single pixel spectra}
Supplementary data of HCO$^+$ 4--3 were taken at the APEX telescope
during July 2007, using the APEX-2a receiver.  All sources were
observed for which no HARP-B data were taken, except GSS 26 and GY
51. The APEX observations were done in excellent weather conditions
with $\tau_{225\rm{GHz}}$ ranging from 0.01 to 0.04 (PWV 0.2 to 0.8
mm). Single spectra were taken with a spectral resolution of 0.4 km
s$^{-1}$ down to an rms of 0.3 K. Calibration errors are estimated to be
$\sim 20$\%. The APEX 12-m dish is slightly smaller than the JCMT 15-m
dish, producing a beam of 18$''$ instead of 15$''$. Pointing errors
are $<$4$''$. Beam efficiency is 0.70.

For the sources GSS 30, VLA 1623, WL 20S, IRS 42, IRS 43, Elias 32 and Elias 33, C$^{18}$O 3--2 spectra were obtained from the CADC archive\footnote{See http://www.cadc-ccda.hia-iha.nrc-cnrc.gc.ca/jcmt/}.
These spectra were taken with the RxB receiver during September 2005.

\begin{figure*}[!htp]
\begin{center}
\includegraphics[width=460pt]{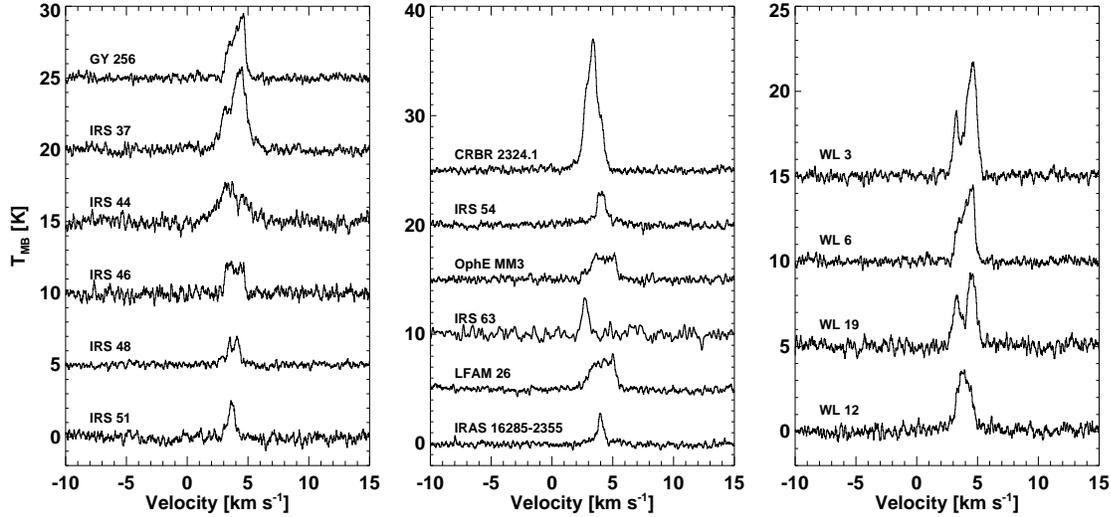}
\end{center}
\caption{C$^{18}$O 3--2 spectra extracted from the HARP-B maps at the
source positions. Spectra are arranged in random order to illustrate the independence of C$^{18}$O to the environment instead of the protostellar envelope.}
\label{Fig:spec3}
\end{figure*}

\begin{table*}[!htp]
\caption{Results from HARP-B (HCO$^+$ 4--3 and C$^{18}$O 3-2), APEX-2a
(HCO$^{+}$ 4--3) and RxB (C$^{18}$O 3--2) observations at the source position. 
}  \small
\begin{center}

\begin{tabular}{l c c c c  l}

\hline \hline 
Source  	& HCO$^+$ 4-3$^a$ &         &C$^{18}$O 3--2&      & Notes\\ 
                &$\int$ $T_{\rm{MB}}$ dV& $T_{\rm{MB}}$   &$\int$ $T_{\rm{MB}}$ dV & $T_{\rm{MB}}$   & \\ 
               & [K km s$^{-1}$] & [K] & [K km s$^{-1}$] & [K] \\
\hline
C2D-162527.6 &-  & $<$0.09 & -         & -     & See Fig. \ref{Fig:harp4}  \\
GSS 26	        &      -   & -	     & -         & -     & See Fig. \ref{Fig:scuba1}\\
CRBR 2315.8-1700         &      6.3 & 5.2*    & -	 & -	 & See Fig.\ref{Fig:harp4}\\
CRBR 2317.3-1925       &  1.5$^b$ & 2.0$^b$ & -	 & -	 & See Fig. \ref{Fig:scuba1}\\
VSSG 1	        &      -   & $<$0.4  & -	 & -	 & See Fig. \ref{Fig:scuba1}\\
GSS 30	        &      16.7& 9.1     & 10.7$^c$	 &6.3$^c$& See Fig.\ref{Fig:harp1} and \ref{Fig:sharc1}\\ 
LFAM 1	        &      18.4& 10.8    & -	 & -	 & See Fig.\ref{Fig:harp1} and \ref{Fig:sharc1}, GSS 30 maps\\
CRBR 2324.1-1619     &      -   & -       & 14.5	 & 12.2  & See  Fig.\ref{Fig:harp1} and \ref{Fig:sharc1}, VLA1623 maps \\
VLA 1623        &      16.2 & 9.0    & 10.4	 &17.9$^d$ & See Fig.\ref{Fig:harp1} and \ref{Fig:sharc1}\\
GY51            &      -   & -	     & -         & -     & See Fig. \ref{Fig:scuba1}\\
CRBR 2339.1-2032          &      0.55& 0.45*   & -	 & -	 & See Fig.\ref{Fig:harp4}\\
WL 12	        &      0.87& 1.4     & 4.8	 & 3.9   & See Fig.\ref{Fig:harp2} and \ref{Fig:sharc1}\\
WL 2            &      0.3 & 0.18*   & -	 & -	 & See Fig.\ref{Fig:harp4}\\
LFAM 26         &      1.95& 1.8    & 6.4	 & 3.3    & See Fig.\ref{Fig:harp2}\\
WL 17	        &      0.6 & 0.45    & -	 &-	 & See Fig.\ref{Fig:harp3}\\
Elias 29    &  4.5$^b$ & 2$^b$   & 10.4$^e$	 & 4.3$^e$   & See Fig.\ref{Fig:harp3}, WL17 map and Fig. \ref{Fig:scuba1} and \ref{Fig:sharc1}\\
GY 224	        &      -   &$<$0.27$^b$& -	 & -	 & See Fig. \ref{Fig:scuba1}\\
 WL19           &      -   & $<$ 0.1 & 6.3	 & 4.5   & See Fig.\ref{Fig:harp3}\\
WL 20S	        &       -&$<$0.26$^b$& 5.0$^c$  &2.3$^c$& See Fig. \ref{Fig:sharc1}\\
IRS 37	        &      3.7 & 2.5     & 9.55	 & 5.7   & See Fig.\ref{Fig:harp2}\\
WL 3            &      3.7 & 2.2     & 8.8	 & 6.3   & See Fig.\ref{Fig:harp2}, IRS 37 map\\
IRS 42	        &      0.45& 0.7*    & 4.0 &1.2*& See Fig.\ref{Fig:harp3}\\
WL 6	        &      0.92& 0.6     & 5.25     & 4.5*& See Fig. \ref{Fig:scuba1} and \ref{Fig:sharc1}\\ 
GY 256	        &      0.75& 0.45*  & 5.4   & 4.5*& See Fig. \ref{Fig:scuba1} and \ref{Fig:sharc1}, WL6 maps\\
IRS 43	        &   4.5$^d$ & 1.6$^d$ & 6.5$^c$	 &3.2$^c$ & See Fig. \ref{Fig:harp1},\ref{Fig:scuba1} and \ref{Fig:sharc1}\\
IRS 44	        &     3.4$^d$& 0.5$^d$     & 5.8	 & 2.8    & See Fig.\ref{Fig:harp2}\\
Elias 32        &      5.6 & 4.3     & 3.1$^c$	 &4.5$^c$& See Fig.\ref{Fig:harp1} and \ref{Fig:sharc1}, Elias 33 maps\\
Elias 33        &      6.3 & 5.7     & 6.0$^c$	&3.0$^c$  &See Fig.\ref{Fig:harp1} and \ref{Fig:sharc1}\\
IRS 48	        &      -   & $<$0.09 & 2.6	 & 2.3   & See Fig.\ref{Fig:harp3}\\
GY 312          &      -   & $<$0.1  & -	 & -	 & See Fig.\ref{Fig:harp4} \\
IRS 51	        &      0.75& 0.75    & 2.0	 &2.8*& See Fig.\ref{Fig:harp3} and \ref{Fig:sharc1}\\
C2D-162741.6  &   -      & $<$ 0.1 & -& -     & See Fig.\ref{Fig:harp4}\\
C2D-162748.1	&  -   & $<$ 0.34$^b$& - & -     & See Fig. \ref{Fig:scuba1}\\
IRS 54	        &      0.53& 0.45    & 3.4	 & 3.3   & See Fig.\ref{Fig:harp2}\\
IRAS 16285-2355 &      2.1 & 3.0     & 2.3	 & 3.0   & See Fig.\ref{Fig:harp2}\\
C2D-162857.9   &      -&$<$ 0.34$^b$& - & -   & See Fig. \ref{Fig:scuba1}\\
IRS 63	        &      0.75& 1.2     & 1.7       & 3.3   & See Fig.\ref{Fig:harp3}\\ \hline
\multicolumn{6}{c}{\textbf{Disks}}  \\ \hline
Haro 1-4        &      -   &$<$0.29$^b$ & -         & -   & See Fig. \ref{Fig:scuba1}\\  
DoAR 25	        &      -   &$<$0.28$^b$ & -         & -   & See Fig. \ref{Fig:scuba1}\\  
OphE MM3        &      1.7 & 1.6*    & 5.5	 & 3.0*& See Fig.\ref{Fig:harp2}, LFAM 26 map\\
SR 21           &      -   &$<$0.28$^b$ & -         & -   & See Fig. \ref{Fig:scuba1}\\  
CRBR 2422.8-3423       &      1.9 & 1.4*    & -	 & -	  & See Fig. \ref{Fig:harp1}, IRS 43 map, and Fig.\ref{Fig:harp3},NW corner of IRS 42\\
IRS 46	        &      -   & $<$0.09 & 2.9	 & 2.1   & See Fig.\ref{Fig:harp2}, IRS 44 map\\
SR 9            &      -   &$<$0.24$^b$ & -         & -   & See Fig. \ref{Fig:scuba1}\\  
2Mass 16282     &  -     &$<$0.35$^b$& -	 & -	  & See Fig. \ref{Fig:scuba1} and \ref{Fig:sharc1}\\ \hline 
\end{tabular} \\
\end{center}
$^a$ Intensities marked with a * do not peak at source position\\
$^b$ APEX-2a receiver. Upper limits (2$\sigma$) in 0.4 km s$^{-1}$ bin \\
$^c$ JCMT RxB data \\
$^d$ Outflowing gas detected (width $\sim$20 km s$^{-1}$)\\
$^e$ C$^{18}$O data taken from \citet{Boogert02}, obtained with the CSO.\\
\label{table:lines}
\end{table*}
\begin{figure*}[!htp]
\begin{center}
\includegraphics[width=460pt]{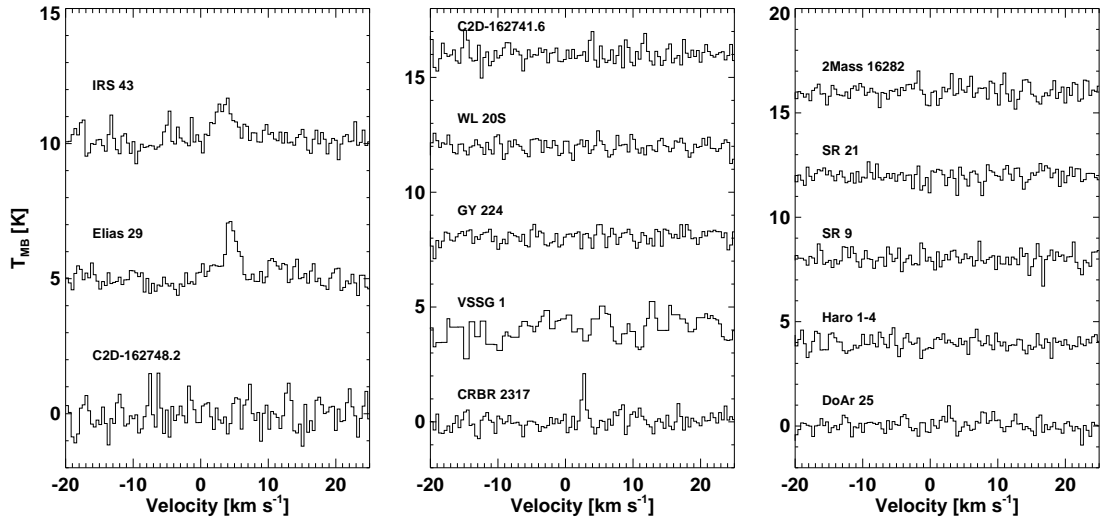}
\end{center}
\caption{HCO$^+$ 4--3 spectra at the source positions obtained at APEX. Elias 29 and IRS 43, with clear detections, are the only sources found to be embedded in $\S$ 6.2. The detection seen for CRBR 2317 is attributed to the cloud material and is classified as a disk. 
}
\label{Fig:spec4}
\end{figure*}
\begin{figure*}[!htp]
\begin{center}
\includegraphics[width=380pt]{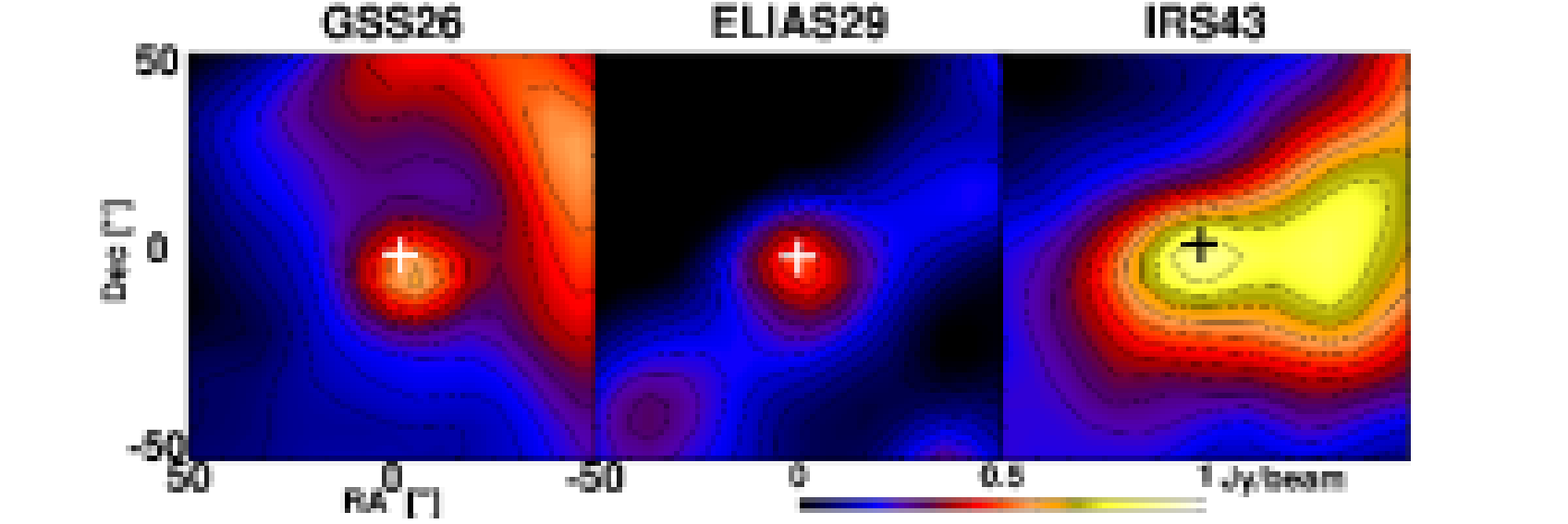}
\includegraphics[width=380pt]{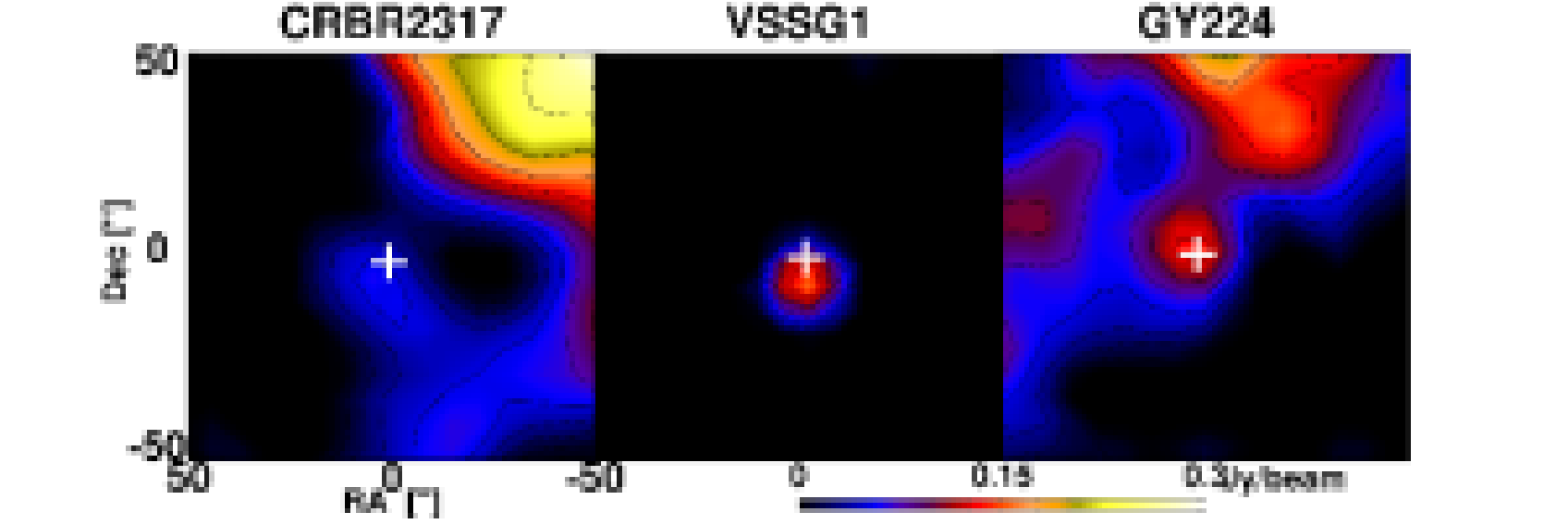}
\includegraphics[width=380pt]{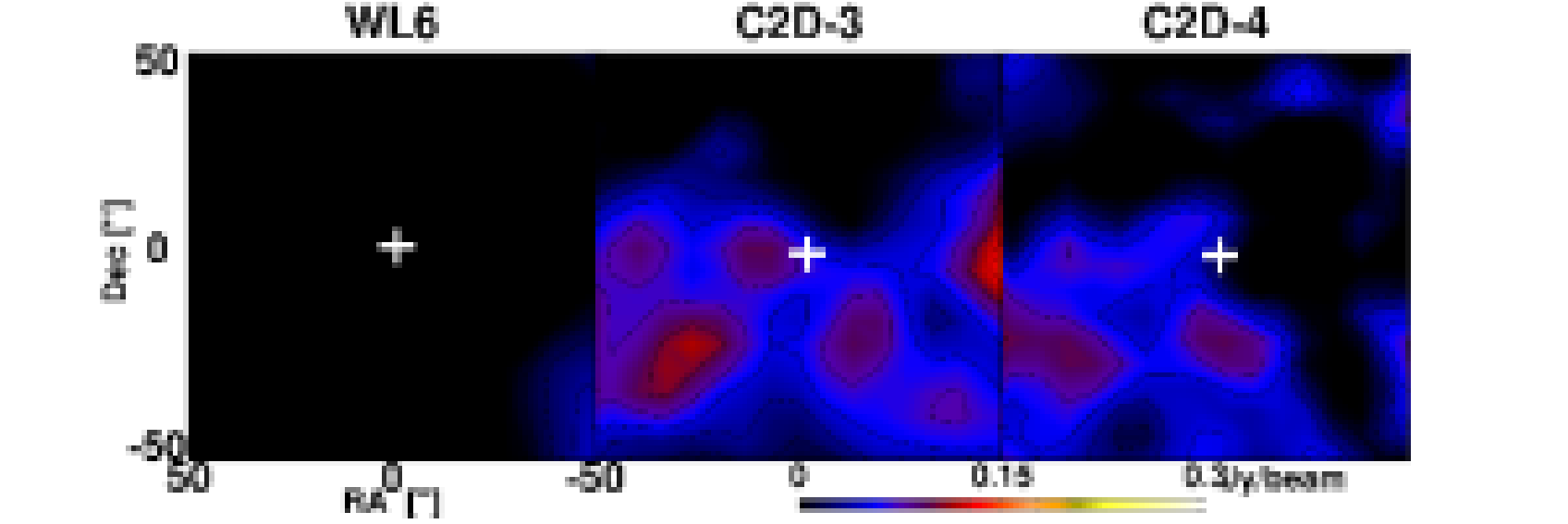}
\includegraphics[width=380pt]{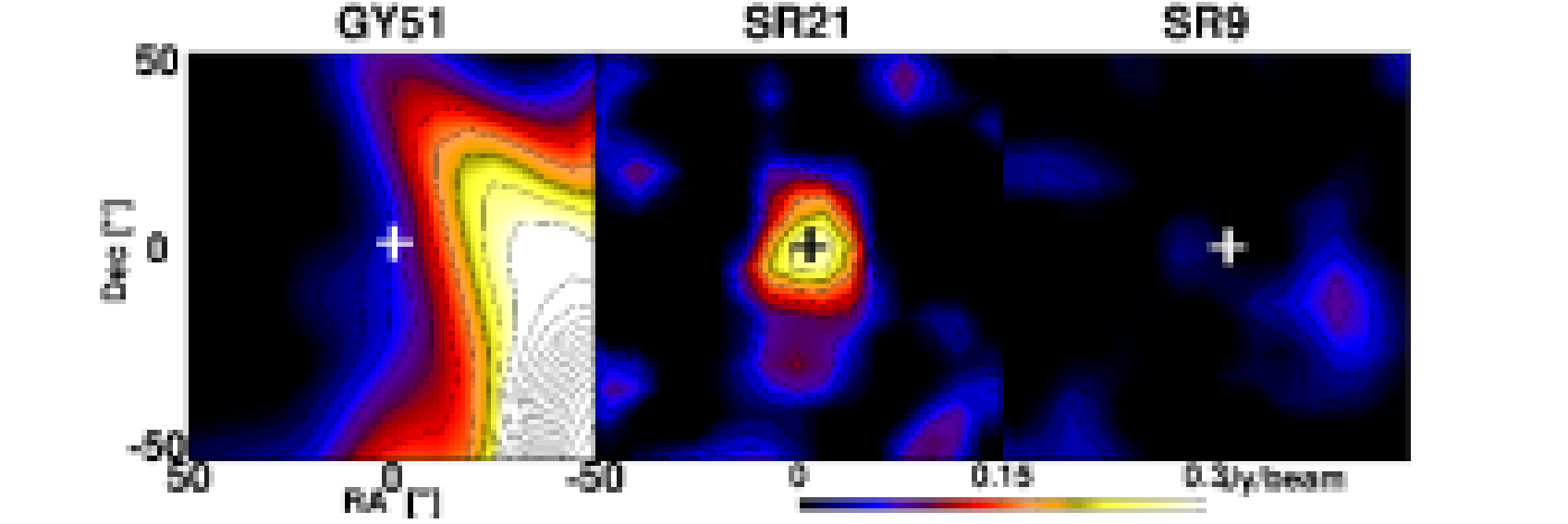}
\includegraphics[width=380pt]{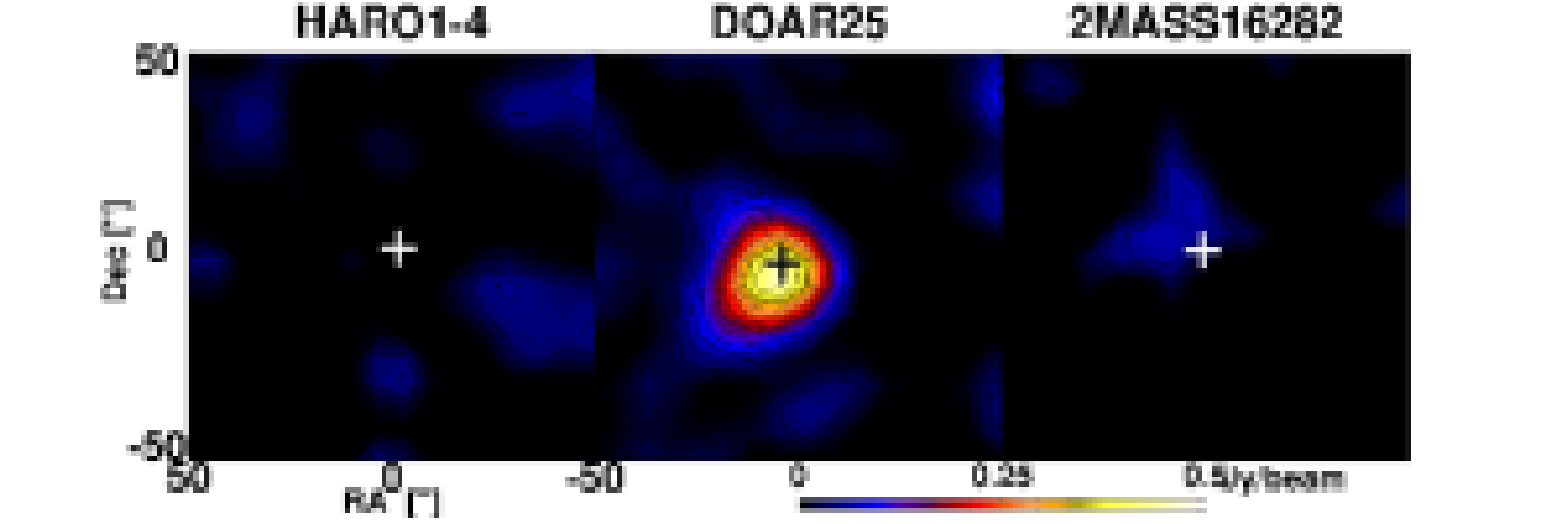}
\end{center}
\caption{Dust maps at 850 $\mu$m extracted from the JCMT-SCUBA
COMPLETE map of GSS 26, Elias 29 and IRS 43; CRBR 2317.3-1925, VSSG 1
and GY 224; WL 6, C2D-162748.2 (C2D-3) and C2D-162857.9 (C2D-4); GY 51,
SR 21 and SR 9; Haro 1-4, DoAr 25 and 2MASS 16282.  Note that the
emission toward  GY 51 is scaled down by a factor of 4.  }
\label{Fig:scuba1}
\end{figure*}
\begin{figure*}[!htp]
\begin{center}
\includegraphics[width=400pt]{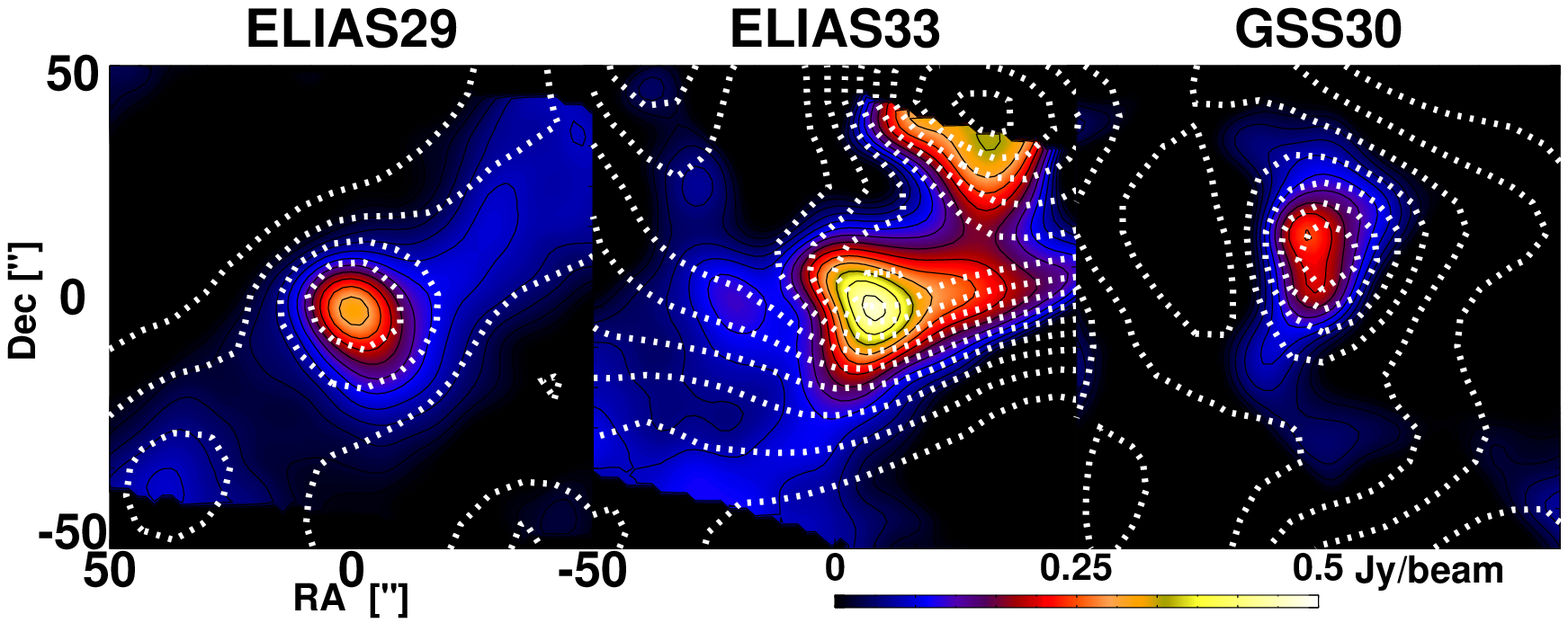}
\includegraphics[width=400pt]{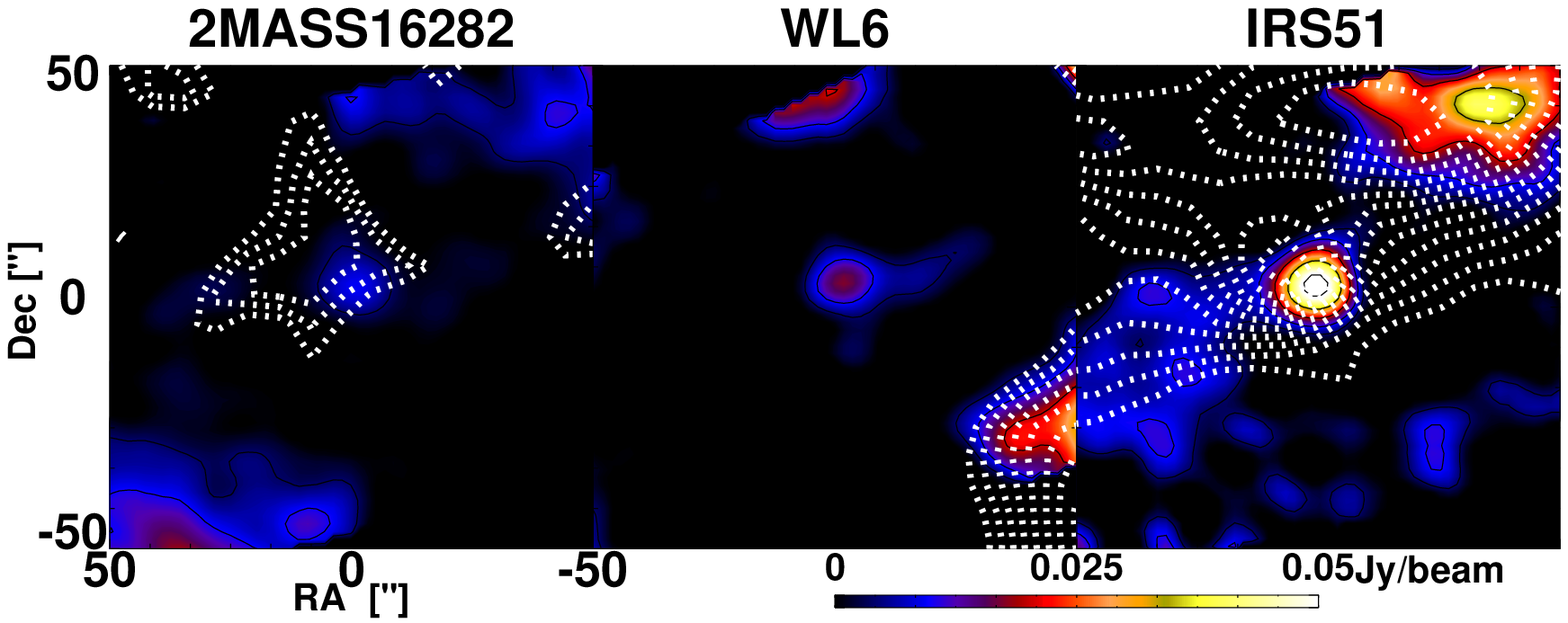}
\includegraphics[width=400pt]{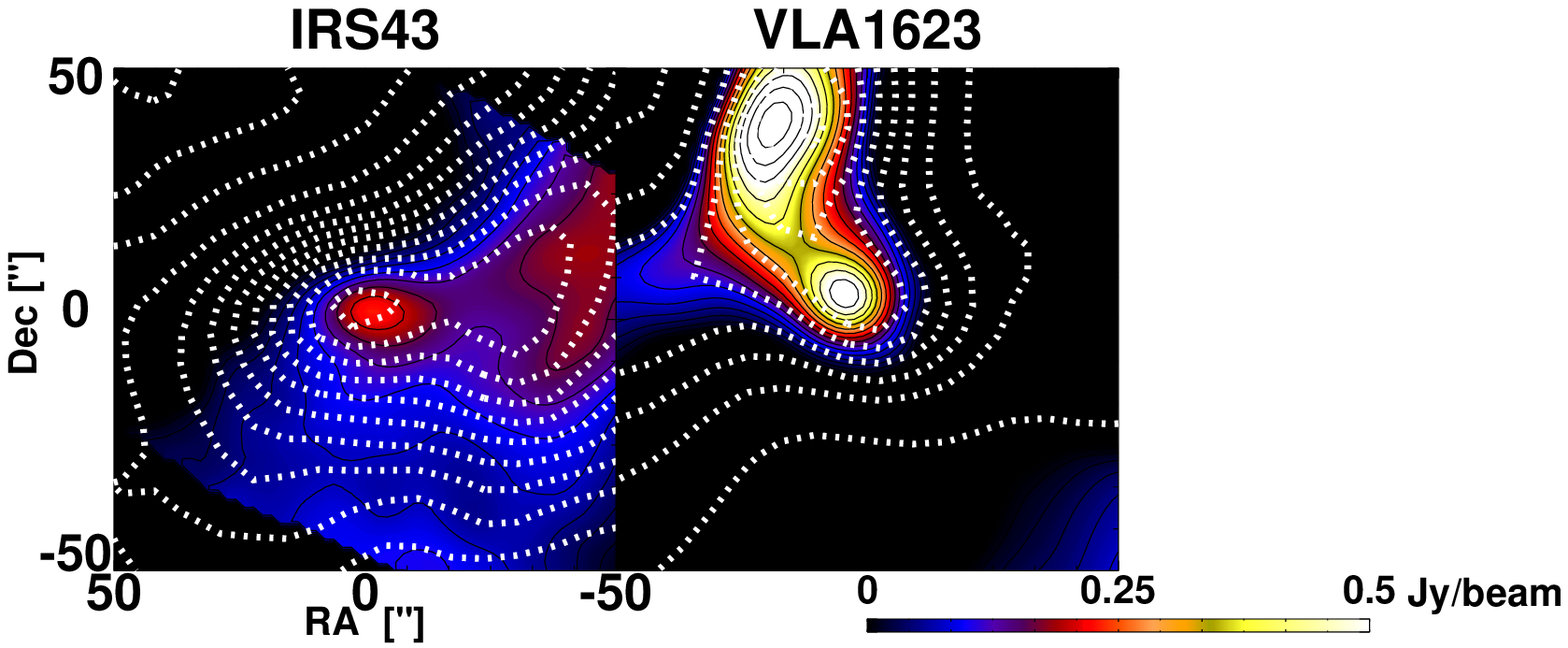}
\includegraphics[width=400pt]{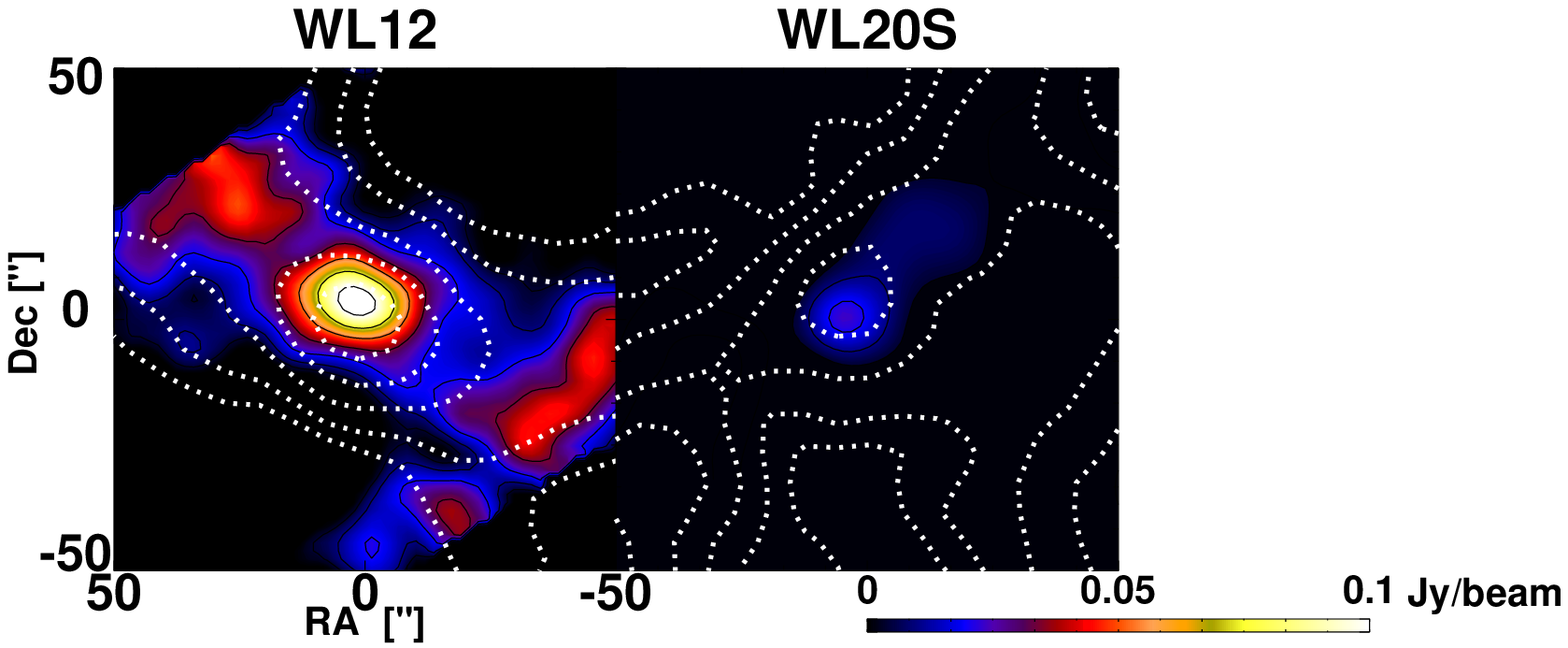}
\end{center}
\caption{Dust maps at 350 $\mu$m observed with the SHARC-II instrument
at the CSO. In white, the contours for the JCMT-SCUBA 850 $\mu$m data
are over-plotted, normalized for the maximum emission in the
map in contours of 10$\%$, 30$\%$, 50$\%$ and 70 $\%$. The shape of the contour levels at 850 $\mu$m agrees well with those found at 350$\mu$m.}
\label{Fig:sharc1}
\end{figure*}
\subsection{Dust maps}

The 850 $\mu$m continuum data of the Ophiuchus region, obtained within
the scope of the COMPLETE project using the SCUBA instrument on the
JCMT, were used to characterize the dust in the environments around
all sources of the sample \citep{Johnstone00,Ridge06,diFrancesco08}.
The most recent map, version 3, was used to extract the information
(see Fig.\ \ref{fig:complete}). This version includes a correction for the chopped out-emission.  Integrated
fluxes within regions with radii ranging from 25$''$ to 40$''$ were
extracted from the map 
(see Table \ref{table:flux}). The exact
radii were selected by calculating the FWHM to the peak flux of each source.  Comparison with the map and published fluxes by
\citet{Johnstone00} within similar radii found by a clumpfind routine showed that fluxes agreed within
5$\%$, significantly better than the calibration error of 20$\%$ for
SCUBA. The sensitivity of the SCUBA map is such that sources down to
90 mJy (3$\sigma$) can be
detected.  \\

In addition, 11 sources were observed using the SHARC-II 350 $\mu$m
continuum instrument at the Caltech Submillimeter Observatory
(CSO)\footnote{http://www.submm.caltech.edu/cso/} during April 2003
\citep{dowell03}. The array has 12$\times32$ pixels, spanning 4.85$''$
per pixel, which results in a footprint of 1$'\times2.6'$ and a beam
size of $9''$.  The sources observed with the CSO were GSS~30, WL~12,
Elias 29, VLA~1623, OphE MM3, WL~20S, WL~6, IRS 43, VSSG 17, IRS 51,
2MASS 16282.  The data were reduced with the CRUSH package and
calibrated using observations of Mars and Saturn. If no planets were
available during the specific night, IRAS 16293-2422 was used to
calibrate the data. The calibration errors are estimated to be on the
order of 30--40$\%$ and assumed to dominate the error in the flux
estimate over other instrumental errors or $S/N$ considerations. The
sensivity of the SHARC-II maps is such that sources down to 400 mJy (3$\sigma$) can be detected. For a typical dust
flux law $\propto \nu^{3.5}$, this implies that the SHARC-II data
are a factor of 4 more sensitive to low-mass sources than the SCUBA
data. 

\subsection{SED and IRS spectra}

The L~1688 core has been targeted by a large number of continuum
surveys, covering wavelengths from 2 to 1300 $\mu$m.
2MASS (1.25, 1.66 and 2.2 $\mu$m) and Spitzer-IRAC (3.6, 4.5, 5.8 and
8 $\mu$m) and Spitzer-MIPS (24 $\mu$m) fluxes were obtained from the
c2d delivery document \citep{Evans07}. Recently, \citet{Padgett08} presented
Spitzer-MIPS 70 and 160 $\mu$m fluxes. The MIPS 160 $\mu$m data were
not used in our analysis since a large part of the map was confused by
either striping or saturation. For the MIPS 70 $\mu$m data, we chose
to go back to the original data and retrieve the 70 $\mu$m fluxes
source-by-source in a typical 10$''$ radius using PSF photometry without the extended contribution subtracted. This avoids potential
errors in the flux estimates from automatic extraction in crowded or
confused regions such as the Ophiuchus core.
In addition to the SCUBA
(850 $\mu$m) and/or SHARC-II (350 $\mu$m) fluxes, a limited number of
sources was also observed using SHARC-II and/or SCUBA at 450 $\mu$m in
a recent survey of disks in the L~1688 core, with fluxes given in the
central 15$''$ beam \citep{Andrews07}. Where available, archival data
at millimeter wavelengths were used from SEST and IRAM 30m,
as reported in \citet{Andrews07} and originally
obtained by \citet{Andre94,Jensen96, Nurnberger98,Motte98} and
\citet{Stanke06}. \\ 

A number of sources in our sample have also been observed with various
observing programs using the IRS instrument on {\it Spitzer} at 5-40 $\mu$m (PID= 2,
172 and 179). These spectra are included in this study to confirm the
continuum fluxes in the wavelength range of IRS, and to investigate
the presence of ice absorption and silicate absorption or emission. \\

\section{Results}

\subsection{Gas maps}

Fig.~\ref{Fig:harp1} to \ref{Fig:harp4} present the 21 observed HARP-B
fields for HCO$^+$ 4--3 and C$^{18}$O
3--2, together with the SCUBA 850 $\mu$m continuum map. 
Note that the sources in Fig.~\ref{Fig:harp1} are much brighter and
have different intensity scales. The integrated intensity and peak
temperatures for all sources are listed in Table \ref{table:lines},
columns 2 to 5.  Column 7 states in which figure(s) the sources are
mapped or in which figure(s) its spectra are
plotted. Fig.~\ref{Fig:spec1} (HCO$^+$) and \ref{Fig:spec3}
(C$^{18}$O) show spectra extracted from the maps at the positions of
the sources. About 70$\%$ of the sample was observed in C$^{18}$O
3--2. The sources for which APEX-2a HCO$^+$ spectra were taken can be
found in Fig. \ref{Fig:spec4}.  \\

Using the molecular line emission maps, the sample can be divided
roughly into three groups.  In the first (see Fig. \ref{Fig:harp1} and \ref{Fig:harp2}), HCO$^+$ peaks at or close to the source
positions, with peak intenstities up to a few K. For these sources, GSS~30 IRS1, LFAM~1, VLA 1623,
WL 12, LFAM 26, IRS 37, WL 3, WL 17, Elias 33, Elias 32, IRS 44, IRAS
16285 and IRS 54, HCO$^+$ seems to be extended on scales of a few
10$''$. C$^{18}$O is always present in these maps, and also extended
on scales of a few 10$''$ to 1$'$.\\

Note that the HCO$^+$ maps in some case show minor offsets w.r.t. the IR positions. For offsets $<$ 8$''$ this can be attributed to either the pointing accuracy of the JCMT or the differences between IR positions from \citet{Bontemps01} and c2d. For a few sources, in particular VLA 1623, IRS 44, larger offsets up to 12$''$ were found. These can be accounted for by potentially three effects. First, self-absorption of HCO$^+$. Second, these sources show strong outflow emission, which peaks off-source and significantly influences the spectrally integrated maps. Third, molecular emission can be suppressed by the freeze-out of molecules in the cold part of the cloud.

A second group (Fig. \ref{Fig:harp3} and \ref{Fig:harp4}) lacks detections in HCO$^+$, down to a limit 0.1 K.
The sources, C2D-162527.6, WL 19, IRS 46, IRS 48, GY 312 and
C2D-162741.6 belong to this group.  The third group (Fig. \ref{Fig:harp3} and \ref{Fig:harp4}) of sources
contains CRBR 2315.8-1700, CRBR 2339.1-2032, WL 2, OphE MM3, IRS 51
and IRS 42.  These show HCO$^+$ detections at the source position. The
HCO$^+$ is extended, but there is no sign of a peak at the source
positions.  For IRS 63 and WL 6, detections of HCO$^+$ are marginal at
$\sim$3 $\sigma$.\\

The HCO$^+$ 4--3 integrated intensities $\int T_{\rm{MB}} $dV in
protostellar envelopes range from as high as 18.4 K km s$^{-1}$ for
LFAM 1 to 0.75 K km s$^{-1}$ for IRS 63. In $\S$ 5.3
the HCO$^+$ 4--3 integrated intensities are compared with the column
densities derived from the C$^{18}$O 3--2. This shows that
most embedded sources with strong HCO$^+$ emission are located in
regions with high column densities. The sole exception is IRAS
16285-2355, which is not located in a region with high column density
(see Fig. 1 and \ref{Fig:harp2}). However, it is possible to find
sources with little or no HCO$^+$ 4--3 emission, such as WL 19, in
regions with similarly high column density.\\

\subsection{Dust maps}

For sources for which molecular emission was taken with the HARP-B
array, the spatial extent of the dust is shown in the left column in
Fig.~\ref{Fig:harp1}-\ref{Fig:harp4}. The sources for which no HARP-B
maps were obtained can be found in Fig.~\ref{Fig:scuba1}.  The
SHARC-II maps are shown in Fig.~\ref{Fig:sharc1}. In these maps, the
white contours show the 850 SCUBA emission.  The integrated fluxes in
a 25$''$ beam are given in Tables \ref{table:sharc} and
\ref{table:flux}. \\

\begin{table}[htp]
\caption{Source fluxes at 350 $\mu$m in a 25$''$ beam
observed with SHARC-II at the CSO.  }
\begin{center}
\begin{tabular}{l l l}
\hline \hline 
Source & Flux & Concentration$^a$ \\ 
  & (Jy) &  \\ \hline
GSS 30 & 2.5 &0.58 \\
 WL12& 6.7 & 0.59\\
  Elias 29& 4.3 & 0.72 \\
 VLA1623& 40 & 0.8 \\
 WL20S& 0.82& 0.64\\
 WL6& $<$0.4 & U   \\
 IRS 43& 2.1& 0.62 \\
 Elias 33& 4.4 & 0.55\\
 IRS 51& 1.4 & 0.64\\
 2MASS 16282& $<$0.2& U\\ \hline
\end{tabular}\\
\end{center}
$^a$ See $\S$5.1; U=not determined since only flux upper limits obtained. \\
\label{table:sharc}
\end{table}

Both the SHARC-II and SCUBA images clearly show that the dust is
extended on scales of at least a few arcminutes at many positions in
Ophiuchus.  The smaller beam and higher frequency of the
SHARC-II observations is able to both resolve smaller envelopes
(e.g., Elias 33) or confirm that other sources (e.g., WL~20S) are not
resolved at 350 $\mu$m down to 9$''$.  The extended emission originates from
cold dust in the parental cloud, as mapped by \citet{Motte98,
Johnstone00} and \citet{Stanke06}. This cloud material can exist close
to or in the line of sight of many of the Class I sources, but is not
necessarily associated with a protostellar envelope, as is often
assumed. A good example is the source WL 19 (Fig.~\ref{Fig:harp3}),
where dust emission is found close to the position of the source, but
does not peak at the position of WL~19. This emission comes from a
prestellar core within the Ophiuchus ridge and not a protostellar
envelope.\\

Based on the spatial extent of the dust, the sample can be divided
into four groups.  First there are sources with spatially extended
dust emission profiles, peaking at the source position. Examples are
GSS~30 IRS1, Elias 29 and IRS 43. A second group shows extended dust
emission, but with no central peak at the position of the source. A
good example is GY~51 (see Fig. \ref{Fig:scuba1}), located close to the
Oph-A core. Some Class I sources only show unresolved dust emission at
the source position, such as IRS~51 (see Fig.\ref{Fig:sharc1}). A final
group shows no emission at 350 or 850 $\mu$m down to our
sensitivity limits. Good examples is Haro 1-4. \\

Cloud material can contribute significantly or even dominate the
emission on the scales probed by SCUBA and SHARC-II. Two fluxes at 850
$\mu$m are therefore given in Table~\ref{table:flux}. The first flux
is within the 15$''$ beam obtained from \citet{Andrews07}. If no value
was given by them, it was extracted from the COMPLETE map for the central
beam. Our extracted fluxes agree to within 15\% for sources also
listed in \citet{Andrews07}, even though we include estimates to
reconstruct the chopped-out extended emission.  The second flux is
extracted from the COMPLETE map for larger apertures encompassing the
envelopes, up to a 25$''$ ($\sim$3,000 AU) radius, the typical
envelope extent where the temperature and density drop to that of the
surrounding cloud. The first number should thus be considered a lower limit
on the total envelope and disk emission, while the second is an
equivalent upper limit. For disk sources, all emission originating
from the source is located within 15$''$ ($<$1,800 AU). 
\\

\subsection{SED and IRS spectra}

\begin{figure*}[!htp]
\begin{center}
\includegraphics[width=530pt]{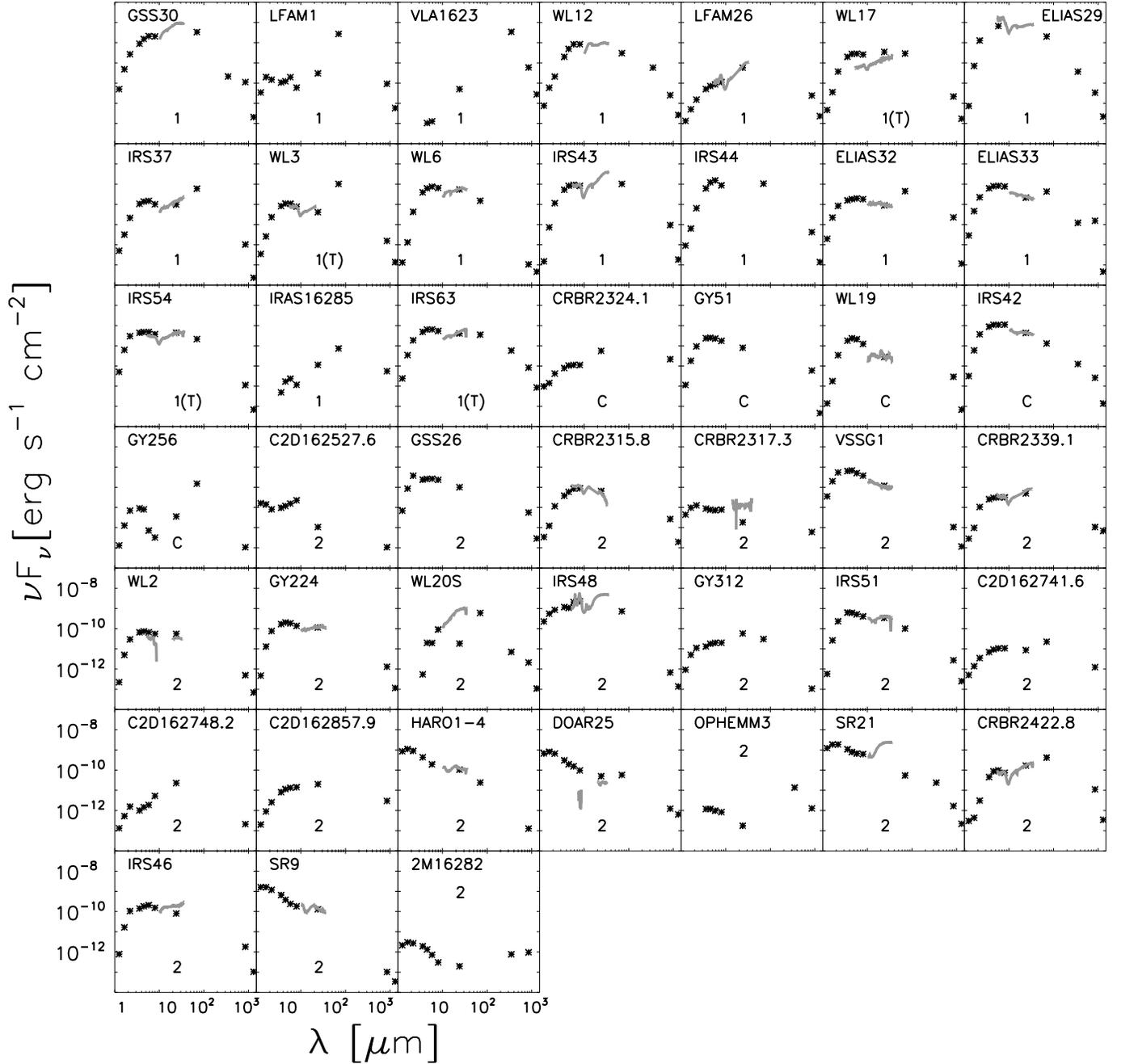}
\end{center}
\caption{The spectral energy distributions of all sources,
supplemented by spectra taken with IRS. Our classification can be
found near the center and bottom of each SED. 1 = Embedded Stage 1,
1(T) = Transitional from embedded to pre-main sequence, C = confused
and 2 = (edge-on) disk.}
\label{Fig:seds}
\end{figure*}

Table \ref{table:flux} summarizes all the Spitzer and (sub-)millimeter
fluxes.  The 2 $\mu$m to 1.3 mm flux densities for all 45 sources are
plotted in Fig.~\ref{Fig:seds}. In addition, low-resolution IRS
spectra are over-plotted between $\sim$5-30 $\mu$m, where
available. In the on-line appendix, blow-ups of the IRS spectra are
presented.

\begin{table*}[!htp]

\caption{Source properties derived from the gas and dust. The table
has been rearranged compared with Table~\ref{label:source} to reflect
the classification in $\S$6.2 } \small
\begin{center}
\begin{tabular}{l l l l l l r l l}
\hline \hline 
Source  	& $\alpha_{\rm{IR}}$ & $N_{\rm{H}_2}^a$& $\Delta\alpha^b$ &Mass$^c$ & $L_{\rm{bol}}$ & $T_{\rm{bol}}$ & $C_{850}$$^d$ & $C_{\rm{HCO^+}}$$^d$ \\ 
	        &  &(10$^{22}$ cm$^{-2})$&  &(10$^{-2}$ $M_{\rm{\odot}})$&(10$^{-1}$ L$_{\rm{\odot}}$) &(K) & &\\ \hline
\multicolumn{5}{c}{{\bf Embedded sources}}\\ \hline
GSS 30$^e$               &1.46   &23.3/19.4*& 2.6 &  20.5  & 33  & 123 & 0.55& 0.69\\ 
LFAM 1$^e$               &0.73   &50*       & 2.6 &  17.1  & 8.3 & 86  & 0.59& 0.69\\ 
VLA 1623                 & no    &441*      & -   & 111    & 2.7 & 12  & 0.7 & 0.72 \\ 
WL 12	                 &2.49   &10.5/15.2*& 3.1 & 4.6    & 34  & 155 & 0.26& 0.81\\ 
LFAM 26                  &1.27   &14.0/18.5*& 4.4 & 4.5    & 0.44& 238 & 0.40& 0.69\\ 
WL 17	                 &0.61   &8.4*      &  1.0& 4.0    & 6.7 & 323 & 0.22& 0.71\\ 
Elias 29                 &1.69 &22.7/33.6*  & 1.5 &6.2     & 25  & 424 & 0.49& -$^f$ \\ 
IRS 37$^e$	         &0.25   &20.8      & 1.0 &1.2     &3.8  & 243 & 0.58    & 0.72\\ 
WL 3$^e$                 &-0.03  &19.2/9.7* & 2.3 &2.9     & 4.6 & 192 & 0.25    & 0.67\\ 
WL 6	                 &0.72   &11.4      & - &$<$0.4    & 8.5 & 394 & U       & 0.77\\ 
IRS 43	                 &1.17   &14.2/20.7* & 2.3 &17.1   & 10  & 134 & 0.58& 0.79\\ 
IRS 44	                 &2.29   &12.6/30.2D* & 2.9 &8.0   & 11  & 140 & 0.33& 0.80\\ 
Elias 32                 &-0.03  &6.8/140*  & 3.7 &41.9    &5    & 321 & 0.37& 0.70\\ 
Elias 33                 &-0.12  &13.1/96* & 3.1 &28.2     &12   & 460 & 0.6 & 0.64\\ 
IRS 54	                 & 0.03  &7.4       & 1.2 &3.1     & 7.8 & 486 & 0.8  & 0.80\\ 
IRAS 16285-2355          & 1.23  &5.0/41.9D* & 3.1 &10.3   & 0.57& 77  & 0.45&0.80\\ 
IRS 63	                 & 0.14  &3.7/84.0D* & -  &16.4    & 13  & 363 & 0.75& 0.60\\ \hline
\multicolumn{5}{c}{{\bf Confused sources}}\\ \hline
CRBR 2324.1-1619         &0.87   & 31.6/70.5*& 5.2 &39.3      & 0.44& 283 & U& U\\ 
GY51                     & 0.05  &27.7*     & 4.5 &11.1      & 2.3 & 706 & U&U\\ 
 WL19                    &-0.43  & 13.7/16.8*& 4.9 &5.5     & 1.5 & 730 & 0.15& 0.53\\ 
IRS 42	                 &-0.03  & 8.7/14.2* & 1.3 &4.7     &14   & 540 & 0.44& 0.60\\ 
GY 256	                 &-0.05  & 11.8& - &$<$0.2 &1.1  & 104 & U& 0.70\\ \hline
\multicolumn{5}{c}{{\bf Disks}}\\ \hline
C2D-162527.6             & 0.36  & -   &  - &$<$0.4 & 0.16& 1051& U&U\\  
GSS 26	                 &-0.46  &24.3D*& 1.8  &10.2     & 3.2 & 877 & 0.65&U\\ 
CRBR 2315.8-1700         & 0.69  &17.6*& 1.0  &5.1     &0.93 & 456 & 0.38&0.74 \\ 
CRBR 2317.3-1925         &-0.56  &12.6*& 0.3  &1.1     &0.11 & 1105& 0.85 &S\\ 
VSSG 1	                 &-0.73  &20.1*& -  &2.0     &5.8  & 930 & 0.74&SU\\ 
CRBR 2339.1-2032         &0.45   &6.7* & -  &2.0     & 0.55& 426 & 0.55& 0.58\\ 
WL 2                     &0.02   & -   & -  &1.0    & 0.88& 573 &0.3 & 0.58\\ 
GY 224	                 &-0.05  &10.9*& 1.0  &2.4     & 2.1 & 614 & 0.63&SU\\ 
 WL 20S	                 &2.75   &10.9/15.3D*& 1.0  &3.9     & 4.5 & 85  & 0.57&SU\\ 
IRS 48	                 &0.88   & 5.7/15.2D*& 2.1  &1.2     &81   & 238 &0.70&U\\ 
GY 312                   &0.64   & -  &  - &$<$0.2 & 0.73&285  &U & 0.3\\  
IRS 51	                 &-0.15  & 4.4/18.4D*& 2.3  &5.0    & 8   & 548 & 0.27&0.59\\ 
C2D-162741.6             &0.32   & -   &  - &2.3     & 0.28&279  & U&U\\ 
C2D-162748.2             &1.55   & -   &  - &0.4     & 0.17&241  & U&SU\\ 
C2D-162857.9             & 0.67  & -    & -  &5.6    & 0.22& 402 & U&SU\\ 
Haro 1-4                 &-0.89 & -    & -  &0.45    & 5.6 &762  & U&SU\\ 
DoAR 25	                 &-1.12 &38.6*     & -  &3.4     & 4.7 &1422 & 0.90&SU\\ 
OphE MM3                 &-0.33 & 12.0/13.4*& 3.4  &2.3     & 0.17&45   & U& 0.68\\ 
SR 21                    &-0.79 &33.6*     &  - &3.1     & 1.8 & 1070& 0.89&SU\\ 
CRBR 2422.8-3423         &1.01  &106.6*     & 2.3  &20.5      & 4.4 & 157 & U &-$^f$\\ 
IRS 46	                 &0.18  & 6.3/15.1* & 2.8  &3.3     & 1.9 & 684 & U &SU\\ 
SR 9                     &-1.07 & -     & -  &$<$0.4 & 8.7 & 1588& U&SU\\ 
2MASS 16282              &-1.55 & -     & -  &1.7     & 0.84& 66  & U&SU\\ \hline
\end{tabular} \\
\end{center}
$^a$ Column densities marked with * are calculated from the SCUBA dust maps. Embedded sources where a large disk contribution is likely are indicated with a D. Note that no column densities could be calculated from the dust below 1$\times10^{23}$ cm$^{-2}$ .\\
$^b$ $\Delta \alpha$ is an {\it upper limit} to change in 2-24 $\mu$m spectral slope due to the reddening by foreground clouds at this source. This is derived by considering the mean column density at 45$''$ off the source position.\\
$^c$ Upper limit derived from the 850 $\mu$m data using the relation from \citet{Shirley00}, with the assumption that all 850 $\mu$m emission originates within a protostellar envelope.\\
$^d$ U=Undetected or Unobserved, S=single position spectrum available only\\
$^e$ Circumbinary envelope.\\
$^f$ Although both Elias 29 and CRBR 2422.8-3423 were observed, these sources are located near the edge of a field and do not have sufficient coverage to obtain a concentration parameter for the HCO$^+$. \\
 \label{table:prop}
\end{table*}

\section{Analysis}

\subsection{Concentration}

One method to distinguish embedded YSOs from edge-on disks, starless
cores or background sources such as AGB stars and galaxies, is to use
a measure of how centrally concentrated the emission is. This method has been
used previously in the analysis of SCUBA data
\citep{Johnstone01, Walawender05,Jorgensen07} by defining a
concentration parameter:
\begin{equation}
C=1-\frac{1.13 B^2 S_{850}}{\pi R^2_{\rm{obs}}F_0}
\end{equation}
where $B$ is the beam-size, $S_{850}$ is the total flux at 850 $\mu$m
within the envelope, $R^2_{\rm{obs}}$ is the observed radius of the
envelope and $F_0$ is the peak flux of the envelope. The observed
radii were selected by calculating the FWHM to the peak flux of each source. The level of
concentration potentially provides an indication of the evolutionary
stage. Highly concentrated cores ($C>$ 0.75) have a very high
probability to contain protostars, while cores with low concentration
($C<$ 0.4) are often starless, as indicated by the absence of near-IR
emission.  The observed radii in the 850 $\mu$m maps were defined by \citet{Kirk06} as $R=(A/\pi)^{1/2}$, with A the area of the core within the lowest contour of their {\it clumpfind} routine.
  The SHARC-II maps also allow a concentration
at 350 $\mu$m to be made, as seen in Table \ref{table:sharc}. These
concentrations are found to be within 10$\%$  to
those found for the 850 $\mu$m data, with the exception of WL12 which is more than a factor 2 higher in the SHARC-II 350 $\mu$m data. \\

\begin{figure}[thp]
\begin{center}
\includegraphics[width=220pt]{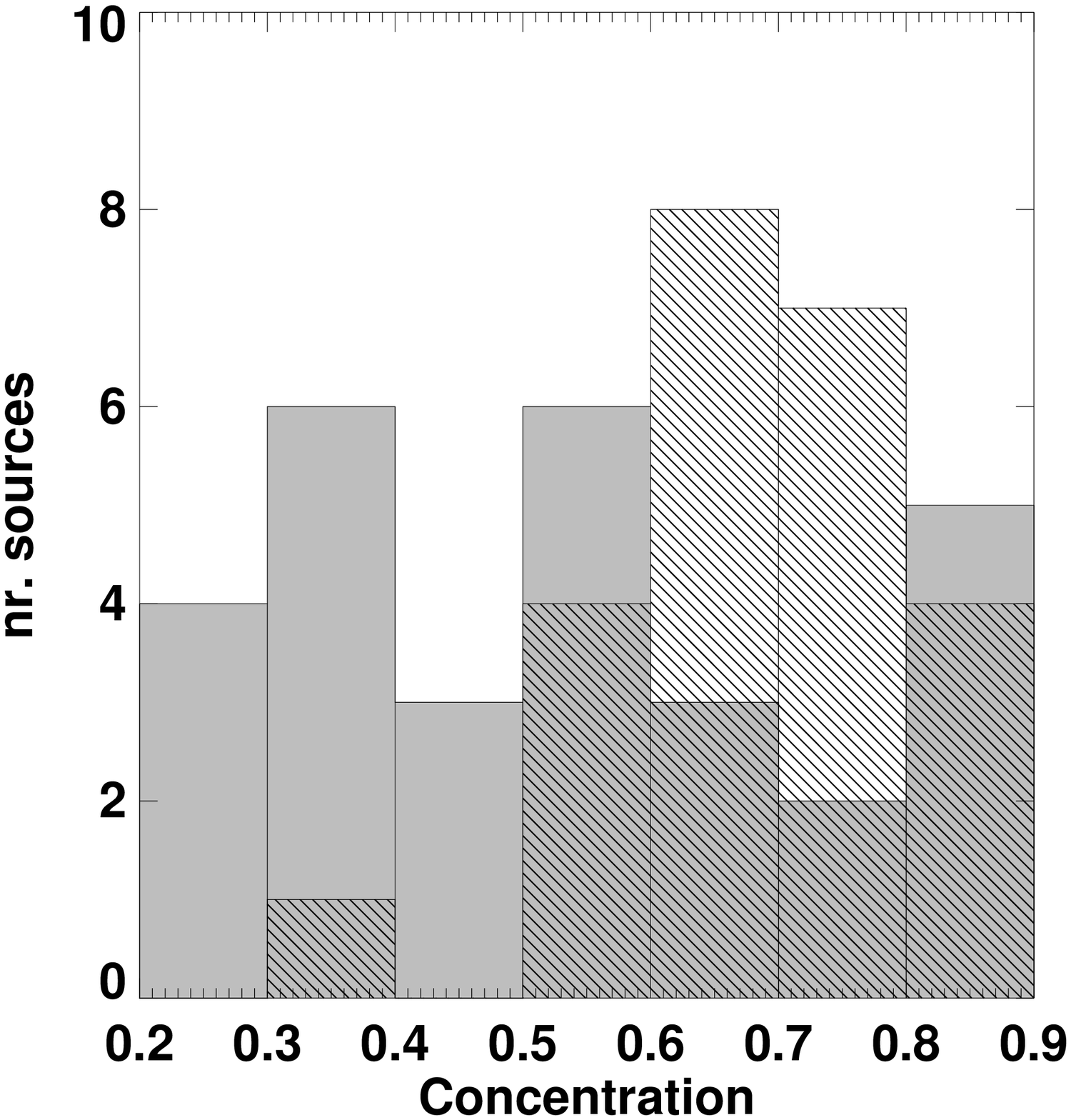}
\end{center}
\caption{The distribution of concentration of SCUBA (gray) and HCO$^+$
4--3 (striped). Only sources with spatially resolved profiles have
been taken into account. Note that the total number of sources is
greater for SCUBA than for HCO$^+$. }
\label{Fig:concentration}
\end{figure}

 A similar concentration parameter can also be calculated for the
 distribution of HCO$^+$, using integrated intensities between
 $V_{\rm{LSR}}-5$ to +5 km s$^{-1}$:
\begin{equation}
C_{\rm{HCO^+}}=1-\frac{1.13 B^2 S_{\rm{HCO^+}}}{\pi R^2_{\rm{obs}} T_{\rm{MB}}},
\end{equation}
with $S_{\rm{HCO^+}}$ the spatially and spectrally integrated
intensity $\int T_{\rm{MB}}dV$ of HCO$^+$ over the entire envelope
with radius $R_{\rm{obs}}$, and $T_{\rm{MB}}$ the peak intensity in the
central beam.  Since no {\it clumpfind} routine was used to characterize the spectral maps, the observed radii of the HCO$^+$ concentration were set equal to the FWHM of each  spectrally integrated core map. Although different methods to define radii result in different values, tests showed that the corresponding changes in concentration factor are less than 5 $\%$.  \\

Tables \ref{table:sharc} and \ref{table:prop} list
the concentrations of both the 850 $\mu$m data SCUBA maps and the
HCO$^{+}$ HARP-B maps, where available. 
Although the number of sources with a concentration
parameter determined by SCUBA maps is greater than those for HCO$^+$,
it is clear that HCO$^+$ is more concentrated than the continuum.
Fig. \ref{Fig:concentration} shows the distribution of the
concentration parameters for spatially resolved sources.

\subsection{Environment}

\begin{figure*}[!htp]
\begin{center}
\includegraphics[width=400pt]{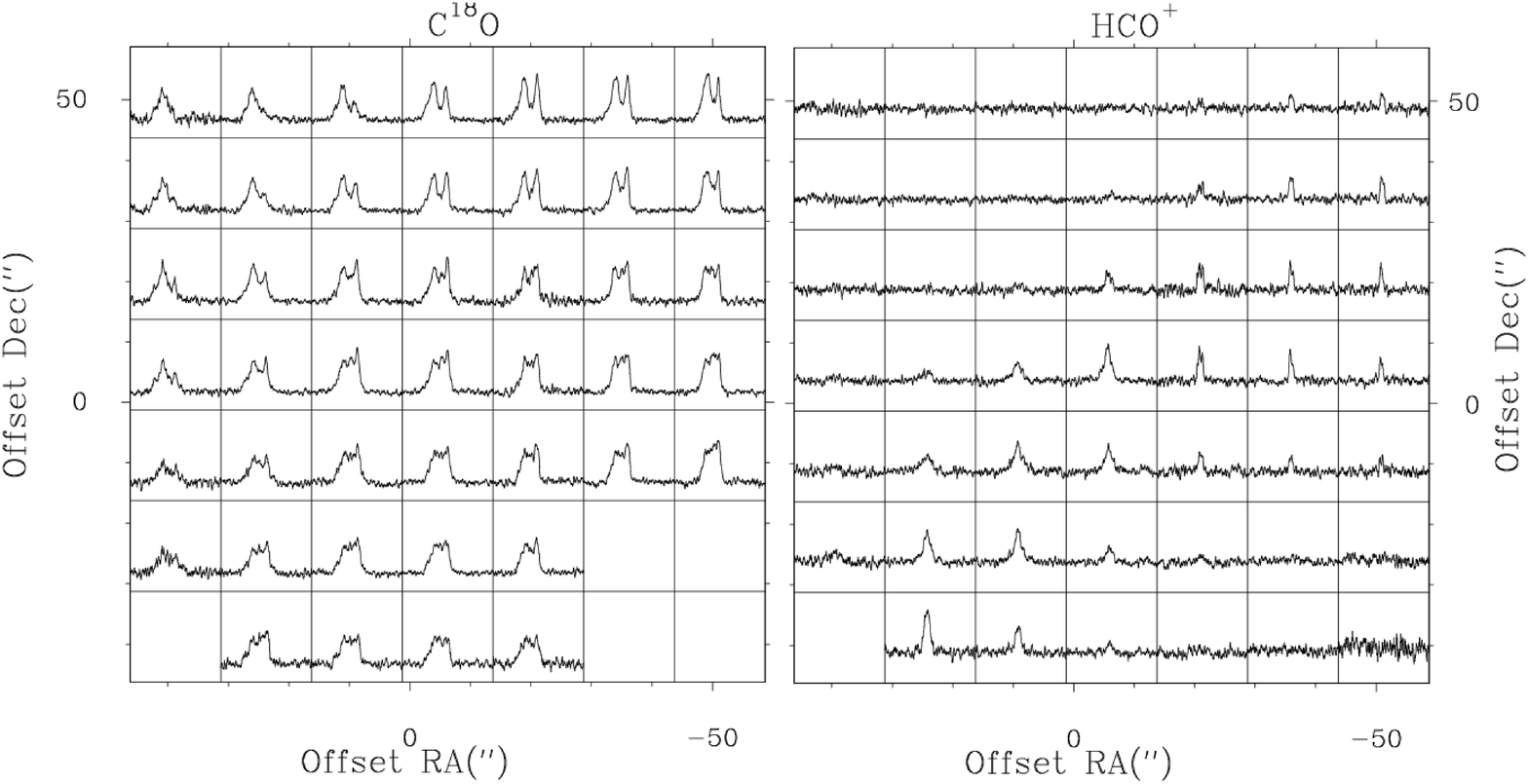}
\end{center}
\caption{The spectra of C$^{18}$O 3--2 and HCO$^+$ 4--3 of the LFAM 26
field. The intensity scale is -1 to 6 K in $T_{\rm{MB}}$ for C$^{18}$O
and -1 to 3 K for HCO$^+$.  OphE-MM3 is located at $\Delta$RA = +5$''$ and $\Delta$Dec = -40$''$ in
this field and has no associated HCO$^+$ emission, except from the
extended cloud.}
\label{Fig:maplfam26}
\end{figure*}
\begin{figure}[!htp]
\begin{center}
\includegraphics[width=200pt]{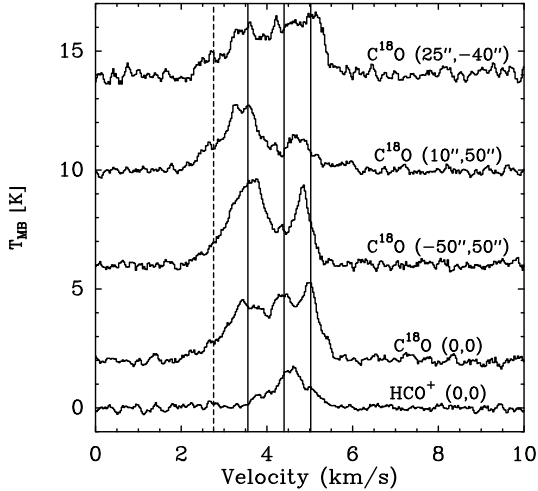}
\end{center}
\caption{Individual spectra at offset positions from LFAM 26 for
HCO$^+$ (0$''$,0$''$) and C$^{18}$O (0,0), (-50,50), (10,50) and
(25,-40). The three main components at 3.5, 4.4 and 5.0 km s$^{-1}$
are shown with solid lines. The component at 2.7 km s$^{-1}$
identified by \citet{Boogert02} is shown with a dashed line and is
generally weaker or absent. The main component of the HCO$^{+}$
emission is located between the components at 4.2 and 4.9 km s$^{-1}$
and peaks at 4.6 km s$^{-1}$.}
\label{Fig:spectrlfam26}
\end{figure}

\begin{figure*}[!htp]
\begin{center}
\includegraphics[width=450pt]{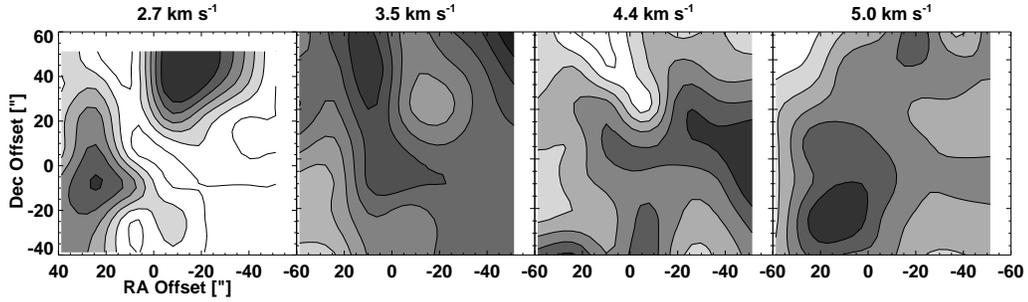}
\end{center}
\caption{The spatial distribution of the integrated gaussians fitted
to the C$^{18}$O 3--2 spectra for each of the four layers seen in the
field of LFAM 26.}
\label{Fig:layer_lfam26}
\end{figure*}

\begin{table*}[!htp]
\caption{The number and properties of the (foreground) layers for the
fields of LFAM 26, IRS 44 and WL~19 in the Ophiuchus ridge and IRS 37
and IRS 4, located near Oph-B cores, see Fig.~\ref{Fig:spec3}.
} \tiny
\begin{center}
\begin{tabular}{l l l l l l l l}
\hline \hline 
Field  	& Position &Nr.\ comp & $V_{\rm{LSR}}$ (km s$^{-1}$) & C$^{18}$O 
$\int T_{\rm{MB}}dV$ (K km s$^{-1}$ & $\Delta V$ (km s$^{-1}$)\\ \hline
LFAM 26 & 0,0 & 3 & 3.5, 4.4, 5.0 & 1.8, 0.9, 1.1 & 0.8,0.7,0.8\\
   & -20,+35 & 4 & 2.7, 3.5, 4.4, 5.0 & 0.6, 1.5, 0.8, 1.1 & 0.9, 0.7, 0.7, 0.7 \\
   & +35, -22 & 3 & 3.5, 4.4, 5.0 &  1.3, 0.4, 0.5, 1.2, 0.7,0.5\\
IRS 44 & 0,0 $^a$& 3 & 2.6, 3.5, 4.5 & 0.5, 2.3, 2.3 & 1.9,1.1, 0.8\\
       & +20,+20 (IRS 46) & 2 & 3.5, 4.3 & 1.9, 1.3 & 0.6,0.5\\
       & -50,20 & 2 & 3.3, 4.0 & 1.1,2.5 & 0.4, 1.1\\
IRS 48 & 0,0 &3 & 2.8, 3.5, 4.1 & 0.6, 1.7, 2.6 & 0.6, 0.4, 0.5 \\
       & -36,+36 & 3 & 2.9, 3.5, 4.1 & 0.8, 2.1, 4.3 & 0.4, 0.4, 0.5 \\
       &  +38, 6 & 3 & 2.8, 3.5, 4.1 & 0.4, 1.1, 3.5 & 0.3, 0.5, 0.5 \\
WL 19 & 0,0 & 2& 3.3, 4.6 & 2.7, 4.7 & 0.8,0.8 \\
      & -6, 36 & 2& 3.4, 4.6 & 2.8, 5.6, 1.1,0.8\\
      & -20,-38 & 2 & 3.2, 4.2 & 1.9, 3.0 & 0.7, 1.0 \\
     & -50, 50 & 4 & 2.6, 3.5, 4.3, 5.1 & 1.4, 3.3, 3.5, 3.2 &0.5, 0.6, 0.7, 0.7\\
IRS 37 & 0,0 & 2 & 3.0, 4.4 & 2.3, 5.6 & 0.7, 1.1\\
       & +23, +7 (WL3) & 2 & 3.2, 4.4 & 2.7, 5.4 & 0.8, 1.0\\
       & -36, 21 & 3 & 3.0, 3.6 , 4.2 & 2.4, 0.8, 2.8 & 0.9, 0.2,0.8 \\ \hline
\end{tabular} \\
\end{center}
$^a$An outflow is detected at 0,0 in HCO$^+$ 4--3 which is probably the component seen at 2.6 km s$^{-1}$\\

\label{table:c18oprop}
\end{table*}

In contrast with deeply embedded Class 0 YSOs, the environment around
the embedded Class I YSOs has a large influence on the analysis of the
generally weaker source data. Fig.~7 in \citet{Boogert02} illustrates
how complex the environment can become. Elias 29, an embedded YSO, is
located in a dense ridge of material, but in front of that ridge, two
foreground layers were identified using the emission from a dozen
different molecules. Continuum emission and ice absorption originate
in all these layers.  For most of our sources, however, the situation appears
to be less complex. Fig. \ref{Fig:spec1} and \ref{Fig:spec3} show that
the HCO$^+$ 4--3 profiles are mostly single peaked. However, C$^{18}$O
3--2 often shows a more complex profile with multiple peaks. 

LFAM 26, located at the top of the Ophiuchus ridge only a few arcmin
north-east of Elias 29, is an especially complex case.
Fig. \ref{Fig:maplfam26} shows the spectral map in 17$''$ spatial bins
for HCO$^+$ 4--3 and C$^{18}$O 3--2 maps. It is clearly seen that
besides the integrated emission (Fig.~\ref{Fig:harp2}), the line
profiles of HCO$^+$ and C$^{18}$O both vary over the entire map. The
HCO$^+$ spectra show mostly single gaussians. At the position of LFAM
26, the HCO$^+$ profile has a small outflow signature on the blue side
of the spectrum. The C$^{18}$O emission requires at least four
distinct components (Fig.~\ref{Fig:spectrlfam26}) at 2.7, 3.5, 4.4 and
5.0 km s$^{-1}$.  The three components at 2.7, 3.5 and 5.0 km s$^{-1}$
are the same as found in \citet{Boogert02} at the position of Elias
29. Most lines observed in \citet{Boogert02} peak short-wards of
5.0 km s$^{-1}$ component, which could be indicative of the 4.4 km
s$^{-1}$ component . However, the difference between these two components is
very pronounced in the spectra in Fig.~\ref{Fig:maplfam26} and
\ref{Fig:spectrlfam26} at the position of LFAM 26 so that we treat
them as separate. HCO$^+$ is aligned with the 4.4 km s$^{-1}$ component.



Using the molecular excitation program RADEX \citep{vanderTak07}, one
can calculate the expected average contribution from cloud material to
the HCO$^+$ 4--3 emission.  Using a HCO$^+$ abundance of 10$^{-8}$
with respect to H$_2$, a density of 10$^4$ cm$^{-3}$, a cloud
temperature of 15 K, a width of 1 km s$^{-1}$ and a H$_2$ column
density of 10$^{22}$ cm$^{-2}$ ($A_V\approx$ 10), a contribution of $\sim$ 0.1
K km s$^{-1}$ is found. Only at higher column ($\sim$10$^{23}$ cm$^{-2}$)  or volume densities ($\sim$1$\times$10$^5$ cm$^{-3}$) are contributions found from the cloud on the order of 1
K km s$^{-1}$, comparable to the observed 4.4 km
s$^{-1}$ emission  at off-positions such as (-50,50). This is in agreement with the numbers given in Table 1 of \citet{Evans99}. Thus, LFAM 26 is embedded in a high-density ridge.

Table \ref{table:c18oprop} shows the results of gaussian fitting at
other positions where multiple peaks in the C$^{18}$O 3--2 profiles
were found.  In the WL~19 field, south of the Elias 29 position, the
same four components are again detected, although the 5.0 km s$^{-1}$
layer is only seen north of WL 19.  In the IRS 44 field along the same
ridge to the south-west, the three layers at 2.6, 3.5 and 4.5 km
s$^{-1}$ are found.

Three layers are found in the Oph-B2 core, which contains the IRS 48
and IRS 37 sources. The first layer is at 2.8 km s$^{-1}$, the second
at 3.5 km s$^{-1}$ and the third at 4.2 km s$^{-1}$.
Outside these two regions, spectra of C$^{18}$O can be fitted with a
single gaussian, almost always at $\sim$ 3.6 km s$^{-1}$.  The spatial
variation in this layer seems to be minor compared with the other
components and it is thus the most
likely candidate for the foreground component with an extinction of
$A_{\rm{v}}\sim$5.5 mag, as suggested by \citet{Dickman90}.
IRS 63, located far from the L~1688 core is at 2.8 km
s$^{-1}$, the same velocity as one of the layers in both Oph-B2 and
Ophiuchus ridge region.

\subsection{Gas column density}

The total column density, $N_{\rm{H_2}}$, toward each source was
calculated from the integrated intensity of the C$^{18}$O 3--2
line,$\int T_{MB}$ dV, using the following formula :
\begin{equation}
N({\rm H_2})=X\times0.75\times10^{14}\frac{e\:^{(16.6/T)}\int T_{MB} dV}{1-e\:^{(-33.2/T)}}
\end{equation}
with $X$, the abundance of H$_2$ with respect to C$^{18}$O, assumed to
be 5.6$\times 10^{6}$ \citep{Wilson94} and $T$, the temperature of the cloud material,
assumed to be 30 K. The temperature is chosen to be higher than that
of the surrounding cloud material, $\sim$15~K, and reflects the warmer
gas associated with the YSO where CO is not frozen out. This formula
assumes the molecular excitation to be in LTE and isothermal. The
C$^{18}$O 3--2 line is assumed to be completely optically thin. The
resulting column densities can be found in Table \ref{table:prop},
column 3.  They range from 3.7$\times$10$^{22}$ cm$^{-2}$ for IRS 63
and the Oph F core to 31.6 $\times$10$^{22}$ cm$^{-2}$ in the Oph A
core. These correspond to visual extinctions $A_V$ of 40-400 mag
, assuming $N_{\rm H}/$$A_V$ =1.8$\times10^{21}$ cm$^{-2}$
mag$^{-1}$ \citep{rachford02} and $N_{\rm H}\approx
2N($H$_2$) (all hydrogen in molecular form).  This range of column densities
is similar to that found by \citet{Motte98} and \citet{Stanke06}
using 1.3 mm continuum imaging, as well as the column densities found
from the 850 $\mu$m dust, see $\S$5.4.

\begin{figure*}[!htp]
\begin{center}
\includegraphics[width=150pt]{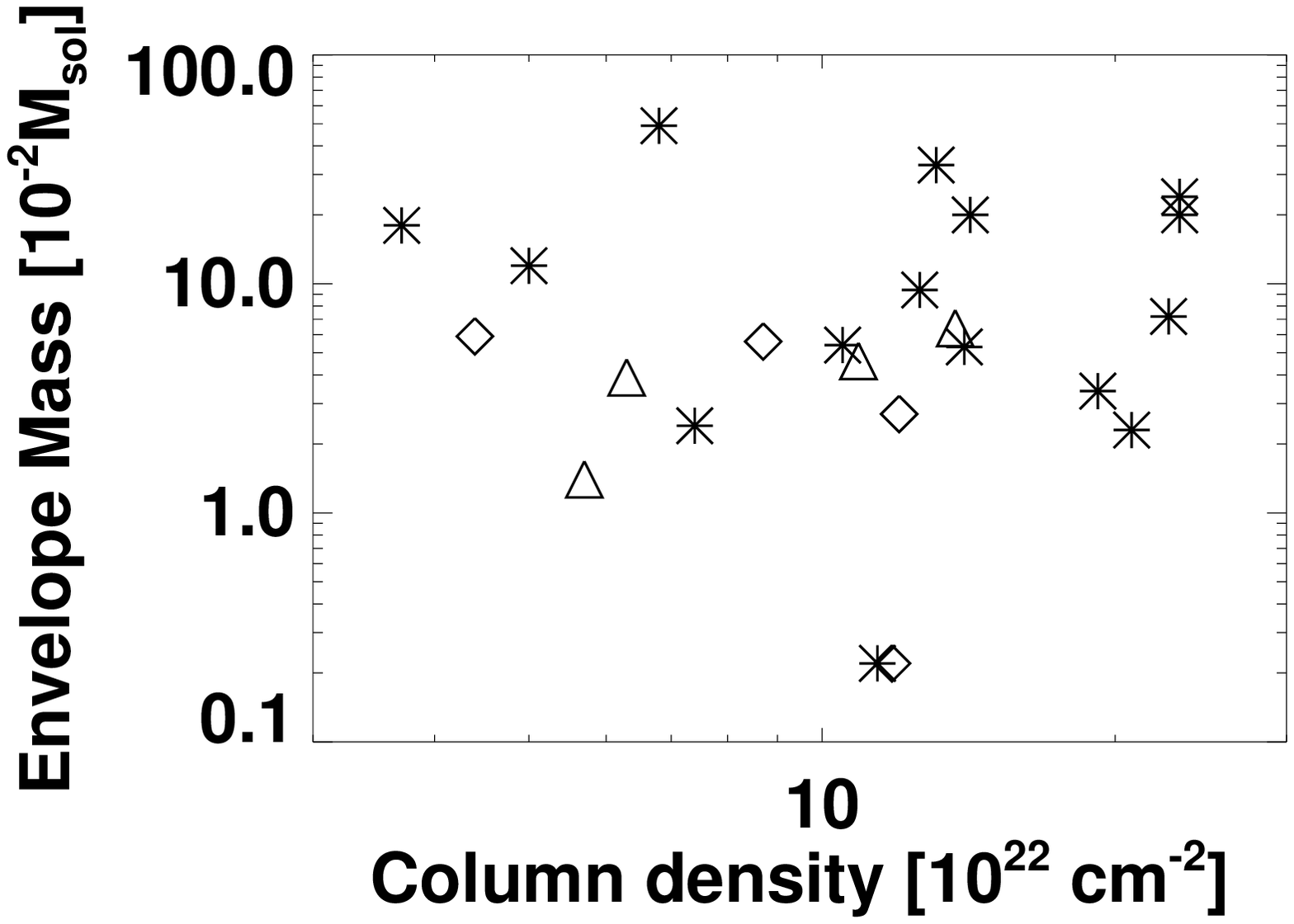}
\includegraphics[width=150pt]{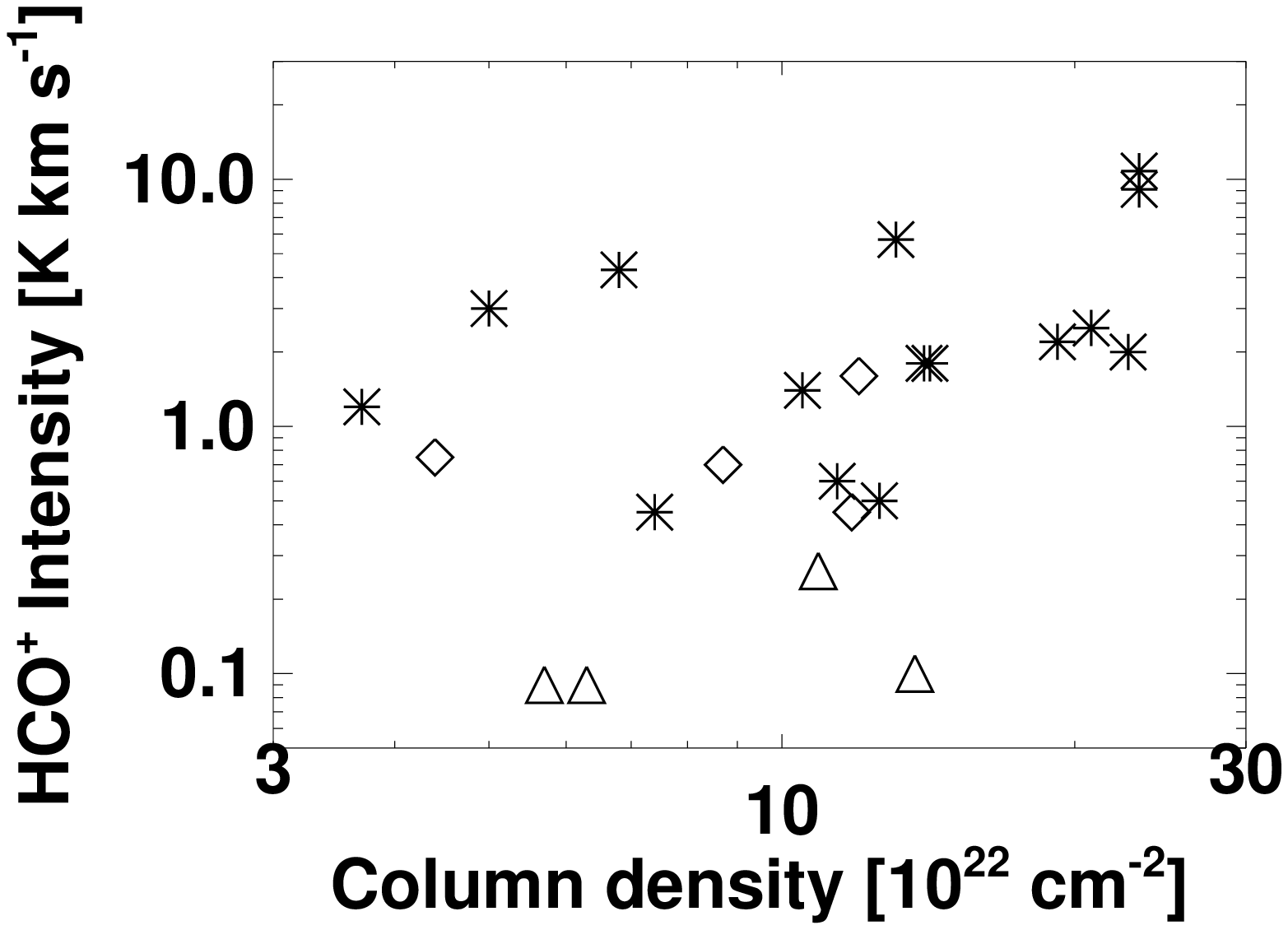}
\includegraphics[width=150pt]{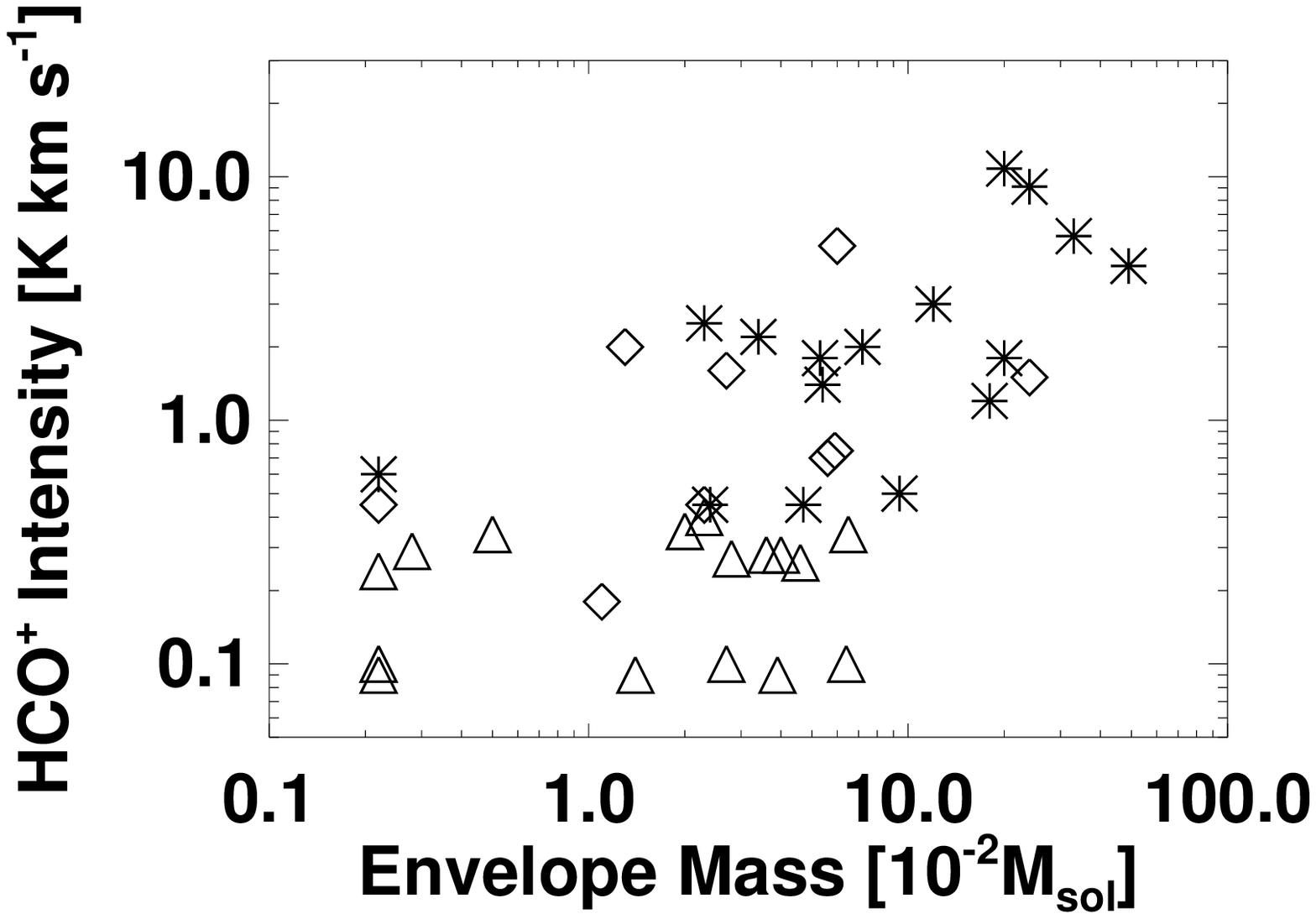}

\end{center}
\caption{{\it Left} : the gas column density (in 10$^{22}$ cm$^{-2}$) derived from
C$^{18}$O 3--2 vs.\ the envelope mass (in 10$^{-2}$ M$_{\rm{\odot}}$),
derived from the SCUBA maps; {\it middle} : the column density vs.\ the
integrated intensity of HCO$^+$ 4--3 in K km s$^{-1}$ 
and {\it right} : the integrated intensity of HCO$^+$ 4--3 vs.\ the envelope mass
. In all figures, sources with a centrally condensed
HCO$^+$ emission are shown with stars, sources with an upper limit on
HCO$^+$ 4--3 with triangles, and sources with HCO$^+$ emission not
peaking on the source with diamonds. }
\label{Fig:nh2_hco}
\end{figure*}

\subsection{Dust}

Table \ref{table:prop} includes the properties that can be derived
from the continuum emission.  The dust column density was calculated
with the following equation.
\begin{equation}
N({\rm H_2})=\frac{S_{850}}{\Omega \;\mu \;m_H \;\kappa_{850}\;B_{850}(T_d)}
\end{equation}
where $S_{850}$ is the flux in the central beam, $\Omega$ is the main beam solid angle
of 15$''$, $\mu = 2.33$ the mean molecular weight, $m_{\rm H}$ the
mass of atomic hydrogen, $\kappa_{850}$ the dust opacity per gram of
gas and dust at 850 $\mu$m and $B_{850}(T)$ the
Planck function at 850 $\mu$m for a temperature $T$, assumed to be 30
K. A $\kappa_{850}$ = 0.01 cm$^2$ gr$^{-1}$ is adopted from
\citet{Ossenkopf94} for dust with thin ice mantles (Table 1, column
4). 

Two assumptions affect the comparison between the column densities
derived from C$^{18}$O and those from dust continuum.  C$^{18}$O can
be frozen out onto the dust grains below $\sim$30 K at sufficiently
high densities, so that C$^{18}$O column densities are expected to be
lower than those obtained from dust data. Second, all dust emission is
assumed to originate in envelope or cloud material, whereas for
several sources the disk contributes significantly to the continuum.
These combined effects likely eplain the lack of correlation seen in Fig.\ref{Fig:nh2_hco} ({\it left}) \\

If one assumes that all emission detected at 850 $\mu$m is associated
with an isothermal envelope, a source mass can be
determined following Equation 4 from \citet{Shirley00}:\\
\begin{equation}
\begin{split}
M_D\; = \;\frac{S_{850} \; D^2}{B_\nu (T_d)\;\kappa_\nu}= 3.69 \;\times \;10^{-6}\;M_{\odot} \\
\times \; S_{850} (\rm{Jy})\;D^2(\rm{pc})\;\times \; (e^{16.9/T_d}\; -\;1)
\end{split}
\end{equation}

Since the 3$\sigma$ sensitivity of the COMPLETE map  is 80 mJy within a 15$''$ beam, the lower limit to the detectable mass is 0.04 M$_\odot$. Such a low mass implies either a source that has nearly shed its envelope and is close to the pre-main sequence phase or an intrinsically very low luminosity object \citep{Dunham08}. 

Fig. \ref{Fig:nh2_hco} shows the relation between the different column
densities, envelope masses and the presence of dense gas as traced by
HCO$^+$. These figures clearly show that at higher envelope masses,
the amount of dense gas and the total envelope mass correlate. The
third figure illustrates that nearly all sources with centrally
concentrated HCO$^+$ emission have larger envelope masses. Above an
envelope mass of 0.1 M$_{\rm{\odot}}$, only a single source that is classified as confused in $\S$ 6
(indicated with diamonds) is seen  with HCO$^+$ $>$ 2 K km/s. This is CRBR 2422.8-3423, which is located in the
line of sight with the envelope of IRS 43. The high mass found for
CRBR 2422.8-3423 can be fully attributed to this envelope.

\subsection{Effect of reddening on $\alpha_{\rm IR}$}
Foreground material reddens the YSO and steepens the spectral slope
$\alpha_{\rm{2-24 \mu m}}$. An estimate can be made of the foreground
column density based on our C$^{18}$O maps, by assuming that the
average column or $A_V$ at 45$''$ offset from the source position in
all directions originates entirely from this cloud. Although it is not known {\it a priori} whether this
material is foreground, back-ground, or a different YSO, we assume here that
it is all foreground (see also \S 5.2) which provides an {\it upper limit} to the reddening. Note that the column densities at these positions were calculated with a temperature of 15 K, a temperature associated with the surrounding cloud, instead of 30 K, a temperature associated with protostellar envelopes.
For typical dust opacities and the $A_V$=5.5 mag layer as suggested by \citet{Dickman90},
$\alpha_{\rm 2-24 \mu m}$ is only increased by 0.18 (see also
\citet{Crapsi08}). However, for higher $A_V$ of $\sim$ 40 mag, as
commonly found in the Ophiuchus ridge, this correction can be as high
as 1.1, enough to change the classification of a Class II T Tauri star
with disk to an embedded Class I source.
Table \ref{table:prop} includes the (maximum) change in $\alpha_{\rm
2-24 \mu m}$ due to foreground reddening for each source derived by
this method. \citet{Chapman08} probe the $A_K$ in the northern part of Ophiuchus, including the northern part of the Oph-A area, where $A_V$ values of 40 mag. and higher are found. In the Oph-A, Oph-B2 and Ridge regions, $A_V$ of this magnitude  were found from C$^{18}$O column densities, reinforcing the conclusion that a strong extinction affects these regions on scales of only a few arcminutes, a resolution often not reached by extinction studies \citep{Cambresy99}.
 Although these numbers have large uncertainties, they do
illustrate the point that this potential extreme foreground reddening can lead to incorrect
classifications for a significant number of sources in Ophiuchus.

 An alternative is to use the velocity resolved C$^{18}$O spectrum at the source position but subtract the layer that includes the source. For example, LFAM 26 ($\Delta\alpha_{\rm{IR}}$=4.4), resides in a dense ridge (see $\S$ 5.2). Subtracting the gaussian profile of this layer at $\sim$4.4 km s$^{-1}$ gives an integrated intensity of 2.9 K km s$^{-1}$, which corresponds to a column density of 10.7$\times10^{22}$ cm$^{-2}$, an $A_V$ of $\sim$60 mag and $\Delta\alpha_{\rm{IR}}$ = 3.2. If one assumes that the layers are arranged in distance in order of increasing $V_{\rm{LSR}}$, only the front layer at 3.5 km s$^{-1}$ will redden and a column density of 6.6$\times10^{22}$ cm$^{-2}$, an $A_V$ of $\sim$40 and $\Delta\alpha_{\rm{IR}}$=1.9 are found. Thus, $\Delta\alpha_{\rm{IR}}$ is likely to be in between the maximum value listed in Table 4 and $\Delta\alpha_{\rm{IR}}/2$, although exceptional cases, such as OphE-MM3, an edge-on disk in front of the cloud, do exist.

\subsection{SEDs: $L_{\rm Bol}$ and $T_{\rm Bol}$}

The IR and submillimeter fluxes have also been used to calculate
the bolometric luminosity, $L_{\rm{bol}}$, and temperature,
$T_{\rm{bol}}$, of each source. These numbers were calculated using the
prismodial or the midpoint methods, with the inclusion of the 2MASS,
IRAC, MIPS-24, MIPS-70, SHARC-II, SCUBA and 1.3 mm fluxes. Only fluxes
with a $S/N >$ 5 were included. For a more thorough discussion on the
useage of these two methods, see Dunham et al. (in prep.) and \citet{Enoch08}.\\

Table \ref{table:enoch} compares the values found for $T_{\rm{bol}}$
and $L_{\rm{bol}}$ in our work with those of \citet{Evans08} for
several sources. The differences stem from the fact that
\citet{Evans08} use CSO-Bolocam 1.1 millimeter data in a 30$''$ beam,
together with 2MASS, IRAC, and MIPS (24, 70 and 160 $\mu$m) fluxes. No
SHARC-II or SCUBA fluxes were included, nor were sources with no
detection at 1.1 mm.  As can be seen from the results for Elias 29 and
GSS 30 IRS1, the inclusion of Bolocam and MIPS-160 produces a higher
$L_{\rm{bol}}$ of up to a factor of 6. In addition, $T_{\rm{bol}}$ is
consistently lower for the brighter sources. This can be largely
attributed to the contributions from the surrounding cloud, which add
significantly to the MIPS-160 and Bolocam fluxes. Although cloud
emission is present in the IRAM-30m 1.3 mm and SCUBA-850 $\mu$m
observations used in our studies, the higher resolution of 15$''$
limits such contributions.  The exclusion of MIPS-160 in our work
causes the $L_{\rm{bol}}$ to be underestimated and the $T_{\rm{bol}}$
to be overestimated, due to the lack of a point at far-IR wavelengths
near the peak of the SED.
Isolated, unresolved sources such as IRS 42 show little to no
differences between the two studies, confirming the influence of the
environment.  Future high resolution ($<$10$''$) observations with the
PACS instrument on Herschel covering the 60 to 600 $\mu$m window will
be able to fully constrain the far-IR emission and SED.
\begin{table}[htp]
\caption{Comparison between $L_{\rm{bol}}$  and $T_{\rm{bol}}$ as calculated in \citet{Evans08} (indicated with Evans) and this work (indicated with van K.). }
\begin{center}
\begin{tabular}{l l l l l}
\hline \hline 
Source & $L_{\rm{bol}}$  & $L_{\rm{bol}}$  & $T_{\rm{bol}}$ &   $T_{\rm{bol}}$   \\ 
  & Evans & van K. & Evans & van K. \\
 & ($L_{\rm{\odot}}$)   & ($L_{\rm{\odot}}$) & (K) & (K) \\ \hline
GSS 30 & 8.7  & 3.3 & 150 & 123 \\
CRBR 2339.1  & $<$0.065  & 0.055 & $<$370& 426\\
WL 2 & 0.12  & 0.09 & 428 & 573\\
LFAM 26 & 0.15  & 0.044 & 110 & 238 \\
WL 17 & 0.6  & 0.67 & 310 & 323 \\
Elias 29 & 17.9  & 2.5 & 257 & 424 \\
IRS 42 & 1.2  & 1.4 & 600 & 540 \\ 
CRBR 2422.8 & 0.19  & 0.44 & $<$300 & 157 \\
IRS 43 & 3.8  & 1.0 & 160 & 134\\
IRS 44 & 15  & 1.1 & 110& 140 \\\hline
\end{tabular}
\end{center}
\label{table:enoch}
\end{table}

\subsection{Ice and silicate absorption}

In the online appendix, the IRS spectra for 28 of the sources in our
sample are plotted. 17 sources 
show ice absorption bands at 15.2 $\mu$m due to CO$_2$ ice (see
Table~\ref{table:class}), and 11 do not (WL 3, LFAM 26, VSSG 1, CRBR
2317.3, CRBR 2315.8, IRS 48, CRBR 2339.1, SR 21, Haro 1-4, IRS 46 and
SR 9). A few of the known edge-on disks, such as CRBR 2422.8-3423.2,
show ice absorptions. \\

\citet{Crapsi08} modelled the spectral features at 3 $\mu$m (H$_2$O ice), 10 $\mu$m
(silicate) and 15.2 $\mu$m (CO$_2$ ice) for a large grid of models,
including embedded sources and T Tauri disks, both seen at a wide
variety of angles. The ice absorptions at 3 and 15.2 $\mu$m were found
to be more prevalent for embedded sources. However, edge-on disks can
also show such absorptions in almost equal strength. The relation of
the 3 $\mu$m ice absorptions with the envelope mass of both embedded
and disk sources is given in Fig. 4 of \citet{Crapsi08} and a similar
relation was found for the 15.2 $\mu$m band. Cold foreground clouds
can also contribute
significantly. 

Using the determination of the column densities from the C$^{18}$O and
SCUBA maps in $\S$ 5.3 and 5.4, Eq.\ 3 and 4, the ice absorption can
be compared to the amount of reddening as indicated by
$\Delta\alpha_{\rm{IR}}$. Although ice absorptions are more commonly
found in sources with a higher reddening, it is concluded that the
YSOs must be characterized on a source by source basis to locate the
origin of the ices. The disk source CRBR 2422.8 has deep ice
absorptions \citep{Pontoppidan05}, most likely originating the large
column of foreground material ($\Delta\alpha = 0.9$). However, LFAM
26, which has an even higher reddening ($\Delta\alpha = 1.7$), shows
no sign of ice absorption. Of course, it is possible LFAM 26
is located in front of the cloud and as such the presence or absence of
ice absorption could serve as a useful diagnostic of the geometry of
the YSO-cloud system.
The silicate feature around 10 $\mu$m
is detected for fewer sources than the 15.2 CO$_2$ ice
absorption. Although LFAM 26, WL17, Elias 29, IRS 42, WL 3, IRS 54,
CRBR 2315.8-1700, CRBR 2339.1-2032 and CRBR 2422.8-3423 all have
silicate absorption, many other sources were not observed at the
wavelengths of the silicate feature.

\section{Classification}

 \begin{table}[!htp]
\caption{The classification of the source sample with the recipe outlined in the text. }
\small
\begin{center}
\begin{tabular}{l l l }
\hline \hline 
Source      & Code$^a$ & Stage \\ \hline
\multicolumn{3}{c}{\textbf{Embedded sources}}\\ \hline
GSS 30	       & S$_{HC8}$ E$_{H8}$ P$_{H8}$I			&  1 \\
LFAM 1	       & S$_{H8}$ N$_{C}$ E$_{H8}$ P$_{H}$ O$_8$  		&1 \\
VLA 1623       & S$_{HC8}$ E$_{H8}$ P$_{H8}$  			&0 \\
WL 12	       & S$_{HC8}$ E$_{HC8}$ P$_{HC8}$I  			& 1     \\
LFAM 26        & S$_{HC8}$ E$_{HC8}$ P$_{H8}$ O$_C$   		& 1    \\
WL 17	       & S$_{H8}$ N$_{C}$ P$_{H8}$I 				& 1(T) \\
Elias 29       & S$_{HC8}$ E$_{H8}$ P$_{H8}$I 	  		&1 \\
IRS 37	       & S$_{HC}$ W$_{8}$ E$_{HC8}$ P$_{HC8}$I  		&1  \\
WL 3           & S$_{HC}$ W$_{8}$ E$_{HC8}$ P$_{H}$  		&1   \\
WL 6	       & S$_{HC}$ N$_{8}$I  				&1(T)  \\
IRS 43	       & S$_{HC8}$ E$_{H8}$ P$_{H8}$I					&1 \\
IRS 44	       & S$_{HC8}$ E$_{HC8}$ P$_{HC8}$  			&1 \\
Elias 32       & S$_{HC8}$ E$_{H8}$ P$_{H8}$I  			&1 \\
Elias 33       & S$_{HC8}$ E$_{H8}$ P$_{8}$I	  		&1  \\
IRS 54	       & S$_{HC}$ W$_{8}$ E$_{HC8}$ P$_{HC8}$I 		&1(T) \\
IRAS 16285-2355& S$_{H8}$ W$_{C}$ E$_{HC8}$ P$_{HC8}$ 		&1 \\
IRS 63	       & S$_{H8}$ W$_{C}$ E$_{HC8}$ P$_{HC8}$I 		&1(T)  \\ \hline
\multicolumn{3}{c}{\textbf{Confused Sources}}\\ \hline
CRBR 2324.1-1619    & S$_{HC8}$ E$_{HC8}$ O$_{HC8}$			&C \\
GY51           & S$_{8}$ N$_{HC}$ E$_8$ O$_8$ 			&C  \\
 WL19          & S$_{C8}$ N$_{H}$ E$_{HC8}$ O$_{HC8}$I		&C    \\
IRS 42	       & S$_{HC8}$ E$_{H8}$ O$_{H8}$I			&C   \\
GY 256	       & S$_{HC}$ N$_{8}$					&C  \\\hline
\multicolumn{3}{c}{\textbf{Disks}}  \\ \hline
SSTc2d J162527.6-243648&N$_{HC8}$				& 2   \\
GSS 26	       & S$_8$ N$_{HC}$ P$_8$				& 2  \\
CRBR 2315.8-1700        & S$_{H8}$ N$_{C}$ E$_H$ P$_8$ O$_H$			&2  \\
CRBR 2317.3-1925      & S$_{H}$ W$_8$ N$_{C}$				&2  \\
VSSG 1	       & W$_8$ N$_{HC}$ P$_8$				& 2  \\
CRBR 2339.1-2032         & S$_{H}$ W$_8$ N$_{C}$ P$_8$				&2 \\
WL 2           & W$_{H8}$ N$_{C}$ E$_{H8}$ O$_{H8}$			&2        \\
GY 224	       & W$_8$ N$_{HC}$ P$_8$I				&2 \\
 WL 20S	       & S$_{C8}$ N$_{H}$ P$_8$I				&2 \\
IRS 48	       & W$_{C8}$ N$_{H}$ E$_C$ P$_8$			&2  \\
GY 312         & N$_{HC8}$					&2 \\
IRS 51	       & S$_{H8}$ W$_{C}$ E$_{HC8}$ P$_8$ O$_{HC}$I		&2  \\
SSTc2d J162741.6-244645& W$_8$ N$_{HC}$				& 2  \\
SSTc2d J162748.2-244225&N$_{HC8}$				& 2 \\
SSTc2d J162857.9-244055&S$_8$ N$_{HC}$				& 2 \\
Haro 1-4       & N$_{HC8}$					& 2 \\
DoAR 25	       & S$_8$ N$_{HC}$ P$_8$				& 2 \\
OphE MM3       & S$_{HC}$ W$_8$ E$_{HC8}$ O$_{H8}$			&2   \\
SR 21          & W$_8$ N$_{HC}$ P$_8$				& 2 \\ 
CRBR 2422.8-3423      & S$_{H8}$ N$_{C}$ E$_8$ O$_8$I			&2  \\
IRS 46	       & S$_8$ W$_{C}$ N$_{H}$ E$_{HC8}$ O$_{HC8}$ 		&2 \\
SR 9           & N$_{HC8}$					&2\\
2Mass 16282    & W$_8$ N$_{HC}$					&2 \\ \hline 
\end{tabular} \\
\end{center}
$^a$The coding in column 2 is as follows:\\
S, W and N determine if a line or continuum is Strong, Weak or Not detected/observed.\\
E, P and O determine if a source is Extended, Peaking or Offset peaking. \\
The subscripts H, C and 8 refer to the HCO$^+$, C$^{18}$O and 850 $\mu$m. \\
I is detected CO$_2$ ice absorption.
Column 3 lists the classification. 1 is a Stage 1 embedded YSO, 2 is a disk and C is Confused.
\label{table:class}
\end{table}

\subsection{Physical classification}

A new classification based on physical parameters has gradually been
introduced \citep{Whitney03a,Whitney03b,Robitaille06} but is not yet
as commonly in use as the traditional Class system. This new
classification identifies the evolutionary stage using the physical
characteristics as opposed to the observed characteristics. The
different evolutionary stages are determined from the ratios between
$M_{\rm{disk}}$, $M_{\rm{envelope}}$ and $M_{\rm{star}}$. The total
circumstellar mass, $M_{\rm{circum}}$, is defined as
$M_{\rm{disk}}$+$M_{\rm{envelope}}$. The evolutionary stages of this
classification are:
\begin{itemize}
\item Stage 0, deeply embedded sources with $M_{\rm{disk}}$/$M_{\rm{envelope}}$ $<<$ 1 and $M_{\rm{circum}}$/$M_{\rm{star}}\sim$ 1
\item Stage 1, embedded sources with 0.1 $<$ $M_{\rm{disk}}$/$M_{\rm{envelope}}$ $<$ 2 and $M_{\rm{circum}} < $  $M_{\rm{star}}$
\item Stage 2, classical T Tauri stars with gas-rich disks ($M_{\rm{envelope}}$ = 0 and $M_{\rm{disk}}$/$M_{\rm{star}}$ $<<$ 1 )
\item Stage 3, Pre-main sequence stars with tenuous disks.
\end{itemize}

Stage 0 sources are equivalent to Class 0 sources and can be
identified by their submm characteristics as originally put forward by
\citet{Andre93}.  Stage 2 and Stage 3 sources, corresponding to the
classical T Tauri stars with disks and pre-main sequence stars with
(gas-poor) tenuous disks, are identified using the IR excess combined
with optical/near-IR spectroscopic properties. The amount of IR
excess and the wavelength where this excess starts are often used to
distinguish Stage 2 and 3 sources.

The Stage~1 sources are most problematic to uniquely identify based on
observational characteristics. Both \citet{Whitney03b} and
\citet{Crapsi08} show that the traditional identification methods of
using the IR spectral slope $\alpha_{\rm{IR}}$ and $T_{\rm{bol}}$ are
insufficient to distinguish the edge-on disks from embedded sources,
as well as missing evolved, face-on embedded YSOs that are
(mis)classified as Class II.  One of the main parameters to
unambiguously identify Stage 1 sources is the
$M_{\rm{disk}}$/$M_{\rm{envelope}}$ ratio as determined from
millimeter continuum observations in a large and small ($\sim 2''$)
beam \citep{Crapsi08}. However, the interferometric observations
needed to unambiguously constrain the disk masses are still
time-consuming for large samples of sources.

\subsection{Identifying embedded stage 1 sources with molecular emission}

To uniquely identify embedded sources and distinguish them from
edge-on disks, prestellar cores and possible background AGB stars and
galaxies without having to resort to interferometers, we propose to
use the single-dish molecular emission.  Embedded sources will be
bright and centrally peaked in HCO$^+$ 4--3 and C$^{18}$O 3--2. The
HCO$^+$ line maps will detect all but the most tenuous envelopes. The
concentration parameter shows that HCO$^+$ 4--3 is much more accurate
in identifying protostellar envelopes than continuum dust
emission. The HCO$^+$ maps also spatially resolve larger envelopes
(e.g., the VLA 1623 and GSS 30 envelopes). The nearby environment,
including foreground layers, is in turn characterized by the
C$^{18}$O 3--2 (see $\S$ 4 and 5).  

To distinguish embedded sources from isolated edge-on disks, a good
limit on the HCO$^+$ 4--3 emission is $ T_{\rm{MB}}>$ 0.4 K.
\citet{Thi04} observed four well-studied gas-rich disk sources in a
large range of molecular transitions, including HCO$^+$ 4--3, and
found $T_{\rm{MB}}\sim$ 0.1 K, four times below our adopted
limit. Although disks around embedded sources could be somewhat larger
and more massive, it is assumed that the contribution from even the
largest embedded disks does not exceed our adopted limit. This value is naturally small because of the severe dilution of emission originating in a disk within the 15 $"$ JCMT beam. For example, a 20 K optically thick line from a 2$''$ disk would be diluted to 0.35 K.  Indeed, for
two embedded sources for which interferometric HCO$^+$ data exist (IRS
63 and Elias 29, \citet{Lommen08}), the HCO$^+$ contribution of the
disk does not exceed this limit (see \S 6.3). 
 
The following definitions were therefore used to classify the sample: \\
\begin{itemize}

\item Stage 1 
\begin{itemize} 
\item Extended HCO$^+$ emission, peaking on source, with $C_{\rm HCO^+}>$ 0.7 and $\int T_{\rm{MB}}$dV $>$ 0.4 K km s$^{-1}$ {\it and/or} 
\item an HCO$^+$ profile that is not extended, but $\int T_{\rm{MB}}$dV $>$ 0.4 K km s$^{-1}$, with SCUBA both extended and $C_{850}>$ 0.7. 
\item Ice absorptions are usually prominent.
\end{itemize}
\item Stage 2 
\begin{itemize} 
\item Absence of HCO$^+$, SCUBA and C$^{18}$O down to the rms limits at the IR position {\it or} 
\item no variation of HCO$^+$, SCUBA and C$^{18}$O on scales of 30$''$ ($C <$ 0.4) {\it or } 
\item no C$^{18}$O 3--2 or SCUBA is seen at scales larger than the central beam of 15$''$ with HCO$^+$ 4--3 not extended and $<$ 0.4 K km s$^{-1}$ {\it or} 
\item $\alpha_{\rm{IR}}<$ -0.5.
\end{itemize}

\item Confused 
\begin{itemize}
\item  HCO$^+$, SCUBA or C$^{18}$O peaking more than 20$''$ away from the IR position \\
\end{itemize}
\end{itemize}

In the rare case that an embedded source is viewed face-on (into the outflow cone, see \citet{Whitney03a}), the above restrictions will identify such a source as stage 1, having $\alpha_{\rm{IR}}<-0.3$, but with extended HCO$^+$ emission, peaking on source ($\int T_{\rm{MB}}$dV $>$ 0.4 K km s$^{-1}$).

Table \ref{table:class} shows the results for this classification. In
Appendix C the sources and their classifications are discussed on a
source by source basis.  Sixteen Class I sources were identified as
embedded Stage 1 sources with envelopes varying in mass and
size. The sources GSS 30 IRS 1 and LFAM 1 appear to be embedded
within an approximately spherical envelope encompassing both
sources. The sources IRS 37 and WL 3 are embedded in a highly
non-spherical envelope covering both sources. In addition to these 16
sources, VLA 1623 is classified as embedded, but considered a Stage 0
source due to its high sub-mm flux relative to the total luminosity.
From comparison with \citet{Jorgensen08} (see $\S$ 6.5), we can conclude that all embedded sources were found  down to a mass limit (envelope+disk) of 0.04 M$_{\rm{\odot}}$. Deep searches for even lower luminosity sources have found only two likely candidate sources within our field  \citep{Dunham08}. Deep (interferometric) sub-mm observations are needed to identify if such sources are truly embedded.
 
Five sources were found to be confused  (C in Table
\ref{table:class}). The exact nature of these sources (obscured, disk or other) cannot be
determined without interferometric observations. However, molecular
emission rules out a protostellar envelope. Observations of C$^{18}$O
show that cloud material is located in the line of sight towards these
sources. They will be further discussed in $\S$6.4.

Twenty-three sources were found to be Stage 2 sources (2 in in Table
\ref{table:class}). Of this sample, sources with $\alpha_{\rm{IR}}>0$ and with no correction from extinction (see Table 4)
are likely to be edge-on or close to edge-on disks. Sources that do have a potential correction to $\alpha_{\rm{IR}}$, such as CRBR 2422.8 or IRS 48, do not have to be edge-on. All sources that
were initially known to be disk sources are identified as Stage 2
sources using this method. If one considers that disk sources with $\alpha_{IR}$  $>$-0.3,  when corrected for the maximum $\Delta\alpha_{\rm{IR}}$), are edge-on, only 7 out of 23 Stage 2 sources of our sample are expected to be edge-on. The total amount of Class II sources found in Ophiuchus by c2d is 176, and although this includes L~1689, the bulk of these disks exist close to or are part of the L~1688 region \citep{Evans08}. Thus, our number of edge-on disks appears to be only a small fraction of the total disk sample.

No face-on embedded sources were found that have $\alpha_{\rm{IR}}<-0.3$,  but do show a HCO$^+$ spectrum that is both peaking on source and stronger than 0.4 K km s$^{-1}$. This very low fraction is not surprising, since such sources have strong restrictions on the line of sight.

Using the results from Spitzer (\citet{Evans07}, \citet{Jorgensen06}, \citet{Evans08}, Allen et al. in prep) a comparison can be made between the classification using $\alpha_{\rm{IR}}$ and the new method. It is found that  $\sim$50$\%$ (11 out of 22) of the sources classified as Class I are Stage 2 disks and not embedded. Note that this includes the known disks OphE MM3 and CRBR 2422.8. However, of the Flat spectrum sources, $\sim$50$\%$ (6 out of 13) appear to be embedded, while the other half is either confused or classified as a Stage 2 disk. In the end, the sample of embedded sources found by c2d changes from 22 out of 45 to 17 out of 45. 

\begin{table}[htp]
\caption{Comparison between the classification of YSOs using $\alpha_{\rm{IR}}$ by \citet{Evans08} and our method using molecular emission. }
\begin{center}
\begin{tabular}{l l l l l}
\hline \hline 
     & Class I & Flat Spectrum & Class II & Total\\ \hline
Stage 1  & 11 & 6 & 0 & 17 \\
Confused$^a$ & 1 & 3 & 1 & 5 \\
Stage 2 & 10 & 4 & 9 & 23  \\ \hline
Total & 22 & 13 & 10 & 45 \\ \hline

\end{tabular}
\end{center}
$^a$Note that although it is possible for Flat Spectrum sources to be embedded, confused sources are ruled out to be embedded.
\label{table:statis}
\end{table}

\subsection{Late stage 1 sources}
Of the sixteen embedded Stage 1 sources, four sources (WL 6, WL 17,
IRS 54 and IRS 63) were found to have only marginal envelopes left,
with C$^{18}$O and 850 $\mu$m emission suggesting that at most a few
$\times 0.01 M_{\odot}$ remains in the envelope.  \citet{Lommen08}
observed IRS 63 using the Sub-Millimeter Array (SMA) and found that 50
\% of the continuum emission originates within a disk no bigger than
200 AU in size with a mass of 0.055 $M_{\rm{\odot}}$ and the remainder
in a remnant envelope with a mass of 0.058 $M_{\rm{\odot}}$. HCO$^+$
3--2 was also detected with the SMA, originating mostly in the
disk. When convolved with the JCMT beam, the observed brightness is
0.28 K km s$^{-1}$.  The HCO$^+$ 4--3 single-dish intensity as
observed with HARP-B is 0.75 K km s$^{-1}$, 3 times as bright. 
For typical disk excitation conditions, HCO$^+$ 4--3 is expected to be equal or
weaker compared to the 3--2. It can thus be concluded from the much higher observed intensity in the 4--3 line, that at least half of the 4--3 emission
within the JCMT beam likely originates in the protostellar envelope and would not be seen by the SMA.

In all four sources, little to no extended C$^{18}$O is seen. The
distributions and strengths in the HCO$^+$ and C$^{18}$O lines of the
other three sources are similar to that of IRS 63, but the SCUBA image
of IRS 63 is a factor of 4 brighter. For IRS 54,
the envelope is spatially resolved in HCO$^+$, C$^{18}$O and SCUBA.

Although rare, it is concluded that these sources are Stage 1 sources
in transition to a Stage 2 source (marked with 1(T) in Table
\ref{table:class}). They have accreted or dispersed almost their
entire envelope. Continuum emission in the sub-mm is likely to contain
a large contribution from the protostellar disk. 

\subsection{Confused sources}

For the five confused sources, CRBR 2324.1-1619, GY 51, WL 19, IRS 42
and GY 256, the exact nature cannot be determined using HCO$^+$,
except that such sources are ruled out as embedded YSOs. None of them
have the characteristic peak of HCO$^+$ that most Stage 1 embedded
YSOs have. However, at least two out of the three tracers, HCO$^+$,
C$^{18}$O and SCUBA, are detected so a few possibilities remain. The
first is an (edge-on) disk, in {\it front} of cloud material. This
would be similar to the case of OphE-MM3, which was confirmed to be an
edge-on disk in front of a dense core \citep{Brandner00}. Without the
near-IR imaging, OphE-MM3 would have been classified as confused. A
second possibility is that they are background sources.  The most
likely options are then a T Tauri star with disk {\it behind} the
cloud (possibly edge-on), or an AGB star. Background main sequence stars are
identified within the c2d delivery document \citep{Evans07} and are thus highly
unlikely to be included in our sample. Chances of background AGB stars
or galaxies being aligned with the cloud are small, but according to
\citet{Jorgensen08}, a single background source can be expected. A
final possibility is that the sources are very late Stage 1 embedded
YSOs, very close to the Stage 2 phase. An upper limit of only 0.02
M$_\odot$ is found for their envelope mass.

\subsection{Comparison to other methods}

\begin{figure*}[!htp]
\begin{center}
\includegraphics[width=410pt]{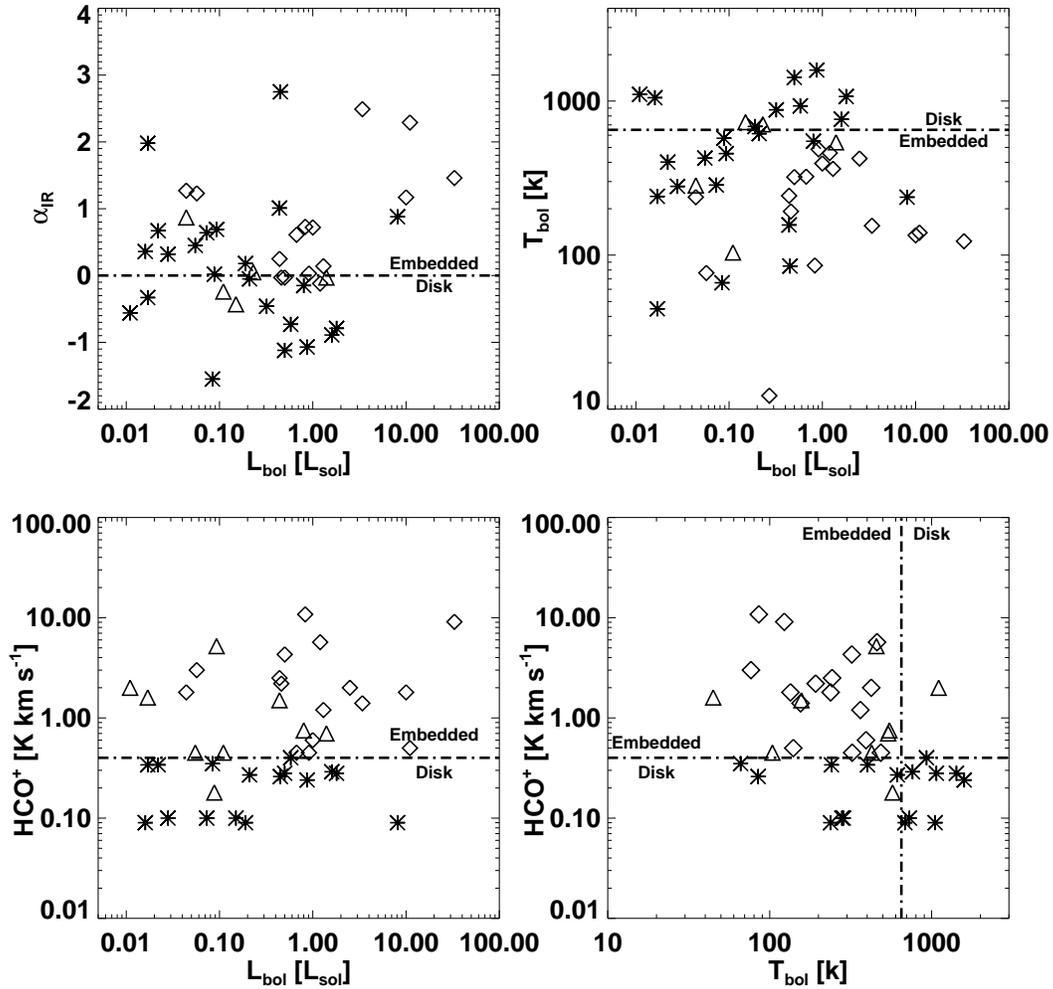}
\end{center}
\caption{Classification of sources using the spectral slope ({\it
    upper left}), bolometric temperature ({\it upper right}), HCO$^+$
  intensity ({\it lower left}) and HCO$^+$ intensity combined with the
  bolometric temperature ({\it lower right}).  In the upper two plots,
  embedded sources are shown with a diamond, disks are shown with a
  star, and confused sources with a triangle. In the lower two plots,
  diamonds are again embedded sources, stars represent sources with an upper
  limit on the HCO$^+$ and triangles are sources with no HCO$^+$ peak
  associated with the source. Note that the confused sources and disk
  sources are both represented with triangles and stars.  The traditional limits
  used for classification, $\alpha_{\rm{IR}}>$0 and $T_{\rm{bol}}<650$ K
  as well as our criterion, HCO$^+ > 0.4 $ K km s$^{-1}$, are shown with dashed lines.}
\label{Fig:prop}
\end{figure*}

Fig. \ref{Fig:prop} shows the effectiveness of the classification of
sources using the method above, as compared to methods using
$\alpha_{\rm{IR}}$ and $T_{\rm{bol}}$. A key parameter is the strength
of the HCO$^+$ integrated intensity. The following limits were adopted
between embedded and disk sources: $\alpha_{\rm{IR}}>0$,
$T_{\rm{bol}}<650$ K and HCO$^+ > 0.4 $ K km s$^{-1}$. 

The advantages of the method using molecular emission are immediately
apparent. Although the classical methods of using $T_{\rm{bol}}$ and
$\alpha_{\rm{IR}}$ are able to identify embedded sources, both methods
also incorrectly identify a number of Stage 2 disks as embedded,
especially for $L_{\rm{bol}}<1$ L$_{\rm{\odot}}$. These are the
edge-on disks. If one uses the limits of the integrated intensity for
HCO$^+$, isolated disk sources are easily identified. With the
additional restriction of having a peak within the HCO$^+$ map,
confused sources or sources in front of the cloud can be readily
identified.

The use of ice absorption, as studied by \citet{Crapsi08}, is limited
by similar constraints. Most embedded sources, with the exception of
LFAM 26 and WL~3, show CO$_2$ ice absorption at 15.2 $\mu$m. However,
4 disk sources IRS 51, WL 19, WL 20S and CRBR 2422.8-3423 also show
strong ice absorptions. These ice absorptions are caused both by the
foreground layers as discussed in $\S$5.2 as well as the material in
the disk itself, if viewed edge-on (as is the case for CRBR
2422.8). Thus, the presence of ice absorption cannot be used to
unambiguously identify embedded sources from other sources. Even if
the origin of the ice absorption can be attributed to a protostellar
envelope, its strength does not seem to be an indication of evolution,
e.g. IRS 63, a small envelope, has a
deeper ice absorption than the large envelope of GSS 30.

\citet{Jorgensen08} published a list of `candidate' embedded YSOs
based on two criteria. First the colors of sources using the IRAC and
MIPS results ($[3.6]-[4.5]$ and $[8]-[24]$). Second, the proximity of
MIPS sources to SCUBA cores. For these cores, the concentration is an
important parameter. Although most candidate embedded objects are
identified by both criteria, several sources are included in their
list based on only a single one. Comparison between their list 
(Table 1 in \citet{Jorgensen08} limited to L~1688,
further referenced as JJ1) and the list in this paper yields the
following results:

\begin{itemize}

\item Four embedded YSOs with SCUBA fluxes below the cut-off adopted by JJ1
  ($<$ 0.15 Jy beam$^{-1}$, the very low envelope masses) are absent from
  the JJ1 list. 
Two of
these sources (WL6 and IRS 54) are classified by us as late Stage 1
  sources with little to no envelope left. 
The other two embedded YSOs not included in JJ1
are IRS 37 and WL 3.

\item Four sources are included in the JJ1 list that have been classified as
  Stage 2 disks by our method. These are GSS 26, CRBR 2315.8, CRBR
  2339.1-2032 and IRS 51. All four sources have associated SCUBA cores
  and MIPS detections. See the appendix for the classification
  reasoning of each of these sources.

\end{itemize}

\section{Conclusions}

A sample of young stellar objects in L~1688 was analyzed using gas
mapping obtained with the new HARP-B heterodyne array receiver in the
HCO$^+$ 4--3 and C$^{18}$O 3--2 lines. Complementary dust maps were
obtained from the COMPLETE project, as observed by JCMT-SCUBA, and
with SHARC-II on the CSO.  The original sample consisted of 45
sources, mostly classified as embedded YSOs or flat-spectrum sources
using their spectral slope $\alpha_{\rm{IR}}$ 
in previous work. Of this sample a few sources were recently
discovered to be edge-on disks. As a control sample, 4 known disk
sources in L~1688 were included. The observations were supplemented by
single-pixel observations from APEX, Spitzer-IRS spectroscopy and
continuum photometry ranging from 1 $\mu$m to 1.3 mm, using a variety
of space-based and ground-based observatories.

The main conclusions are:
\begin{itemize}

\item The concentration of the dense gas, as traced by the HCO$^+$
  4--3 line mapping, provides an excellent tool to characterize the dense
  gas in the inner regions of protostellar envelopes. Material in the
  cold outer envelopes, (edge-on) disks, prestellar cores or cloud
  material does not emit strongly in HCO$^+$ 4--3.

\item Most envelopes in L~1688 have low masses, ranging from 0.05 to
  0.5 M$_{\rm{\odot}}$. The main accretion phase onto the star has
  already taken place. The only exception is the Stage 0 source VLA
  1623 which contains nearly 1 M$_{\rm{\odot}}$.

\item A new classification tool based on molecular emission to
  uniquely identify embedded sources is proposed. Of the total sample,
  17 sources were found to be embedded, 5 sources were confused and 23
  sources were identified as (edge-on) disks. Due to the foreground layers found through the C$^{18}$O, it is concluded that a significant fraction of the sources previously classified as Class I or Flat Spectrum are disks and have been strongly reddened. Combined with the analysis in \citet{Jorgensen08} all embedded sources in L~1688 were included and characterized down to a mass limit of  0.04 M$_{\rm{\odot}}$ (envelope+disk).

\item Four embedded YSOs are found that are in transition from the
  Stage 1 embedded phase to the Stage 2 T Tauri phase. These sources,
  IRS 63, IRS 54, WL 6 and WL~17, have little envelope left
  ($M_{\rm{env}} \sim 0.05 M_{\rm{\odot}}$).

\item Five sources are so confused by cloud material that identifying
  them as T Tauri disks or background sources is not possible with our
  data. The (lack of) variation and concentration of the molecular
  emission rules them out as embedded YSOs, however.


\item All sources previously identified as disks using traditional
  methods such as $T_{\rm{bol}}$ and $\alpha_{\rm{IR}}$ are indeed
  recovered as disks as are all edge-on disk sources, identified by
  near-IR mapping (e.g. IRS 46, CRBR 2422.8-3423 and OphE-MM3).

\item Spectral line mapping of C$^{18}$O reveals that foreground
  layers are predominantly present in the Ophiuchus ridge and the
  Oph-B core. Outside these regions, foreground layers are
  absent. These foreground layers are responsible for heavy reddening with typical $A_V > 20$, confusing classification schemes based on IR colors.

\end{itemize}

The YSOs in the Ophiuchus clouds show a wide variety of
characteristics, ranging from the rare Stage 0 deeply embedded source
VLA 1623 and Stage 1 embedded YSOs such as Elias 29, IRS 43 and WL 12
to the embedded YSOs with wimpy envelopes in transition to the T Tauri
stage like IRS 63.  The results above imply that the molecular
emission maps uniquely characterize the envelopes associated with
embedded YSOs, while simultaneously identifying disks erroneously classified as embedded YSOs. Such
characterization is needed before statistical studies can be used to
estimate time scales and star formation efficiencies within L~1688.\\

 One characteristic that has yet to be determined is the ratio of the disk to envelope mass. 
Interferometric studies with advanced facilities such as the SMA, CARMA or the PdB interferometer will allow the disks to be characterized  and, through comparison with continuum single dish studies, the envelope \citep[e.g.][]{Jorgensen05,Lommen08}. 
Such detailed
information will be necessary as a starting point
for the interpretation of future
observations with  Herschel and ALMA
of the embedded YSOs in
Ophiuchus and other clouds. A central role is envisioned for molecular
line data.


\begin{acknowledgements}

  TvK and astrochemistry at Leiden Observatory are supported by a
  Spinoza prize and by NWO grant 614.041.004. Remo Tilanus and Jan Wouterloot are
  thanked for the extensive support on the observing and reduction of
  HARP-B data and Arno Kockx for carrying out a large part of the
  HARP-B observations. The support from the APEX and MPIfR staff on
  the APEX2a data is greatly appreciated. The assistance on the usage
  of the COMPLETE data from Doug Johnstone and Helen Kirk is
  recognized. The authors are grateful to the c2d team, especially
  Mike Dunham (general), Luisa Rebull (MIPS) and Fred Lahuis (IRS),
  for their help on the Spitzer data and useful discussion.

\end{acknowledgements}

\bibliographystyle{aa}
\bibliography{biblio}

\begin{thebibliography}{83}
\expandafter\ifx\csname natexlab\endcsname\relax\def\natexlab#1{#1}\fi

\bibitem[{{Adams} {et~al.}(1987){Adams}, {Lada}, \& {Shu}}]{Adams87}
{Adams}, F.~C., {Lada}, C.~J., \& {Shu}, F.~H. 1987, \apj, 312, 788

\bibitem[{{Alexander} {et~al.}(2003){Alexander}, {Casali}, {Andr{\'e}},
  {Persi}, \& {Eiroa}}]{Alexander03}
{Alexander}, R.~D., {Casali}, M.~M., {Andr{\'e}}, P., {Persi}, P., \& {Eiroa},
  C. 2003, \aap, 401, 613

\bibitem[{{Andr\'e} \& {Montmerle}(1994)}]{Andre94}
{Andr\'e}, P. \& {Montmerle}, T. 1994, \apj, 420, 837

\bibitem[{{Andr\'e} {et~al.}(1993){Andr\'e}, {Ward-Thompson}, \&
  {Barsony}}]{Andre93}
{Andr\'e}, P., {Ward-Thompson}, D., \& {Barsony}, M. 1993, \apj, 406, 122

\bibitem[{{Andrews} \& {Williams}(2007)}]{Andrews07}
{Andrews}, S.~M. \& {Williams}, J.~P. 2007, \apj, 671, 1800

\bibitem[{{Barsony} {et~al.}(1997){Barsony}, {Kenyon}, {Lada}, \&
  {Teuben}}]{Barsony97}
{Barsony}, M., {Kenyon}, S.~J., {Lada}, E.~A., \& {Teuben}, P.~J. 1997, \apjs,
  112, 109

\bibitem[{{Barsony} {et~al.}(2005){Barsony}, {Ressler}, \& {Marsh}}]{Barsony05}
{Barsony}, M., {Ressler}, M.~E., \& {Marsh}, K.~A. 2005, \apj, 630, 381

\bibitem[{{Blake} {et~al.}(1995){Blake}, {Sandell}, {van Dishoeck},
  {Groesbeck}, {Mundy}, \& {Aspin}}]{Blake95}
{Blake}, G.~A., {Sandell}, G., {van Dishoeck}, E.~F., {et~al.} 1995, \apj, 441,
  689

\bibitem[{{Blake} {et~al.}(1994){Blake}, {van Dishoek}, {Jansen}, {Groesbeck},
  \& {Mundy}}]{Blake94}
{Blake}, G.~A., {van Dishoek}, E.~F., {Jansen}, D.~J., {Groesbeck}, T.~D., \&
  {Mundy}, L.~G. 1994, \apj, 428, 680

\bibitem[{{Bontemps} {et~al.}(2001){Bontemps}, {Andr{\'e}}, {Kaas}, {Nordh},
  {Olofsson}, {Huldtgren}, {Abergel}, {Blommaert}, {Boulanger}, {Burgdorf},
  {Cesarsky}, {Cesarsky}, {Copet}, {Davies}, {Falgarone}, {Lagache},
  {Montmerle}, {P{\'e}rault}, {Persi}, {Prusti}, {Puget}, \&
  {Sibille}}]{Bontemps01}
{Bontemps}, S., {Andr{\'e}}, P., {Kaas}, A.~A., {et~al.} 2001, \aap, 372, 173

\bibitem[{{Boogert} {et~al.}(2008){Boogert}, {Pontoppidan}, {Knez}, {Lahuis},
  {Kessler-Silacci}, {van Dishoeck}, {Blake}, {Augereau}, {Bisschop},
  {Bottinelli}, {Brooke}, {Brown}, {Crapsi}, {Evans}, {Fraser}, {Geers},
  {Huard}, {Jorgensen}, {Oberg}, {Allen}, {Harvey}, {Koerner}, {Mundy},
  {Padgett}, {Sargent}, \& {Stapelfeldt}}]{Boogert08}
{Boogert}, A., {Pontoppidan}, K., {Knez}, C., {et~al.} 2008, ArXiv e-prints,
  801

\bibitem[{{Boogert} {et~al.}(2002){Boogert}, {Hogerheijde}, {Ceccarelli},
  {Tielens}, {van Dishoeck}, {Blake}, {Latter}, \& {Motte}}]{Boogert02}
{Boogert}, A.~C.~A., {Hogerheijde}, M.~R., {Ceccarelli}, C., {et~al.} 2002,
  \apj, 570, 708

\bibitem[{{Boogert} {et~al.}(2000){Boogert}, {Tielens}, {Ceccarelli},
  {Boonman}, {van Dishoeck}, {Keane}, {Whittet}, \& {de Graauw}}]{Boogert00}
{Boogert}, A.~C.~A., {Tielens}, A.~G.~G.~M., {Ceccarelli}, C., {et~al.} 2000,
  \aap, 360, 683

\bibitem[{{Brandner} {et~al.}(2000){Brandner}, {Sheppard}, {Zinnecker},
  {Close}, {Iwamuro}, {Krabbe}, {Maihara}, {Motohara}, {Padgett}, \&
  {Tokunaga}}]{Brandner00}
{Brandner}, W., {Sheppard}, S., {Zinnecker}, H., {et~al.} 2000, \aap, 364, L13

\bibitem[{{Brown} {et~al.}(2007){Brown}, {Blake}, {Dullemond}, {Mer{\'{\i}}n},
  {Augereau}, {Boogert}, {Evans}, {Geers}, {Lahuis}, {Kessler-Silacci},
  {Pontoppidan}, \& {van Dishoeck}}]{Brown07}
{Brown}, J.~M., {Blake}, G.~A., {Dullemond}, C.~P., {et~al.} 2007, \apjl, 664,
  L107

\bibitem[{{Cambr{\'e}sy}(1999)}]{Cambresy99}
{Cambr{\'e}sy}, L. 1999, \aap, 345, 965

\bibitem[{{Chapman} {et~al.}(2008){Chapman}, {Mundy}, {Lai}, \&
  {Evans}}]{Chapman08}
{Chapman}, N., {Mundy}, L., {Lai}, S.-P., \& {Evans}, N. 2008, \apj, {in press}

\bibitem[{{Comeron} {et~al.}(1993){Comeron}, {Rieke}, {Burrows}, \&
  {Rieke}}]{Comeron93}
{Comeron}, F., {Rieke}, G.~H., {Burrows}, A., \& {Rieke}, M.~J. 1993, \apj,
  416, 185

\bibitem[{{Crapsi} {et~al.}(2008){Crapsi}, {van Dishoeck}, {Hogerheijde},
  {Pontoppidan}, \& {Dullemond}}]{Crapsi08}
{Crapsi}, A., {van Dishoeck}, E.~F., {Hogerheijde}, M.~R., {Pontoppidan},
  K.~M., \& {Dullemond}, C.~P. 2008, ArXiv e-prints, 801

\bibitem[{{Di Francesco} {et~al.}(2008){Di Francesco}, {Johnstone}, {Kirk},
  {MacKenzie}, \& {Ledwosinska}}]{diFrancesco08}
{Di Francesco}, J., {Johnstone}, D., {Kirk}, H., {MacKenzie}, T., \&
  {Ledwosinska}, E. 2008, \apjs, 175, 277

\bibitem[{{Dickman} \& {Herbst}(1990)}]{Dickman90}
{Dickman}, R.~L. \& {Herbst}, W. 1990, \apj, 357, 531

\bibitem[{{Dowell} {et~al.}(2003){Dowell}, {Allen}, {Babu}, {Freund},
  {Gardner}, {Groseth}, {Jhabvala}, {Kovacs}, {Lis}, {Moseley}, {Phillips},
  {Silverberg}, {Voellmer}, \& {Yoshida}}]{dowell03}
{Dowell}, C.~D., {Allen}, C.~A., {Babu}, R.~S., {et~al.} 2003, Presented at the
  Society of Photo-Optical Instrumentation Engineers (SPIE) Conference, 4855,
  73

\bibitem[{{Dunham} {et~al.}(2008){Dunham}, {Crapsi}, {Evans}, {Bourke},
  {Huard}, {Myers}, \& {Kauffmann}}]{Dunham08}
{Dunham}, M.~M., {Crapsi}, A., {Evans}, II, N.~J., {et~al.} 2008, ArXiv
  e-prints, 806

\bibitem[{{Elias}(1978)}]{Elias78}
{Elias}, J.~H. 1978, \apj, 224, 453

\bibitem[{{Enoch} {et~al.}(2008){Enoch}, {Evans}, {Sargent}, \&
  {Glenn}}]{Enoch08}
{Enoch}, M.~L., {Evans}, N.~J., {Sargent}, A.~I., \& {Glenn}, J. 2008, \apj,
  {submitted}

\bibitem[{{Evans}(1999)}]{Evans99}
{Evans}, N.~J. 1999, \araa, 37, 311

\bibitem[{{Evans} {et~al.}(2003){Evans}, {Allen}, {Blake}, {Boogert}, {Bourke},
  {Harvey}, {Kessler}, {Koerner}, {Lee}, {Mundy}, {Myers}, {Padgett},
  {Pontoppidan}, {Sargent}, {Stapelfeldt}, {van Dishoeck}, {Young}, \&
  {Young}}]{Evans03}
{Evans}, N.~J., {Allen}, L.~E., {Blake}, G.~A., {et~al.} 2003, \pasp, 115, 965

\bibitem[{{Evans} {et~al.}(2007){Evans}, {Harvey}, {Dunham}, {Huard}, {Mundy},
  {Lai}, {Chapman}, {Brooke}, {Enoch}, \& {Stapelfeldt}}]{Evans07}
{Evans}, N.~J., {Harvey}, P.~M., {Dunham}, M.~M., {et~al.} 2007, c2d delivery
  document

\bibitem[{{Evans} {et~al.}(2009){Evans}, {Dunham}, {J{\o}rgensen}, {Enoch},
  {Mer{\'{\i}}n}, {van Dishoeck}, {Alcal{\'a}}, {Myers}, {Stapelfeldt},
  {Huard}, {Allen}, {Harvey}, {van Kempen}, {Blake}, {Koerner}, {Mundy},
  {Padgett}, \& {Sargent}}]{Evans08}
{Evans}, II, N.~J., {Dunham}, M.~M., {J{\o}rgensen}, J.~K., {et~al.} 2009, ApJ
  in press

\bibitem[{{Greene} \& {Meyer}(1995)}]{Greene95}
{Greene}, T.~P. \& {Meyer}, M.~R. 1995, \apj, 450, 233

\bibitem[{{Greene} {et~al.}(1994){Greene}, {Wilking}, {Andre}, {Young}, \&
  {Lada}}]{Greene94}
{Greene}, T.~P., {Wilking}, B.~A., {Andre}, P., {Young}, E.~T., \& {Lada},
  C.~J. 1994, \apj, 434, 614

\bibitem[{{Hogerheijde} {et~al.}(1997){Hogerheijde}, {van Dishoeck}, {Blake},
  \& {van Langevelde}}]{Hogerheijde97}
{Hogerheijde}, M.~R., {van Dishoeck}, E.~F., {Blake}, G.~A., \& {van
  Langevelde}, H.~J. 1997, \apj, 489, 293

\bibitem[{{Jensen} {et~al.}(1996){Jensen}, {Mathieu}, \& {Fuller}}]{Jensen96}
{Jensen}, E.~L.~N., {Mathieu}, R.~D., \& {Fuller}, G.~A. 1996, \apj, 458, 312

\bibitem[{{Johnstone} {et~al.}(2001){Johnstone}, {Fich}, {Mitchell}, \&
  {Moriarty-Schieven}}]{Johnstone01}
{Johnstone}, D., {Fich}, M., {Mitchell}, G.~F., \& {Moriarty-Schieven}, G.
  2001, \apj, 559, 307

\bibitem[{{Johnstone} {et~al.}(2000){Johnstone}, {Wilson}, {Moriarty-Schieven},
  {Joncas}, {Smith}, {Gregersen}, \& {Fich}}]{Johnstone00}
{Johnstone}, D., {Wilson}, C.~D., {Moriarty-Schieven}, G., {et~al.} 2000, \apj,
  545, 327

\bibitem[{{J{\o}rgensen}(2004)}]{Jorgensen04}
{J{\o}rgensen}, J.~K. 2004, \aap, 424, 589

\bibitem[{{J{\o}rgensen} {et~al.}(2007){J{\o}rgensen}, {Bourke}, {Myers}, {Di
  Francesco}, {van Dishoeck}, {Lee}, {Ohashi}, {Sch{\"o}ier}, {Takakuwa},
  {Wilner}, \& {Zhang}}]{Jorgensen07}
{J{\o}rgensen}, J.~K., {Bourke}, T.~L., {Myers}, P.~C., {et~al.} 2007, \apj,
  659, 479

\bibitem[{{J{\o}rgensen} {et~al.}(2008){J{\o}rgensen}, {Johnstone}, {Kirk},
  {Myers}, {Allen}, \& {Shirley}}]{Jorgensen08}
{J{\o}rgensen}, J.~K., {Johnstone}, D., {Kirk}, H., {et~al.} 2008, \apj, {in
  press}

\bibitem[{{J{\o}rgensen} {et~al.}(2006){J{\o}rgensen}, {Johnstone}, {van
  Dishoeck}, \& {Doty}}]{Jorgensen06}
{J{\o}rgensen}, J.~K., {Johnstone}, D., {van Dishoeck}, E.~F., \& {Doty}, S.~D.
  2006, \aap, 449, 609

\bibitem[{{J{\o}rgensen} {et~al.}(2002){J{\o}rgensen}, {Sch{\"o}ier}, \& {van
  Dishoeck}}]{Jorgensen02}
{J{\o}rgensen}, J.~K., {Sch{\"o}ier}, F.~L., \& {van Dishoeck}, E.~F. 2002,
  \aap, 389, 908

\bibitem[{{J{\o}rgensen} {et~al.}(2005){J{\o}rgensen}, {Sch{\"o}ier}, \& {van
  Dishoeck}}]{Jorgensen05}
{J{\o}rgensen}, J.~K., {Sch{\"o}ier}, F.~L., \& {van Dishoeck}, E.~F. 2005,
  \aap, 437, 501

\bibitem[{{Kirk} {et~al.}(2006){Kirk}, {Johnstone}, \& {Di Francesco}}]{Kirk06}
{Kirk}, H., {Johnstone}, D., \& {Di Francesco}, J. 2006, \apj, 646, 1009

\bibitem[{{Knude} \& {H{\o}g}(1998)}]{Knude98}
{Knude}, J. \& {H{\o}g}, E. 1998, \aap, 338, 897

\bibitem[{{Lada} \& {Wilking}(1984)}]{Lada84}
{Lada}, C.~J. \& {Wilking}, B.~A. 1984, \apj, 287, 610

\bibitem[{{Lahuis} {et~al.}(2006){Lahuis}, {van Dishoeck}, {Boogert},
  {Pontoppidan}, {Blake}, {Dullemond}, {Evans}, {Hogerheijde}, {J{\o}rgensen},
  {Kessler-Silacci}, \& {Knez}}]{Lahuis06}
{Lahuis}, F., {van Dishoeck}, E.~F., {Boogert}, A.~C.~A., {et~al.} 2006, \apjl,
  636, L145

\bibitem[{{Liseau} {et~al.}(1999){Liseau}, {White}, {Larsson}, {Sidher},
  {Olofsson}, {Kaas}, {Nordh}, {Caux}, {Lorenzetti}, {Molinari}, {Nisini}, \&
  {Sibille}}]{Liseau99}
{Liseau}, R., {White}, G.~J., {Larsson}, B., {et~al.} 1999, \aap, 344, 342

\bibitem[{{Loinard} {et~al.}(2008){Loinard}, {Torres}, {Mioduszewski}, \&
  {Rodr{\'{\i}}guez}}]{Loinard08}
{Loinard}, L., {Torres}, R.~M., {Mioduszewski}, A.~J., \& {Rodr{\'{\i}}guez},
  L.~F. 2008, \apjl, 675, L29

\bibitem[{{Lommen} {et~al.}(2008){Lommen}, {J{\o}rgensen}, {van Dishoeck}, \&
  {Crapsi}}]{Lommen08}
{Lommen}, D., {J{\o}rgensen}, J.~K., {van Dishoeck}, E.~F., \& {Crapsi}, A.
  2008, \aap, 481, 141

\bibitem[{{Looney} {et~al.}(2000){Looney}, {Mundy}, \& {Welch}}]{Looney00}
{Looney}, L.~W., {Mundy}, L.~G., \& {Welch}, W.~J. 2000, \apj, 529, 477

\bibitem[{{Loren}(1989)}]{Loren89}
{Loren}, R.~B. 1989, \apj, 338, 902

\bibitem[{{Loren} {et~al.}(1990){Loren}, {Wootten}, \& {Wilking}}]{Loren90}
{Loren}, R.~B., {Wootten}, A., \& {Wilking}, B.~A. 1990, \apj, 365, 269

\bibitem[{{Luhman} \& {Rieke}(1999)}]{Luhman99}
{Luhman}, K.~L. \& {Rieke}, G.~H. 1999, \apj, 525, 440

\bibitem[{{Maret} {et~al.}(2004){Maret}, {Ceccarelli}, {Caux}, {Tielens},
  {J{\"o}rgensen}, {van Dishoeck}, {Bacmann}, {Castets}, {Lefloch}, {Loinard},
  {Parise}, \& {Sch{\"o}ier}}]{Maret04}
{Maret}, S., {Ceccarelli}, C., {Caux}, E., {et~al.} 2004, \aap, 416, 577

\bibitem[{{Maret} {et~al.}(2005){Maret}, {Ceccarelli}, {Tielens}, {Caux},
  {Lefloch}, {Faure}, {Castets}, \& {Flower}}]{Maret05}
{Maret}, S., {Ceccarelli}, C., {Tielens}, A.~G.~G.~M., {et~al.} 2005, \aap,
  442, 527

\bibitem[{{Mezger} {et~al.}(1992){Mezger}, {Sievers}, {Zylka}, {Haslam},
  {Kreysa}, \& {Lemke}}]{Mezger92}
{Mezger}, P.~G., {Sievers}, A., {Zylka}, R., {et~al.} 1992, \aap, 265, 743

\bibitem[{{Motte} {et~al.}(1998){Motte}, {Andre}, \& {Neri}}]{Motte98}
{Motte}, F., {Andre}, P., \& {Neri}, R. 1998, \aap, 336, 150

\bibitem[{{Myers} \& {Ladd}(1993)}]{Myers93}
{Myers}, P.~C. \& {Ladd}, E.~F. 1993, \apjl, 413, L47

\bibitem[{{Nuernberger} {et~al.}(1998){Nuernberger}, {Brandner}, {Yorke}, \&
  {Zinnecker}}]{Nurnberger98}
{Nuernberger}, D., {Brandner}, W., {Yorke}, H.~W., \& {Zinnecker}, H. 1998,
  \aap, 330, 549

\bibitem[{{Ossenkopf} \& {Henning}(1994)}]{Ossenkopf94}
{Ossenkopf}, V. \& {Henning}, T. 1994, \aap, 291, 943

\bibitem[{{Padgett} {et~al.}(2008){Padgett}, {Rebull}, {Stapelfeldt},
  {Chapman}, {Lai}, {Mundy}, {Evans}, {Brooke}, {Cieza}, {Spiesman},
  {Noriega-Crespo}, {McCabe}, {Allen}, {Blake}, {Harvey}, {Huard},
  {J{\o}rgensen}, {Koerner}, {Myers}, {Sargent}, {Teuben}, {van Dishoeck},
  {Wahhaj}, \& {Young}}]{Padgett08}
{Padgett}, D.~L., {Rebull}, L.~M., {Stapelfeldt}, K.~R., {et~al.} 2008, \apj,
  672, 1013

\bibitem[{{Pontoppidan} {et~al.}(2005){Pontoppidan}, {Dullemond}, {van
  Dishoeck}, {Blake}, {Boogert}, {Evans}, {Kessler-Silacci}, \&
  {Lahuis}}]{Pontoppidan05}
{Pontoppidan}, K.~M., {Dullemond}, C.~P., {van Dishoeck}, E.~F., {et~al.} 2005,
  \apj, 622, 463

\bibitem[{{Pontoppidan} {et~al.}(2003){Pontoppidan}, {Fraser}, {Dartois},
  {Thi}, {van Dishoeck}, {Boogert}, {d'Hendecourt}, {Tielens}, \&
  {Bisschop}}]{Pontoppidan03}
{Pontoppidan}, K.~M., {Fraser}, H.~J., {Dartois}, E., {et~al.} 2003, \aap, 408,
  981

\bibitem[{{Rachford} {et~al.}(2002){Rachford}, {Snow}, {Tumlinson}, {Shull},
  {Blair}, {Ferlet}, {Friedman}, {Gry}, {Jenkins}, {Morton}, {Savage},
  {Sonnentrucker}, {Vidal-Madjar}, {Welty}, \& {York}}]{rachford02}
{Rachford}, B.~L., {Snow}, T.~P., {Tumlinson}, J., {et~al.} 2002, \apj, 577,
  221

\bibitem[{{Ridge} {et~al.}(2006){Ridge}, {Di Francesco}, {Kirk}, {Li},
  {Goodman}, {Alves}, {Arce}, {Borkin}, {Caselli}, {Foster}, {Heyer},
  {Johnstone}, {Kosslyn}, {Lombardi}, {Pineda}, {Schnee}, \&
  {Tafalla}}]{Ridge06}
{Ridge}, N.~A., {Di Francesco}, J., {Kirk}, H., {et~al.} 2006, \aj, 131, 2921

\bibitem[{{Robitaille} {et~al.}(2007){Robitaille}, {Whitney}, {Indebetouw}, \&
  {Wood}}]{Robitaille07}
{Robitaille}, T.~P., {Whitney}, B.~A., {Indebetouw}, R., \& {Wood}, K. 2007,
  \apjs, 169, 328

\bibitem[{{Robitaille} {et~al.}(2006){Robitaille}, {Whitney}, {Indebetouw},
  {Wood}, \& {Denzmore}}]{Robitaille06}
{Robitaille}, T.~P., {Whitney}, B.~A., {Indebetouw}, R., {Wood}, K., \&
  {Denzmore}, P. 2006, \apjs, 167, 256

\bibitem[{{Sch{\"o}ier} {et~al.}(2002){Sch{\"o}ier}, {J{\o}rgensen}, {van
  Dishoeck}, \& {Blake}}]{Schoeier02}
{Sch{\"o}ier}, F.~L., {J{\o}rgensen}, J.~K., {van Dishoeck}, E.~F., \& {Blake},
  G.~A. 2002, \aap, 390, 1001

\bibitem[{{Sch{\"o}ier} {et~al.}(2004){Sch{\"o}ier}, {J{\o}rgensen}, {van
  Dishoeck}, \& {Blake}}]{Schoeier04}
{Sch{\"o}ier}, F.~L., {J{\o}rgensen}, J.~K., {van Dishoeck}, E.~F., \& {Blake},
  G.~A. 2004, \aap, 418, 185

\bibitem[{{Shirley} {et~al.}(2000){Shirley}, {Evans}, {Rawlings}, \&
  {Gregersen}}]{Shirley00}
{Shirley}, Y.~L., {Evans}, N.~J., {Rawlings}, J.~M.~C., \& {Gregersen}, E.~M.
  2000, \apjs, 131, 249

\bibitem[{{Smith} {et~al.}(2003){Smith}, {Hills}, {Withington}, {Richer},
  {Leech}, {Williamson}, {Gibson}, {Dace}, {Ananthasubramanian}, {Barker},
  {Baldwin}, {Stevenson}, {Doherty}, {Molloy}, {Quy}, {Lush}, {Hales}, {Dent},
  {Pain}, {Wall}, {Hastings}, {Graham}, {Baillie}, {Laidlaw}, {Bennett},
  {Laidlaw}, {Duncan}, {Ellis}, {Redman}, {Wooff}, {Yeung}, {Fitzsimmons},
  {Avery}, {Derdall}, {Josephson}, {Anthony}, {Atwal}, {Chylek}, {Shutt},
  {Friberg}, {Rees}, {Philips}, {Kroug}, {Klapwijk}, \& {Zijlstra}}]{Smith00}
{Smith}, H., {Hills}, R.~E., {Withington}, S., {et~al.} 2003, Presented at the
  Society of Photo-Optical Instrumentation Engineers (SPIE) Conference, 4855,
  338

\bibitem[{{Stanke} {et~al.}(2006){Stanke}, {Smith}, {Gredel}, \&
  {Khanzadyan}}]{Stanke06}
{Stanke}, T., {Smith}, M.~D., {Gredel}, R., \& {Khanzadyan}, T. 2006, \aap,
  447, 609

\bibitem[{{Thi} {et~al.}(2004){Thi}, {van Zadelhoff}, \& {van
  Dishoeck}}]{Thi04}
{Thi}, W.-F., {van Zadelhoff}, G.-J., \& {van Dishoeck}, E.~F. 2004, \aap, 425,
  955

\bibitem[{{van der Tak} {et~al.}(2007){van der Tak}, {Black}, {Sch{\"o}ier},
  {Jansen}, \& {van Dishoeck}}]{vanderTak07}
{van der Tak}, F.~F.~S., {Black}, J.~H., {Sch{\"o}ier}, F.~L., {Jansen}, D.~J.,
  \& {van Dishoeck}, E.~F. 2007, \aap, 468, 627

\bibitem[{{Walawender} {et~al.}(2005){Walawender}, {Bally}, {Kirk}, \&
  {Johnstone}}]{Walawender05}
{Walawender}, J., {Bally}, J., {Kirk}, H., \& {Johnstone}, D. 2005, \aj, 130,
  1795

\bibitem[{{Whitney} {et~al.}(2003{\natexlab{a}}){Whitney}, {Wood}, {Bjorkman},
  \& {Cohen}}]{Whitney03b}
{Whitney}, B.~A., {Wood}, K., {Bjorkman}, J.~E., \& {Cohen}, M.
  2003{\natexlab{a}}, \apj, 598, 1079

\bibitem[{{Whitney} {et~al.}(2003{\natexlab{b}}){Whitney}, {Wood}, {Bjorkman},
  \& {Wolff}}]{Whitney03a}
{Whitney}, B.~A., {Wood}, K., {Bjorkman}, J.~E., \& {Wolff}, M.~J.
  2003{\natexlab{b}}, \apj, 591, 1049

\bibitem[{{Wilking} \& {Lada}(1983)}]{Wilking83}
{Wilking}, B.~A. \& {Lada}, C.~J. 1983, \apj, 274, 698

\bibitem[{{Wilking} {et~al.}(1989){Wilking}, {Lada}, \& {Young}}]{Wilking89}
{Wilking}, B.~A., {Lada}, C.~J., \& {Young}, E.~T. 1989, \apj, 340, 823

\bibitem[{{Wilson} \& {Rood}(1994)}]{Wilson94}
{Wilson}, T.~L. \& {Rood}, R. 1994, \araa, 32, 191

\bibitem[{{Wootten}(1989)}]{Wootten89}
{Wootten}, A. 1989, \apj, 337, 858

\bibitem[{{Young} {et~al.}(2004){Young}, {J{\o}rgensen}, {Shirley},
  {Kauffmann}, {Huard}, {Lai}, {Lee}, {Crapsi}, {Bourke}, {Dullemond},
  {Brooke}, {Porras}, {Spiesman}, {Allen}, {Blake}, {Evans}, {Harvey},
  {Koerner}, {Mundy}, {Myers}, {Padgett}, {Sargent}, {Stapelfeldt}, {van
  Dishoeck}, {Bertoldi}, {Chapman}, {Cieza}, {DeVries}, {Ridge}, \&
  {Wahhaj}}]{Young04}
{Young}, C.~H., {J{\o}rgensen}, J.~K., {Shirley}, Y.~L., {et~al.} 2004, \apjs,
  154, 396

\bibitem[{{Young} {et~al.}(2003){Young}, {Shirley}, {Evans}, \&
  {Rawlings}}]{Young03}
{Young}, C.~H., {Shirley}, Y.~L., {Evans}, II, N.~J., \& {Rawlings}, J.~M.~C.
  2003, \apjs, 145, 111

\bibitem[{{Young} {et~al.}(2006){Young}, {Enoch}, {Evans}, {Glenn}, {Sargent},
  {Huard}, {Aguirre}, {Golwala}, {Haig}, {Harvey}, {Laurent}, {Mauskopf}, \&
  {Sayers}}]{Young06}
{Young}, K.~E., {Enoch}, M.~L., {Evans}, II, N.~J., {et~al.} 2006, \apj, 644,
  326

\end{thebibliography}

\Online
\begin{table*}[!htp]
\caption{Source fluxes including
IRAC 1-4, MIPS 1-2, SHARC-II, SCUBA and 1.3 mm data. 
}
\tiny
\begin{tabular}{l c c c c c c c c c c c}
 \hline \hline 
Source  	&3.6$\mu$m$^e$&4.5$\mu$m$^e$&5.8$\mu$m$^e$&8.0$\mu$m$^e$&24$\mu$m$^e$&70$\mu$m&350$\mu$m&450$\mu$m&850$\mu$m&850$\mu$m&1300$\mu$m\\
                & IRAC I&IRAC II&IRACIII&IRAC IV& MIPS I&MIPS II &SHARC & SCUBA & SCUBA& SCUBA& \\
                &       &       &       &       &       &$\sim$10$''$& 9-25$''$$^b$& 10 $''$$^b$ &15$''$& 25$''$& 15$''$ \\
	        & mJy   &mJy    &mJy    &mJy    &mJy    &mJy     &mJy      &  mJy   &  mJy      &   mJy   &  mJy        \\ \hline
C2D-162527.6    & 1.15e1& 1.77e1& 3.00e1& 6.03e1& 8.51e1& -      &   -     & -        &  $<$3e1    & $<$3e1   & \\
GSS 26	        & 2.86e2K& 3.86e2K& 5.09e2& 6.17e2& 7.98e2& -      &         & 6.5e2      &  3.0e2$^b$  & 1.6e3    & 1.e2\\
CRBR 2315.8-1700& 4.53e1& 8.77e1& 1.61e2& 2.37e2& 5.16e2& -      &   -     &          & 2.1e2       & 7.5e2     & 8.5e1)\\
CRBR 2317.3-1925& 1.05e1& 1.15e1& 1.39e1& 2.05e1& 1.46e1& -      &         &          &1.5e2$^b$    & 1.7e2     &\\
VSSG 1	        & 7.57e2& 1.01e3& 9.67e2& 9.96e2& 9.48e2& -      &         & $<$ 6.7e2  &2.4e2$^b$    & 3.0e2     &5e1\\
GSS 30	        & 1.06e3& 2.27e3& 4.1e3 & 5.4e3 & -	& 7.83e4 &2.5e3$^a$ & 1.4e3     & 2.2e2$^b$   & 3.2e2    & 9e1\\
LFAM 1	        & 1.28e1& 1.82e1*& 3.78e1& 1.59e1*& 2.41e2*& 6.31e4 &         & 2.7e3     & 5.7e2$^b$   & 2.6e3    & 1.1e3\\
CRBR 2324.1-1619     & 9.37e0& 1.59e1& 2.17e1& 2.95e1& 4.47e2& -      &         &          & 8.4e2       & 6.1e3    &\\
VLA 1623        &  -	& 1.57e-1*&2.45e-1*& -	& 4.04e1& -      &4e5$^c$  &3.9e5$^c$ & 5.0e3      & 1.7e4   & 1.2e3\\
GY51            & 2.86e2& 3.71e2& 4.38e2& 4.66e2& 6.39e2*& -      &         &          & 3.3e2       & 1.7e3    & 2e1\\
CRBR 2339.1-2032          & 3.04e1&4.48e1 & 6.05e1& 8.28e1& 3.93e2& -      &         &          & 8e1        & 3e2     &3e2\\
WL 12	        & 2.39e2&7.44e2 & 1.61e3& 2.24e3&   -	& 7.0e3  &6.7e3$^a$ & 7.5e2      & 2.2e2$^b$   & 7.1e2     & 1.2e2\\
WL 2            & 7.92e1&1.11e2 & 1.28e2& 1.48e2& 4.46e2& -      &         &          & 4e1        & 1.4e2     & 3e1\\
LFAM 26         & 6.11e0&1.07e1 & 1.66e1& 2.96e1& 4.59e2& -      &         &          & 2.2e2       & 6.9e2     & 1e2\\
WL 17	        &  2.4e2&4.16e2 & 5.53e2& 6.95e2&2.79e3 & 6.8e3  &         & $<$8.7e2   & 1e2$^b$   & 6.1e2     & 7.5e1\\
Elias 29        &  -	& -	& 1.28e4&  -	& -	&$<$4.7e4&4.3e3$^a$ & 2.0e3     & 4.1e2$^b$   & 9.4e2     & 9.5e1\\
GY 224	        & 2.03e2&3.05e2K &3.58e2 & 3.67e2& 9.07e2&        &         &          & 1.3e2       & 3.7e2     & 5e1\\
 WL19           & 2.15e2&3.54e2K & 4.06e2& 3.28e2&2.23e2 & -      &         &          & 2.0e2       & 8.3e2     & 3e1\\
WL 20S	        & 6.53e-1*$^d$&2.97e1& 3.74e1*& 2.43e2&1.46e2* & 1.4e4 & 8.2e2$^a$& 4.0e2      & 1.8e2$^b$   & 6.0e2     & 4.7e1\\
IRS 37	        & 1.27e2& 2.06e2& 2.86e2& 2.68e2& 7.8e2 & 1.4e4  &         & $<$ 5.9e2  & 9.3e1$^b$    & 3.0e2     & $<$ 1e1\\
WL 3            & 9.86e1& 1.63e2& 2.08e2& 2.04e2& 3.25e2&$<$2.4e4&         & $<$ 5.8e2  & 1.2e2$^b$   & 4.4e2     & $<$6e1\\
IRS 42	        & 1.06e3& 1.63e3& 2.1e3 & 2.98e3& 3.45e3& 3.0e3  &1.5e3$^a$ &          & 1.7e2       & 7.3e2     & $<$ 6e1\\
WL 6	        & 4.67e2K& 9.25e2& 1.44e3& 1.73e3& 4.36e3& 3.5e3  &         & 1.4e2      &$<$3e1      & $<$3e1   & $<$ 2e1\\
GY 256	        & 1.09e1& 1.19e1& 1.38e0*& 8.58e-1*&2.86e1*& 3.5e3  &         &          &  $<$3e1    & $<$3e1   &\\
IRS 43	        & 6.29e2& 1.24e3& 1.79e3& 2.19e3& -	& 2.4e4  & 2.1e3$^b$& 1.8e3     &  2.5e2$^b$  & 2.7e3    & 8e1\\
IRS 44	        & 7.31e2& 1.83e3& 2.94e3& 2.32e3& -	& 2.5e4  &         &          & 3.6e2       & 1.2e3    & 6e1\\
Elias 32        & 1.87e2& 2.72e2& 3.82e2& 4.81e2& 7.12e2& 1.1e4  &         &          & 1.7e3      & 6.4e3    & $<$ 5e1\\
Elias 33        & 7.4e2 & 1.19e3& 1.58e3& 2.04e3& 1.72e3&$<$1.0e4&4.4e3$^a$ &          & 1.2e3      & 4.4e3    &$<$ 2e1\\
IRS 48	        & 1.41e3& 1.6e3 & 4.06e3& 6.0e3 & -	& 1.7e4  &         & $<$ 1.6e3 & 1.8e2$^b$   & 1.9e2     & 6e1\\
GY 312          & 1.58e1& 2.67e1& 3.82e1& 5.32e1& 4.61e2& 7.1e2  &         &          & $<$3e1     & $<$3e1   &\\
IRS 51	        & 7.52e2& 9.16e2& 1.00e3& 1.07e3& 2.73e3& 2.4e3  &1.4e3$^a$ &  $<$ 6.8e2 & 2.2e2       & 7.8e2     &1.1e2\\
C2D-162741.6    & 8.31e0& 1.39e1& 2.04e1& 2.86e1& 6.89e1& 5.3e2  &         &          & 1e2       & 3.5e2     &\\
C2D-162748.2    & 1.21e0& 2.21e0& 3.64e0& 1.41e1& 1.84e2&        &         &          & 5e1        & 6e1      &\\
IRS 54	        & 5.25e2K& 7.12e2& 9.31e2& 1.01e3& 3.56e3& 5.0e3  &         &          & 1e2       & 3.2e2     & 3e1\\
IRAS 16285      & 5.89e-1&2.55e0& 4.49e0K& 3.13e0& 9.01e1& 1.7e3  &         &          & 5e2       & 1.6e3    &\\
C2D-162857.9    & 9.73e0 &1.7e1 &2.57e1 & 3.83e1& 1.61e2& -      &         &          & 5e1        & 8.5e2     &\\
IRS 63	        & 5.89e2& 9.57e2& 1.25e3& 1.45e3& 3.23e3& 8.3e3  &6.7e3$^b$ & 4.7e3     & 1.0e3$^b$  & 2.4e3    & 3.7e2\\ \hline
\multicolumn{12}{c}{\textbf{Disks}}  \\ \hline
Haro 1-4        & 5.19e2& -     & 3.73e2& -     & 8.39e2& 5.6e2  &         &          &   - & -      &\\
DoAR 25	        & 3.67e2& 2.92e2& 2.99e2& 2.58e2&3.99e2 & 1.4e3  &         &          & 4.6e2$^b$   &5.2e2      & 2.8e2\\
OphE MM3        & 1.42e0&1.79e0 & 1.88e0*& 2.22e0& 1.41e0*& -      &1.6e3$^a$ & 5.4e2      & 1.6e2       & 3.6e2     & \\
SR 21           & 1.3e3 & 1.17e3& 1.28e3& 1.66e3& -     & 1.3e3  & 2.8e3$^b$& 1.9e3     & 4.0e2$^b$   &4.7e2      & 9.5e1\\
CRBR 2422.8-3423& 5.4e1 & 1.31e2& 1.91e2& 1.98e2& 1.34e3& 9.8e3  &         &          & 1.3e3      & 3.2e3    & 1.5e2\\
IRS 46	        & 1.72e2K& 2.71e2& 4.02e2& 4.11e2& 6.39e2*& -      &1.4e3$^a$ &          & 1.8e2       & 5.1e2     & 4.5e1\\
SR 9            & 7.84e2& 5.75e2& 4.65e2& 4.84e2& 1.04e3& -      &         &          &  $<$25$^b$&2.9e1& 1.5e1\\ 
2Mass 16282     & 2.32e0& 2.00e0& 1.39e0& 8.15e-1&1.58e0& -      &  9e1$^b$ & 8e1       & 5.8e1        & 2.7e2\\\hline 
\end{tabular}\\

$^a$ This paper\\
$^b$ 350/450 $\mu$m flux taken from \citet{Andrews07}. Fluxes obtained with SHARC-II are given either in a 9$''$ beam \citep{Andrews07} or in a 25$''$ radius (This work).  \\ 
$^c$ Data taken from \citet{Andre93}. \\
$^d$ two sources within 4$''$ with a total flux of 2e2 mJy\\
$^e$ IRAC and MIPS-24 Fluxes marked with $'*'$ are band-filled sources with a S/N$<$5 or sources with S/N$<$5. Such limits are caused either by saturation or a very low emission. A 'K' added to the flux indicates that the obtained flux was flagged as 'Complex', with multiple detection within 2-3$''$
\label{table:flux}
\end{table*}
\newpage

\begin{figure*}[!htp]
\begin{center}
\includegraphics[width=140pt]{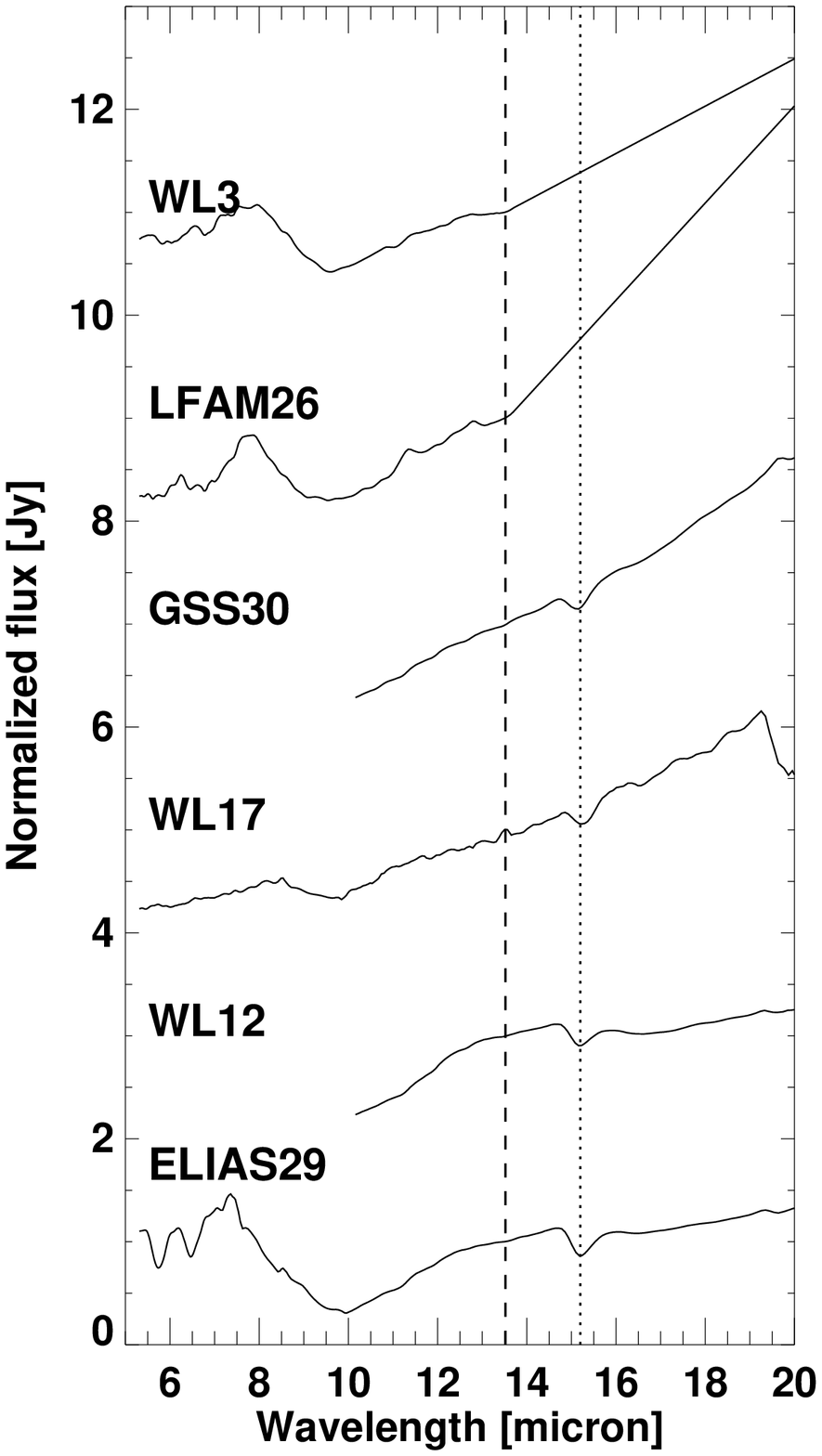}
\includegraphics[width=140pt]{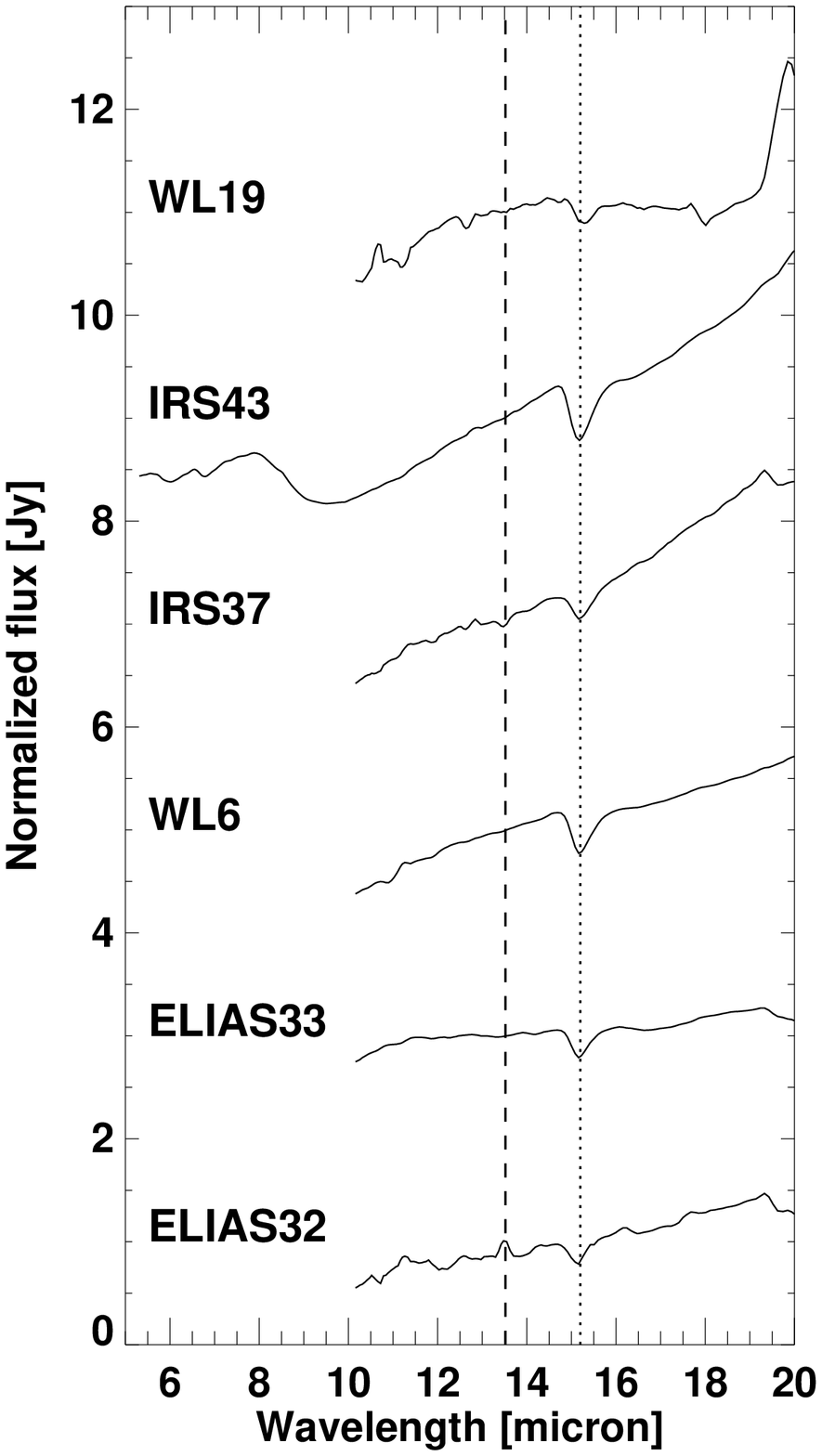}
\includegraphics[width=140pt]{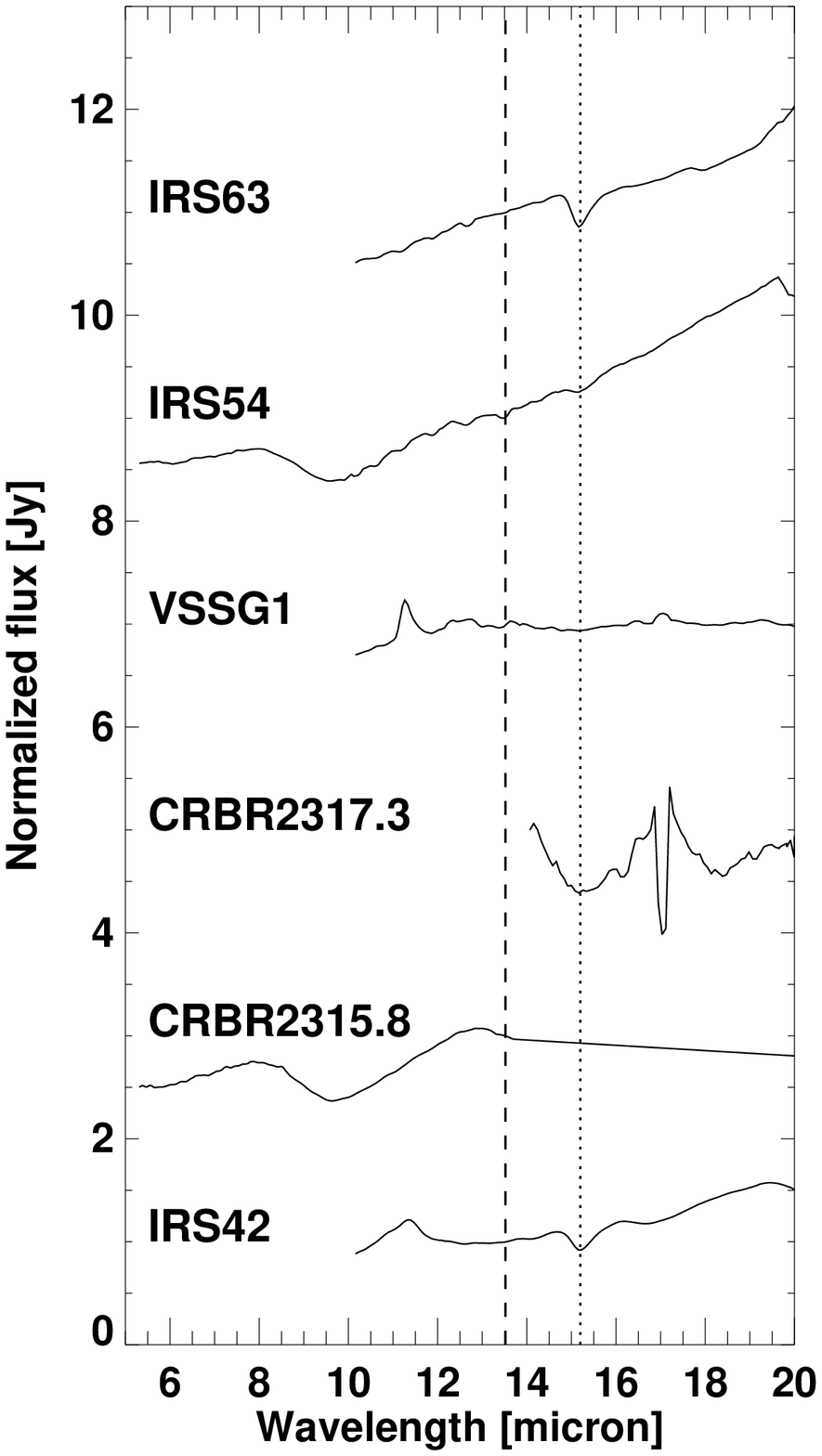}
\includegraphics[width=140pt]{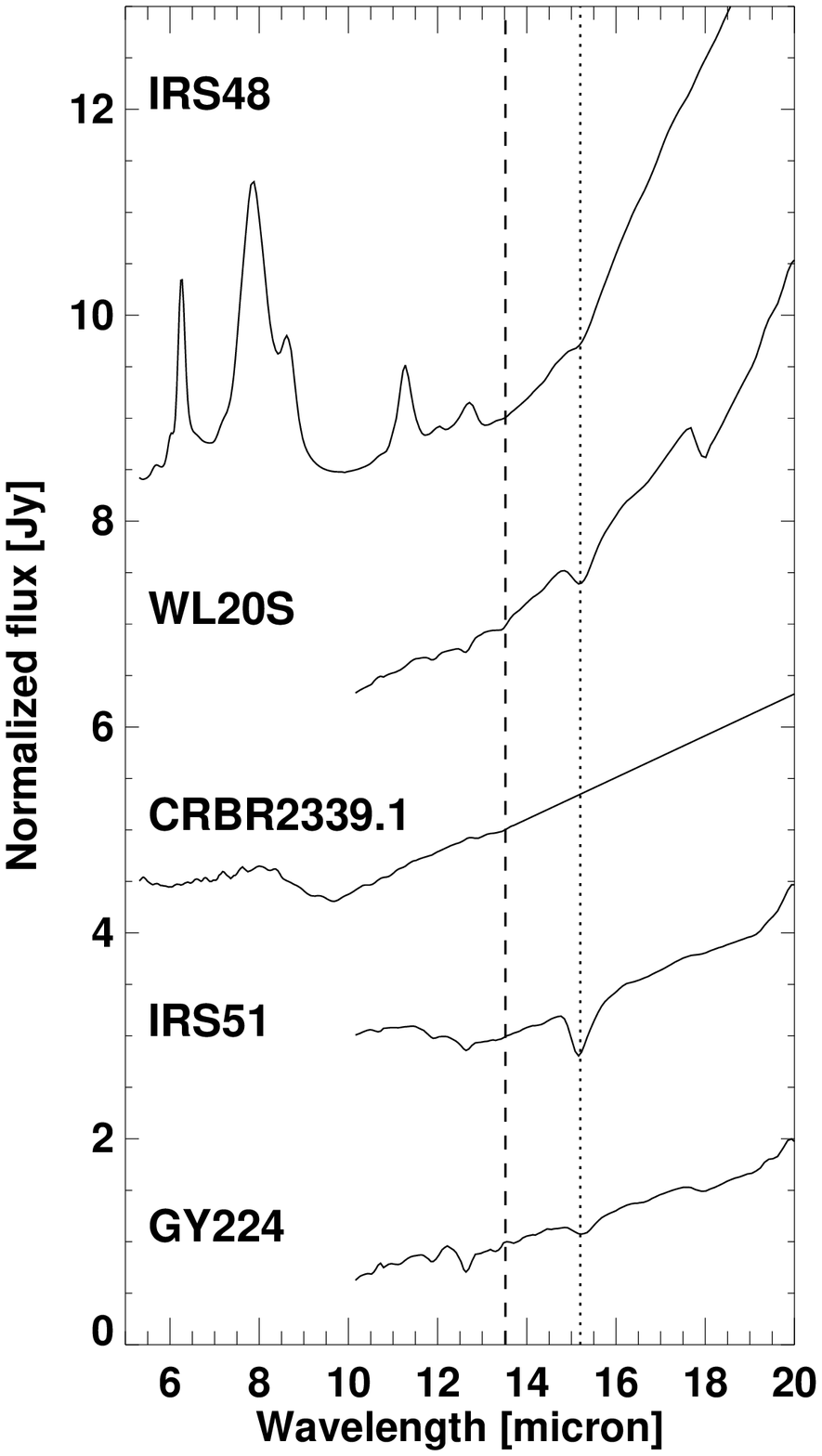}
\includegraphics[width=140pt]{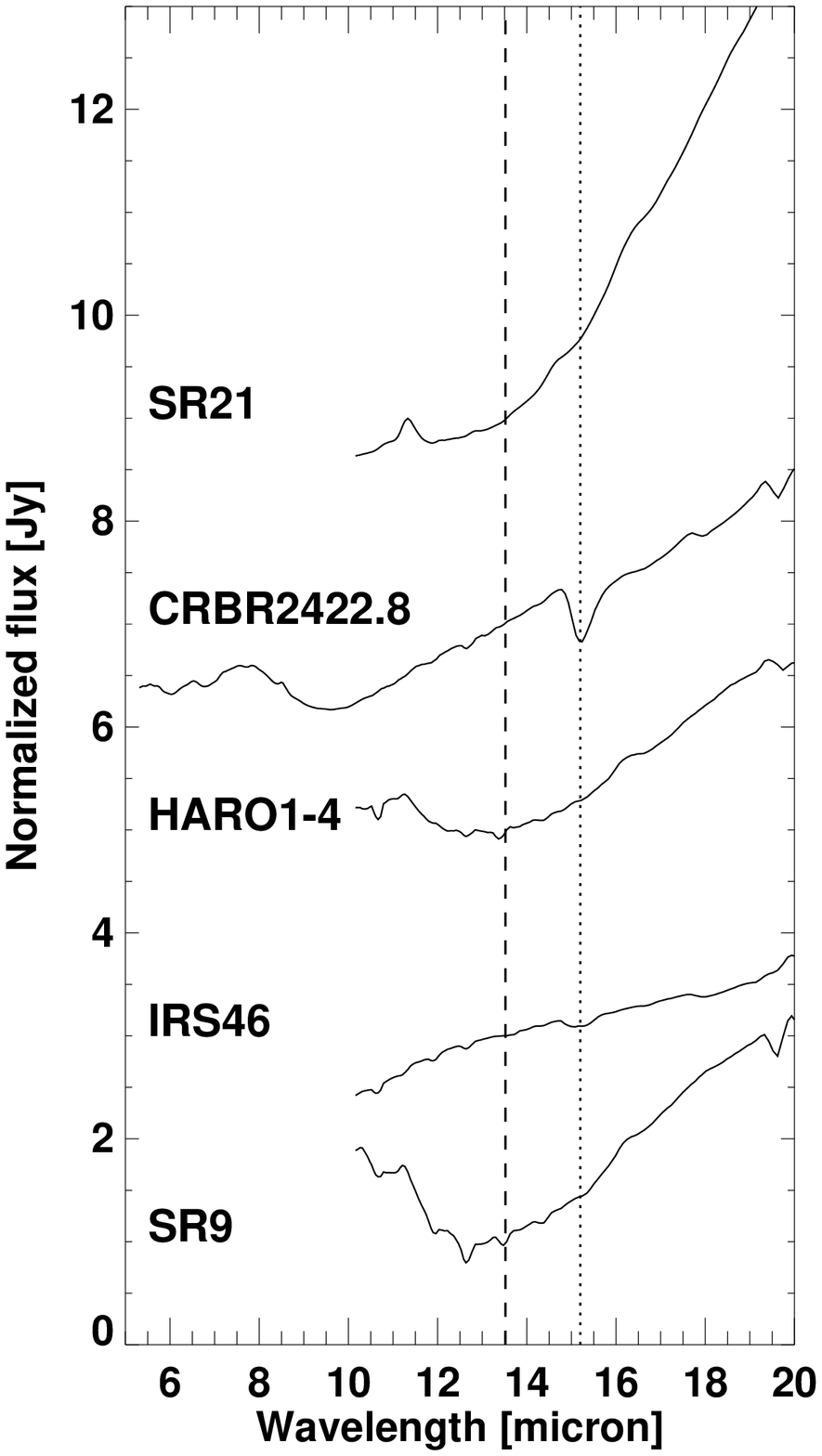}
\end{center}
\caption{The individual IRS spectra (where available). Each spectra is plotted from 5 to 20 $\mu$m. All spectra are normalised to their own emission at 13.5 micron, indicated with a vertical dashed line.}
\label{Fig:irs}
\end{figure*}

\newpage

\large
\textbf {Detailed source description}\\
\\
\normalsize
{\bf GSS 30 and LFAM 1.} These sources are both Stage 1 sources with a separation of 15$''$, but share a common circumbinary envelope on scales up to 40$''$. Strong HCO$^+$ is seen, peaking in between both IR positions. Although outflowing gas was detected, the bulk of the emission is believed to be quiescent.  \\
{\bf VLA 1623.} HCO$^+$ is found to be extended on scales of 30$''$, larger than that found for most sources. The lack of any emission at IR-wavelengths confirms this to be a Stage 0 source. \\
{\bf WL 12.} This source is a good example of a small embedded Stage 1 YSO. It is embedded in a small ridge of material, extending from a north-east to west direction, as can be clearly seen in both dust and C$^{18}$O mapping. This ridge contributes only little to the HCO$^+$ emission, which is concentrated at the source position. It is spatially resolved  and peaks to 1.4 K. This is much too brigth to be associated with a disk. \\
{\bf LFAM 26.} This Stage 1 is located at the tip of the Ophiuchus ridge that extends from LFAM 26 down to IRS 51. See $\S$ 5.2 and appendix B for a discussion on the environment of LFAM 26. The onset of this ridge is clearly seen, extending to the south-east direction, with much less material present in the north-west direction. Both HCO$^+$ and 850 $\mu$m peak at the source position and are bright. Even if the emission originating in the cloud is subtracted from the HCO$^+$ emission at the source position, it is too bright to be associated with a disk. This classification is confirmed by the very high $\alpha_{\rm{IR}}$ of 1.27.\\
{\bf WL 17.} This is a Stage 1 source, but evolving towards a Stage 2, which we classified 'late' Stage 1. The emission of HCO$^+$ is too bright to be associated with a disk. Since little to no extended emission is seen at scales of 40$''$, the source is probably unobscured by cloud material.\\
{\bf Elias 29.} This is a well-studied Stage 1 source \citep{Boogert02,Lommen08}. The analysis of HCO$^+$ and continuum, both from SCUBA and SHARC-II, confirm this source to be embedded. \\
{\bf IRS 37 and WL 3.} The Stage 1 sources IRS 37 and WL 3, separated by 20$''$ share a common envelope, with both C$^{18}$O and HCO$^+$ peaking on both source positions. This envelope is highly non-spherical. The proximity of the two sources is probably responsible for this. Most of the mass is located near IRS 37, although the amount of HCO$^+$ emission and concentration of the cores are equal. \\
{\bf WL 6 and GY~256.} The weak HCO$^+$ seen for these two sources with a separation of 12$''$ is brighter than that assumed to originate within a disk. In the dust maps, no emission is seen. WL~6 is probably a more evolved Stage 1 source with little envelope mass left. It is possible that GY 256 is a second component in a binary due to the rising spectra between 2 and 24 $\mu$m. However, since the flux from GY 256 at 6 and 8 $\mu$m is less than at 4.5, it is more likely that GY 256 is a Stage 2 source. GY 256 is classified as confused. \\
{\bf IRS 43.} This is a Stage 1 source, with a well-defined envelope that is bright in HCO$^+$ and C$^{18}$O. Both the SCUBA and HCO$^+$ maps show a slightly elongated envelope with a possible second component about 30$''$ to the east. This component is relatively much weaker in HCO$^+$ than in the dust. However, no IR source is detected at that position. At the position of IRS 43, dense outflowing gas is prominently detected in HCO$^+$ 4--3.\\
{\bf IRS 44 and IRS 46.} The Stage 2 source IRS 46 was studied by \citet{Lahuis06} in detail using IRS spectroscopy. Supplementary data from the JCMT confirmed that this source was a disk, reddened by a nearby envelope. The outflowing gas around the Stage 1 source IRS 44 is responsible for the reddening of IRS 46, as seen in Fig. \ref{Fig:harp2}.\\
{\bf Elias 32 and 33.} The Oph-B2 region dominates the environment around these two Stage 1 sources. At the position of Elias 33, HCO$^+$ and 850 $\mu$m continuum clearly peaks. Elias 32 is located in the middle of the Oph-B2 region. A more pronounced SCUBA and a less pronounced HCO$^+$ peak are seen within 10$''$ of Elias 32 and $\alpha_{\rm{IR}}$ is in both cases near 0.0. \\
{\bf IRS 54.} Extended emission was found for SCUBA, HCO$^+$ and C$^{18}$O. In all three cases, the emission is weak. IRS 54 is located outside of the main clouds. Classified as a late Stage 1, this source has accreted most of its envelope mass.\\
{\bf IRAS 16285-2355.} This is the only non-late stage 1 source that is located outside of the Ophiuchus ridge, Oph-B2 or Oph-A region. All emission in HCO$^+$, C$^{18}$O and 850 $\mu$m continuum is associated with the source in a spherical envelope.\\
{\bf IRS 63.} This stage 1 source consists of a large disk with little envelope material. The observed HCO$^+$ intensity of 1.2 K km s$^{-1}$ confirms that not all envelope material has accreted onto the disk, as suggested by \citet{Lommen08}.\\

{\bf CRBR 2324.1-1619.} Due to the location of this source, within 2$'$ of both VLA 1623 and LFAM 1/GSS 30, and located in between these bright sources, large amounts of material associated with the Oph A core are present, greatly confusing this source. However, the HCO$^+$, C$^{18}$O or SCUBA emission do not peak at the source position. The spectrum is sharply rising with $\alpha_{\rm IR}$=0.87. However, this is mainly due to the very high flux seen at 24 $\mu$m. \\
{\bf GY 51.} This source was not observed in HCO$^+$. The SCUBA map shows no peak at the position, but bright extended emission in the direction of the Oph-A core. The source is classified as confused, but due to the $\alpha_{\rm{IR}}$ of 0.05, this source is likely a background disk, heavily extincted by the Oph-A cloud material. \\
{\bf WL 19.} In the continuum and molecular emission maps, a large amount of material is found, centered 20$''$ north of WL 19. This material, probably associated with a pre-stellar core, is responsible for obscuring WL 19. Since the $\alpha_{\rm{IR}}$ = -0.43 this source is not embedded. \\
{\bf IRS 42.} Most of the emission from IRS 42, in both 850 $\mu$m and HCO$^+$  originates within the Ophiuchus ridge, near the core of IRS 43 and the location of CRBR 2422.8-3423.2. This source is classified as confused. However, with a $\alpha_{\rm{IR}}$=-0.03 it is more likely that this source is a disk, probably edge-on. The less likely possibility is that this source is a back-ground source.\\

{\bf SSTc2d J162527.6-243648.} This source, located far from the main L~1688 regions, is a disk. There is no detected SCUBA or HCO$^+$ emission. The rising IR spectrum of 0.36, combined with the bolometric temperature of 1051~K, is a strong indication that this source is an edge-on Stage 2 source. \\
 
{\bf GSS 26.} Extended SCUBA emission is seen near the source but is
likely mostly associated with nearby cloud material. If this cloud
emission is subtracted, a small unresolved core remains. The falling
$\alpha_{\rm{IR}}$ of $-0.46$ is the main reason that this source was
classified as Stage 2.\\ {\bf CRBR 2315.8-1700.} Strong HCO$^+$
emission and a strongly rising spectrum ($\alpha_{\rm{IR}}$=0.69) are
seen for this source. The HCO$^+$ map however suggests that this
source is reddened by the large amount of material present in Oph A.
An unresolved source is seen in the 850 $\mu$m data, but that core is
not apparent in the HCO$^+$ map. No variation was found in the HCO$^+$
4--3 at a level of 0.2 K km s$^{-1}$.  Thus, this source is likely an
disk, in line of sight of the cloud
material. \\ {\bf CRBR 2317.3-1925.} CRBR 2317.3-1925 was observed
with APEX-2a in HCO$^+$ and was found to be reasonably strong with a
peak temperature of 1.5 K. However, the source does not have a rising
spectrum and the SCUBA core seen at the source position is very
weak. In addition, the SCUBA emission is extended to the
south-east. This source is classified as Stage 2. \\ {\bf VSSG 1.}
Similar to CRBR 2317.3-1925, VSSG 1 does not have a rising
spectrum. Unlike, CRBR 2317.3-1925, however, no HCO$^+$ was found with
APEX-2a. 850 $\mu$m emission is detected, but unresolved. The very
high $T_{\rm{bol}}$, caused by sharply falling spectra long-ward of 24
$\mu$m, confirms this source to be a Stage 2 source, most likely
edge-on.\\ {\bf CRBR 2339.1-2032.} This Stage 2 source has unresolved
emission at 850 $\mu$m. Although HCO$^+$ was detected, the
contribution is weak, only just above our cut-off for Stage 1, and
does not seem associated with the source but with a patch of cloud
material about 30$''$ to the north-east of CRBR 2339.1-2032.  \\ {\bf
WL 2.} No HCO$^+$ is detected associated with the source. In the dust,
a small enhancement is seen at the source position, but it is not
brighter than 0.15 Jy beam$^{-1}$. All HCO$^+$ seen in the JCMT map is
associated with a small core located between Oph-A and Oph-C. Combined
with $\alpha_{\rm{IR}}=0.02$ and $T_{\rm{bol}}$=573 K, the source is
classified as a disk, probably edge-on as well as extincted by the
cloud material. \\ {\bf GY 224.} This source, located south of the Oph
ridge, shows no rising spectrum and no HCO$^+$. Since no 850 $mu$m
emission was found down to the noise limits of the COMPLETE map, this
source is concluded to be a Stage 2.\\ {\bf WL 20S.} The emission seen
in the SCUBA map does not peak at the position of WL 20S. In the
SHARC-II map, a very weak (0.05 Jy beam$^{-1}$) and unresolved peak is
detected. In addition, no HCO$^+$ was detected down to the limit
adopted for disks. Its location on the Ophiuchus ridge confuses the
environment of this source. The very high $\alpha_{\rm{IR}}=2.7$
suggests that this source must be an edge-on disk, even if located
behind the Ophiuchus ridge. \\ {\bf IRS 48.} The disk source IRS 48 is
isolated, away from the main cores in Ophiuchus. A weak, unresolved
peak is seen in the SCUBA. The absence of any C$^{18}$O and HCO$^+$
confirms the classification to be Stage 2.\\ {\bf GY 312.} GY 312 is
not detected in the dust or gas mapping. In additon, no contribution
from large-scale material is seen. Its clearly rising IR spectrum
identifies this source as an edge-on Stage 2 source.\\ {\bf IRS 51.}
IRS 51 is located at the south end of the Ophiuchus ridge.  The ridge
is clearly seen in SCUBA as well as in C$^{18}$O and HCO$^+$ maps
around IRS 51. Although a clear peak is seen at IRS 51 in the dust
map, no peak was seen in the molecular line mapping at the position of
IRS 51. It is concluded that IRS 51 is a Stage 2 source with strong
disk emission, reddened by the Ophiuchus ridge.\\ {\bf SSTc2d
J162741.6-244645.} This stage 2 source, located south of the Ophiuchus
ridge, was not detected in either HCO$^+$ or SCUBA. The rising
spectrum suggests that this source must be edge-on since no cloud
material is nearby to redden the source. \\ {\bf SSTc2d
J162748.2-244225.} For this source, located east of the southern tip
of the Ophiuchus ridge, no SCUBA or HCO$^+$ emission was found. The
very high $\alpha_{\rm{IR}}$ of 1.98 suggests that this source must be
an edge-on disk.\\ {\bf SSTc2d J162857.9-244055.} This source is
located between the L~1688 and L~1689 regions. The combination of a
steeply rising spectrum and an absence of HCO$^+$ and 850 $\mu$m
continuum indicates that this source is a edge-on Stage 2 source.\\
{\bf Haro 1-4.} Located north of the main Ophiuchus clouds, this
source has no HCO$^+$ or 850 $\mu$m associated with it and clearly is
a disk.\\ {\bf DoAr 25.} Although a bright peak is seen in the 850
$\mu$m map, the $\alpha_{\rm{IR}}$ of -1.12 clearly confirms this to
be a disk source.\\ {\bf OphE MM3.} This source, located within the
Ophiuchus ridge, was first discovered in the mm-regime by \citet{Motte98}. Both the 850 $\mu$m continuum and HCO$^+$ show little material
associated with an envelope at this source. Near-IR observations
\citep{Brandner00} confirm it to be an edge-on disk.\\ {\bf SR 21.}
This source has a very high bolometric temperature of 1070~K. Although
a clear signal is seen in the 850 $\mu$m data, no emission is seen in
HCO$^+$. It has been confirmed as a cold disk with a large inner hole
by \cite{Brown07}. \\ {\bf CRBR 2422.8-3423.} CRBR 2422.8-3423 is
located along the Ophiuchus ridge. From near-IR imaging
\citep{Brandner00,Pontoppidan05} it is known that CRBR 2422.8-3423 is
an edge-on disk and located in line of sight of the IRS 43 envelope. Even at
far-IR wavelengths, emission from the Ophiuchus ridge rather than the source
dominates. HCO$^+$ mapping confirms this.\\ {\bf SR 9.} This source is
not detected at 850 $\mu$m. Combined with $\alpha_{\rm{IR}}<-1$,
it is concluded that SR 9 is a Stage 2 source. \\ {\bf 2MASS 16282.}
Although 2MASS 16282 has a very low bolometric temperature of 66~K, the
$\alpha_{\rm{IR}}<-1$ identifies this source as a disk. \\

\end{document}